\newif\ifhideproofs
\hideproofstrue
\hideproofsfalse
\newif\ifnotodos
\notodostrue
\newif\ifcr
\crtrue
\crfalse

\documentclass{ecai} 
\usepackage{booktabs,colortbl} %
\usepackage[ruled,linesnumbered]{algorithm2e} %

\SetAlFnt{\small}
\SetAlCapFnt{\small}
\SetAlCapNameFnt{\small}
\SetAlCapHSkip{0pt}
\IncMargin{-\parindent}
\usepackage{amsmath,dsfont,amsthm,amssymb}
\newcommand{\mypara}[1]{
  \smallskip
  \noindent \textbf{#1}
}

\newcommand{\contrstitle}[1]{
  \smallskip
  \noindent \textit{#1}
}

\usepackage[colorlinks]{hyperref}
\hypersetup{
  linkcolor=blue,        %
  citecolor=red,         %
}

\usepackage{latexsym}
\usepackage[inline]{enumitem}
\usepackage{graphicx}
\usepackage{color}
\usepackage{cleveref}
\usepackage{wrapfig}
\usepackage{ifthen}
\crefname{table}{Table}{Tables}
\crefname{figure}{Figure}{Figures}
\crefname{theorem}{Theorem}{Theorems}
\crefname{definition}{Definition}{Definitions}
\crefname{corollary}{Corollary}{Corollaries}
\crefname{observation}{Observation}{Observations}
\crefname{independentobservation}{Observation}{Observations}
\crefname{lemma}{Lemma}{Lemmas}
\crefname{example}{Example}{Examples}
\crefname{reduction}{Reduction}{Reductions}
\crefname{construction}{Construction}{Constructions}
\crefname{subsection}{Section}{Sections}
\crefname{section}{Section}{Sections}
\crefname{proposition}{Proposition}{Propositions}
\crefname{algorithm}{Algorithm}{Algorithms}
\crefname{drule}{Rule}{Rules}
\crefname{mainclaim}{Claim}{Claims}
\crefname{claim}{Claim}{Claims}
\crefname{appendix}{Appendix}{Appendix}
\crefname{equation}{Equation}{Equations}

\newcommand{\decprob}[3]{%

  \smallskip   
\begin{minipage}{0.94\linewidth}%
  {\textsc{#1}}\\
  \textbf{Input:} #2\\
  \textbf{Question:} #3
\end{minipage}%
\medskip

\par 
}

\def\rotateclockwise#1{
  \newdimen\xrw
  \pgfextractx{\xrw}{#1}
  \newdimen\yrw
  \pgfextracty{\yrw}{#1}
  \pgfpoint{\yrw}{-\xrw}
}

\def\rotatecounterclockwise#1{
  \newdimen\xrcw
  \pgfextractx{\xrcw}{#1}
  \newdimen\yrcw
  \pgfextracty{\yrcw}{#1}
  \pgfpoint{-\yrcw}{\xrcw}
}

\def\outsidespacerpgfclockwise#1#2#3{
  \pgfpointscale{#3}{
    \rotateclockwise{
      \pgfpointnormalised{
        \pgfpointdiff{#1}{#2}}}}
}

\def\outsidespacerpgfcounterclockwise#1#2#3{
  \pgfpointscale{#3}{
    \rotatecounterclockwise{
      \pgfpointnormalised{
        \pgfpointdiff{#1}{#2}}}}
}

\def\outsidepgfclockwise#1#2#3{
  \pgfpointadd{#2}{\outsidespacerpgfclockwise{#1}{#2}{#3}}
}

\def\outsidepgfcounterclockwise#1#2#3{
  \pgfpointadd{#2}{\outsidespacerpgfcounterclockwise{#1}{#2}{#3}}
}

\def\outside#1#2#3{
  ($ (#2) ! #3 ! -90 : (#1) $)
}

\def\cornerpgf#1#2#3#4{
  \pgfextra{
    \pgfmathanglebetweenpoints{#2}{\outsidepgfcounterclockwise{#1}{#2}{#4}}
    \let\anglea\pgfmathresult
    \let\startangle\pgfmathresult

    \pgfmathanglebetweenpoints{#2}{\outsidepgfclockwise{#3}{#2}{#4}}
    \pgfmathparse{\pgfmathresult - \anglea}
    \pgfmathroundto{\pgfmathresult}
    \let\arcangle\pgfmathresult
    \ifthenelse{180=\arcangle \or 180<\arcangle}{
      \pgfmathparse{-360 + \arcangle}}{
      \pgfmathparse{\arcangle}}
    \let\deltaangle\pgfmathresult

    \newdimen\x
    \pgfextractx{\x}{\outsidepgfcounterclockwise{#1}{#2}{#4}}
    \newdimen\y
    \pgfextracty{\y}{\outsidepgfcounterclockwise{#1}{#2}{#4}}
  }
  -- (\x,\y) arc [start angle=\startangle, delta angle=\deltaangle, radius=#4]
}

\def\corner#1#2#3#4{
  \cornerpgf{\pgfpointanchor{#1}{center}}{\pgfpointanchor{#2}{center}}{\pgfpointanchor{#3}{center}}{#4}
}

\def\hedgeiii#1#2#3#4{
  \outside{#1}{#2}{#4} \corner{#1}{#2}{#3}{#4} \corner{#2}{#3}{#1}{#4} \corner{#3}{#1}{#2}{#4} -- cycle
}

\def\hedgem#1#2#3#4{

  \outside{#1}{#2}{#4}
  \pgfextra{
    \def\hgnodea{#1}
    \def\hgnodeb{#2}
  }
  foreach \c in {#3} {
    \corner{\hgnodea}{\hgnodeb}{\c}{#4}
    \pgfextra{
      \global\let\hgnodea\hgnodeb
      \global\let\hgnodeb\c
    }
  }
  \corner{\hgnodea}{\hgnodeb}{#1}{#4}
  \corner{\hgnodeb}{#1}{#2}{#4}
  -- cycle
}

\def\hedgeii#1#2#3{
  \hedgem{#1}{#2}{}{#3}
}

\def\hedgei#1#2{
  (#1) circle [radius = #2]
}
 
\usepackage{paralist,mathtools}
\ifnotodos
  \usepackage[textsize=tiny,colorinlistoftodos,textwidth=1.5cm,linecolor=green!70!black, backgroundcolor=green!10, bordercolor=black, disable]{todonotes}
\else
  \usepackage[textsize=tiny,colorinlistoftodos,textwidth=1.5cm,linecolor=green!70!black, backgroundcolor=green!10, bordercolor=black]{todonotes}
\fi
\usepackage{tikz}
\usetikzlibrary{graphs,decorations,arrows,petri,topaths,backgrounds,shapes,positioning,fit,calc,decorations.pathreplacing,patterns,intersections,decorations.pathmorphing,matrix}
\pgfdeclarelayer{bg}    %
\pgfdeclarelayer{fg}    %
\pgfsetlayers{bg,main,fg}  %
\tikzstyle{mainS}=[draw, circle, scale=0.75, inner sep=1.8pt]
\tikzstyle{hiddenV}=[opacity=0, scale=0.75, inner sep=1.8pt]
\tikzstyle{firstagent}=[mainS, draw=red, fill = red]
\tikzstyle{secondagent}=[mainS, draw=blue, fill = blue]
\tikzstyle{thirdagent}=[mainS, draw=orange, fill = orange]

\tikzstyle{firstagentedge} = [color=red] 
\tikzstyle{secondagentedge} = [color=blue] 
\tikzstyle{thirdagentedge} = [color=orange]

\tikzstyle{bigvertex} = [draw, circle, scale=0.75, inner sep=3.5pt, draw=black,fill=white]

\tikzstyle{vcvertex} = [mainS, draw=black, fill =black,]

\tikzstyle{mainHedge}=[opacity=.15, fill opacity = .15]

\tikzstyle{multiedgeA}=[white, line width=1pt]
\tikzstyle{multiedgeB}=[black, line width=3pt]

\tikzstyle{firstagentA}=[pattern color = red!75!white, draw=none, pattern=north west lines]
\tikzstyle{secondagentA}=[pattern color = blue!35!white, draw=none, pattern=grid]
\tikzstyle{thirdagentA}=[fill = orange!15!white, draw=none]

\newtheorem{theorem}{Theorem}
\newtheorem{mainclaim}{Claim}[theorem]
\newtheorem{claim}{Claim}[subsection] 
\newtheorem{lemma}{Lemma}

\newtheorem{definition}{Definition}
\newtheorem{example}{Example}
\newtheorem{observation}{Observation}[subsection]

\newtheorem{remark}{Remark}

\allowdisplaybreaks

\newcommand{\todoR}[1]{\todo[inline, linecolor=red!70!black, backgroundcolor=red!30]{#1}}

\newcommand{\todoSinline}[1]{\todo[inline, linecolor=red!70!black, backgroundcolor=yellow!10]{S: #1}}

\newcommand{\ir}{individually rational}

\newcommand{\myemph}[1]{{\color{red!40!black}{\emph{#1}}}}

\newcommand{\PCOG}{\textsc{PCOG}}
\newcommand{\PCOGlong}{\textsc{Partitioned Combinatorial Optimization Game}}
\newcommand{\COG}{\textsc{COG}}
\newcommand{\COGlong}{\textsc{Combinatorial Optimization Game}}

\newcommand{\DS}{\textsc{Dominating Set}\xspace}
\newcommand{\VCM}{\textsc{Vertex Cover Membership}\xspace}

\newcommand{\VC}{\textsc{Vertex Cover}\xspace}

\newcommand{\minDS}{\textsc{MinDS}}
\newcommand{\minVC}{\textsc{MinVC}}
\newcommand{\minST}{\textsc{MinST}}
\newcommand{\maxMatching}{\textsc{MaxMatch}}
\newcommand{\maxbMatching}{\textsc{Max$b$Match}}

\newcommand{\minDSlong}{\textsc{Minimum Dominating Set}}
\newcommand{\minVClong}{\textsc{Minimum Vertex Cover}}
\newcommand{\minSTlong}{\textsc{Minimum Cost Spanning Tree}}
\newcommand{\maxMatchinglong}{\textsc{Maximum Matching}}

\newcommand{\DSab}{\minDS}
\newcommand{\VCab}{\minVC}
\newcommand{\Mab}{\maxMatching}
\newcommand{\STab}{\minST}

\newcommand{\stable}{core-stable}

\newcommand{\posutil}{\ensuremath{\nu}}
\newcommand{\util}{\cost}
\newcommand{\cost}{\ensuremath{c}}

\newcommand{\weight}{\ensuremath{\omega}}

\newcommand{\players}{\ensuremath{\mathsf{N}}}
\newcommand{\pset}{\ensuremath{S}}
\newcommand{\costtxt}{cost}

\newcommand{\thetaC}{\ensuremath{\Theta_2^{\text{P}}}}
\newcommand{\baseDP}{DP}
\newcommand{\DP}{\text{\normalfont{\baseDP}}}
\newcommand{\NP}{\text{\normalfont{NP}}}
\newcommand{\FP}{\text{\normalfont{FP}}}
\newcommand{\FNP}{\text{\normalfont{FNP}}}
\newcommand{\FPNPlog}{\ensuremath{\FP^{\NP[\log]}}}

\newcommand{\coNP}{\text{\normalfont{coNP}}}
\newcommand{\PP}{\text{\normalfont{P}}}
\newcommand{\NPNP}{\ensuremath{\NP^{\NP}}}
\newcommand{\FPT}{\text{\normalfont{FPT}}}

\newcommand{\wrt}{wrt.\ }

\newcommand{\thetaCb}{$\boldsymbol{\Theta_2^{\text{P}}}$}

\newcommand{\DPb}{\textbf{DP}}

\newcommand{\coNPb}{\textbf{{coNP}}}

\newcommand{\FPTb}{\textbf{{FPT}}}

\newcommand{\sigmaC}{\ensuremath{\Sigma_2^{\text{P}}}}
\newcommand{\nbOrg}{\ensuremath{\hat{n}}}
\newcommand{\coln}{\nbOrg}
\newcommand{\Orgsize}{\ensuremath{c_{\text{max}}}}

\newcommand{\Parti}{Partitioned}

\newcommand{\parProb}{P}

\newcommand{\allocation}{allocation}

\newcommand{\preimputation}{pre-imputation}

\newcommand{\alloc}{\ensuremath{\boldsymbol{\alpha}}}
\newcommand{\allocel}[1]{\ensuremath{\alpha_{#1}}}

\newcommand{\trueT}{\ensuremath{\mathsf{T}}}
\newcommand{\falseF}{\ensuremath{\mathsf{F}}}

\newcommand{\supernode}{\ensuremath{\hat{s}}}
\newcommand{\supernodetext}{supply node}
\newcommand{\DSgame}{Dominating Set Game}
\newcommand{\VCgame}{Vertex Cover Game}

\newcommand{\Mgame}{Matching Game}
\newcommand{\STgame}{Spanning Tree Game}
\newcommand{\ParDSgame}{\Parti\ \DSgame}
\newcommand{\ParVCgame}{\Parti\ \VCgame}

\newcommand{\ParMgame}{\Parti\ \Mgame}
\newcommand{\ParSTgame}{\Parti\ \STgame}

\newcommand{\minasgwosatmember}{\textsc{MinAsg 2-SAT Member}}

\newcommand{\abcog}{COG}
\newcommand{\VCgameshort}{\textsc{\abcog-\VCab}}
\newcommand{\DSgameshort}{\textsc{\abcog-\DSab}}
\newcommand{\Mgameshort}{\textsc{\abcog-\Mab}}
\newcommand{\STgameshort}{\textsc{\abcog-\STab}}
\newcommand{\ParVCgameshort}{\textsc{\parProb\VCgameshort}}
\newcommand{\ParDSgameshort}{\textsc{\parProb\DSgameshort}}
\newcommand{\ParMgameshort}{\textsc{\parProb\Mgameshort}}
\newcommand{\ParSTgameshort}{\textsc{\parProb\STgameshort}}

\newcommand{\probPVCCEhint}{\textsc{Hinted}-\probPVCCE}
\newcommand{\probPDSCEhint}{\textsc{Hinted}-\probPDSCE}

\newcommand{\probPVCCEhintshort}{\textsc{Hinted}-VC}

\newcommand{\probCE}{\textsc{CE}}
\newcommand{\probCV}{\textsc{CV}}
\newcommand{\prob}{\ensuremath{\Pi}}

\newcommand{\probCElong}{\textsc{Core Stability Existence}}
\newcommand{\probCVlong}{\textsc{Core Stability Verification}}

\newcommand{\probDSCE}{\textsc{\abcog-\DSab-\probCE}}

\newcommand{\probPDSCE}{\textsc{\ParDSgameshort-\probCE}}
\newcommand{\probPVCCE}{\textsc{\ParVCgameshort-\probCE}}
\newcommand{\probPMCE}{\textsc{\ParMgameshort-\probCE}}
\newcommand{\probPSTCE}{\textsc{\ParSTgameshort-\probCE}}

\newcommand{\probPDSCV}{\textsc{\ParDSgameshort-\probCV}}
\newcommand{\probPVCCV}{\textsc{\ParVCgameshort-\probCV}}
\newcommand{\probPMCV}{\textsc{\ParMgameshort-\probCV}}
\newcommand{\probPSTCV}{\textsc{\ParSTgameshort-\probCV}}

\newcommand{\minvc}{\ensuremath{k^*}}

\newcommand{\svar}[1]{\ensuremath{{x}_{#1}}}
\newcommand{\snegvar}[1]{\ensuremath{\overline{x}_{#1}}}

\newcommand{\BibTeX}{B\kern-.05em{\sc i\kern-.025em b}\kern-.08em\TeX}

\newcommand{\argmin}{\arg \min}
\newcommand{\argmax}{\arg \max}

\newcommand{\fmds}{fractional dominating set value}

\newcommand{\appsymb}{$\star$}

\usepackage{thm-restate}
\usepackage[title]{appendix}

\ifhideproofs
\usepackage{environ}
\NewEnviron{killcontents}{}

\fi

\usepackage{etoolbox} %
\newcommand{\toappendix}[1]{%
  \gappto{\appendixtext}{
    {#1}
   }
}

\newcommand{\toappendixalter}[3]{%
  #1
  \gappto{\appendixtext}{
    #2 {#3}
  }
}

\newcommand{\appendixproofwithstatement}[3]{%
  \gappto{\appendixtext}{
    \subsection{Proof of \cref{#1}}\label{proof:#1}
    #2
    #3
  }
}

\newcommand{\appendixproofwithstatementandsketch}[4]{%
#3
  \gappto{\appendixtext}{
    \subsection{Proof of \cref{#1}}\label{proof:#1}
    #2
    #4
  }
}

\newcommand{\appendixsection}[1]{%
  \gappto{\appendixtext}{
    \section{Additional material for Section~\ref{#1}}
    \label{appsec:#1}
  }
}

\begin{document}

\begin{frontmatter}

\paperid{8738}

\title{Partitioned Combinatorial Optimization Games}

\author[A]{\fnms{Jiehua}~\snm{Chen}\orcid{0000-0002-8163-1327}\footnote{Equal contribution.}
  }
\author[A]{\fnms{Christian}~\snm{Hatschka}\orcid{0000-0002-0881-8259}\footnotemark}
\author[A]{\fnms{Sofia}~\snm{Simola}\orcid{0000-0001-7941-0018}\footnotemark} 

\address[A]{TU Wien, Vienna, Austria}

\begin{abstract}
  We propose a class of cooperative games, %
  called \myemph{\PCOGlong}{s} (\myemph{\PCOG}s).
  The input of \PCOG\ consists of a set of agents and a combinatorial structure (typically a graph) with a fixed optimization goal on this structure (e.g., finding a minimum dominating set on a graph) such that the structure is divided among the agents. 
  The value of each coalition of agents is derived from the optimal solution for the part of the structure possessed by the coalition.
  We study two fundamental questions related to the \myemph{core}: \textsc{Core Stability Verification} and \textsc{Core Stability Existence}.
  We analyze the algorithmic complexity of both questions for four classic graph optimization tasks: 
 minimum vertex cover, minimum dominating set,  minimum spanning tree, and maximum matching. 
\end{abstract}

\end{frontmatter}

\todo{TODONOTES ARE ON}
\section{Introduction}
\todoR{R2: Em dashes aren't used consistently.}
\todoSinline{I think I fixed this but it makes sense to keep eyes open for this}
\todoR{R3: While the paper introduces the notion of individual rationality and discusses its computational implications, this is not discussed largely, which can also be seen as an opportunity for improvement.}
\todoR{R3: This reviewer perhaps desired more motivation, do we want to add some somehow?}
\todoR{R4: In the introduction, four games are mentioned. You could describe each of them in the context of PCOG. You have already provided one example for the vertex cover game.}
\todoSinline{I don't understand this.}
\myemph{\COGlong}s~(\myemph{\COG}s)~\cite{BHH2001ORgames,Deng2009COG} form a rich class of cooperative games with transferable utilities, defined on combinatorial structures such as graphs.
In these games, each agent (aka.\ player) owns a single unit of the structure—typically a vertex or an edge—and the value of each subset of agents (aka.\ coalition) is derived from the optimal solution of the combinatorial optimization problem restricted to the substructure induced by the agents in the coalition. 
This framework has yielded several influential game-theoretic models, including 
\emph{minimum vertex cover game} (where the agents are the edges)~\cite{deng1999algorithmic},
\emph{minimum dominating set game}~\cite{Velzen04}, 
\emph{minimum-cost spanning tree game}~\cite{claus1973cost},
\emph{maximum matching game}~\cite{deng1999algorithmic}, and
\emph{traveling salesperson game}~\cite{FK93CCG}, among others.
For example, in the {maximum matching game}, each agent is a vertex in an undirected graph such that
the utility of any coalition of agents is the size of a maximum matching in the subgraph induced by the coalition.

The restriction that each agent owns exactly one element, however, limits the applicability of \COG{s} to many contemporary collaborative scenarios where participants control multiple resources.
In international multi-organization collaborations such as multi-organizational scheduling~\cite{pascual2009introduction} and international kidney exchange programs~\cite{Biro19}, a single organization may contribute dozens or even hundreds of primitive elements—e.g., cluster servers or patient–donor pairs. 
These \emph{multi-ownership} settings violate the classical assumption, creating fresh challenges for fairness, stability, and computational tractability that the present work seeks to address.

To address the limitation of \COG{s}, we propose 
\myemph{\PCOGlong}s (\myemph{\PCOG}s) as a general framework extending \COG{s}. 
In a \PCOG, each agent owns a part of the given structure (e.g., a subset of vertices or edges), and the value of a coalition is determined by the optimal solution over the combined substructure possessed by its members.
Clearly, \COG\ is a special case of \PCOG\ %
in which either all agents own exactly one vertex or all agents each own exactly one edge. 
Moreover, since these coalition values are derived from the underlying combinatorial optimization problem, the input size of a \PCOG\ remains compact—polynomial in the size of the input graph—analogously to traditional \COG{s}.

We focus on \myemph{core stability}~\cite{Shapley1967}, a central solution concept in cooperative game theory that captures stable utility distributions where no coalition has incentive to deviate.
We investigate two fundamental computational problems: %
(1) \textbf{\probCVlong} (\textbf{\probCV}): Is a given allocation \stable\ (i.e., in the core)? 
(2) \textbf{\probCElong} (\textbf{\probCE}): Does a \stable\ allocation exist?
To gauge the computational effort required to address these problems,
we analyze four classical combinatorial optimization problems on graphs: \minVClong~(\minVC), \minDSlong~(\minDS), \minSTlong~(\minST), and \maxMatchinglong~(\maxMatching).
This selection allows us to systematically compare the computational impact of partitioning between NP-hard problems (the first two) and polynomial-time solvable problems (the latter two).

\paragraph{Main Contributions.}
We study a generalized framework for combinatorial optimization games that naturally accommodates scenarios involving multi-organization collaboration.
Our contributions are summarized as follows:
\begin{compactitem}[--]
  \item We propose the \PCOG\ framework and establish a \emph{sufficient} condition when the core stability always exists. 
  \item \textbf{\probCV:}
  We show that for the two \NP-hard optimization problems—\minDS\ and \minVC—the \probCV\ problem is \DP-complete; see \cref{subsec:complexity} for more information on the complexity class \DP.
  We also use a known result by~\citet{faigle1997complexity} for \COG\ to derive \coNP-completeness for \PCOG-\minST.
  Note that for \PCOG-\maxMatching, the \probCV\ problem is known to be \coNP-complete~\cite{Biro19}.

\item \textbf{\probCE:}
For \PCOG-\minVC\ and \PCOG-\minDS\ we establish that the \probCE\ is $\thetaC$-complete (see \cref{subsec:complexity} for the definition of~$\thetaC$).
While the $\thetaC$-hardness is relatively straightforward, 
the key step in the upper-bound argument is to adapt the classical ellipsoid method~\cite{GLS1988} for an auxiliary problem and show that a separation routine for the problem lies in \coNP.  
Our construction is inspired by the ellipsoid-based approach of \citet{biro2018stable} and \citet{Biro19}, who obtained a straightforward \emph{polynomial-time} algorithm for a more permissive variant of \PCOG-\maxMatching\ in which each agent controls at most two vertices.
In our setting, the underlying optimization problem is already $\NP$-hard, so the oracle must therefore be engineered more carefully.
We believe that the resulting technique can also be applied to other combinatorial optimization games.

A further interesting distinction emerges within the polynomial-time solvable optimization problems.
For \PCOG-\minST, we derive from the results for \COG~\cite{Bird76,Granot81} that a \stable\ allocation always exists and finding one can be done in polynomial time.
In contrast, we complement the \coNP-hardness for \PCOG-\maxMatching\ from the literature~\cite{Biro19} by showing \coNP-containment.

\item We investigate parameterized complexity through two structural parameters: number~\myemph{$\nbOrg$} of agents and maximum number~\myemph{$\Orgsize$} of vertices (or edges) own by any agent.
\end{compactitem}
In summary, our results demonstrate that the flexibility of allowing an agent to possess a larger portion of the combinatorial structure has a more significant computational impact on NP-hard optimization problems than on those that are polynomial-time solvable.

\newcommand{\gudcell}{\cellcolor{green}}
\newcommand{\midcell}{\cellcolor{yellow}}
\newcommand{\ezcell}{\cellcolor{red}}

\newcommand{\paramOrgNb}{\small \#agents~$\nbOrg$}
\newcommand{\paramMaxOrg}{\small max~ag-size~$\Orgsize$}

\newcommand{\DSCVcite}{[\textbf{T}\ref{lem:verifDS}]}
\newcommand{\DSCVparamcite}{[\textbf{T}\ref{lem:verifDS}]}

\newcommand{\DSCEcite}{[\textbf{T}\ref{thm:DSCEcont},\textbf{T}\ref{thm:DStheta}]}
\newcommand{\DSCEparamcite}{[\textbf{T}\ref{thm:DSCEcont},\textbf{T}\ref{thm:DStheta}]}
\newcommand{\DSCScite}{[\textbf{T}\ref{corr:FPNPlog},\ref{thm:DSFPNP}]}

\newcommand{\DSCENbOrgcite}{[\textbf{T}\ref{thm:DStheta}]}
\newcommand{\DSCSNbOrgcite}{[\textbf{T}\ref{corr:FPNPlog},\ref{thm:nbORGFPNPlog}]}

\newcommand{\DSCVOrgsizecite}{[\textbf{T}\ref{lem:verifDS}]}
\newcommand{\DSCEOrgsizecite}{[\textbf{T}\ref{thm:DSCEcont},\textbf{T}\ref{thm:DSthetaorgsize}]}
\newcommand{\DSCSOrgsizecite}{?}

\newcommand{\VCCVcite}{[\textbf{T}\ref{thm:VCverif}]}
\newcommand{\VCCVparamcite}{[\textbf{T}\ref{thm:VCverif}]}
\newcommand{\VCCEcite}{[\textbf{T}\ref{thm:VCCEthetacont},\textbf{T}\ref{thm:VCtheta}]}
\newcommand{\VCCEparamcite}{[\textbf{T}\ref{thm:VCCEthetacont},\textbf{T}\ref{thm:VCtheta}]}
\newcommand{\VCCScite}{[\textbf{T}\ref{corr:FPNPlog},\ref{thm:VCFPNP}]}

\newcommand{\VCCENbOrgcite}{[\textbf{T}\ref{thm:VCtheta}]}
\newcommand{\VCCSNbOrgcite}{[\textbf{T}\ref{corr:FPNPlog},\ref{thm:nbORGFPNPlog}]}

\newcommand{\VCCVOrgsizecite}{[\textbf{T}\ref{lem:orgsizeverifVC}]}
\newcommand{\VCCEOrgsizecite}{[\textbf{T}\ref{thm:VCCEthetacont},\textbf{T}\ref{thm:VCorgsize}]}
\newcommand{\VCCSOrgsizecite}{?}

\newcommand{\MSTCVcite}{\cite{faigle1997complexity}[\textbf{O}\ref{pro:MSTconpc}]}
\newcommand{\MSTCEcite}{\cite{Bird76,Granot81}[\textbf{C}\ref{cor:parSTalwaysY}]}
\newcommand{\MSTCScite}{\cite{Bird76,Granot81}[\textbf{C}\ref{cor:parSTalwaysY}]}

\newcommand{\MSTCVNbOrgcite}{[\textbf{P}\ref{thm:MSTMVerifk}]}
\newcommand{\MSTCSNbOrgcite}{\cite{Bird76,Granot81}[\textbf{C}~\ref{cor:parSTalwaysY}]}

\newcommand{\MSTCVOrgsizecite}{\cite{faigle1997complexity}[\textbf{O}\ref{pro:MSTconpc}]}
\newcommand{\MSTCSOrgsizecite}{--}

\newcommand{\MCVcite}{\cite{Biro19}}
\newcommand{\MCEcite}{\cite{biro2018stable,Biro19}[\textbf{P}\ref{lem:PMGcoNPcontain}]}
\newcommand{\MCScite}{\cite{biro2018stable,Biro19}}

\newcommand{\MCVNbOrgcite}{[\textbf{P}\ref{thm:MSTMVerifk}]}
\newcommand{\MCENbOrgcite}{[\textbf{P}\ref{thm:PMGCFkFPT}]}
\newcommand{\MCSNbOrgcite}{[\textbf{L}\ref{lem:PMGnbOrgFPT}]}

\newcommand{\own}[1]{\textbf{#1}}

\begin{table}[t]
  \centering 
  
  \caption{Overview of our (parameterized) complexity results. Our contributions are in bold.
    ``\PP$^*$'' means that the corresponding game always admits a \stable\ allocation, and finding one can be done in polynomial time. 
    \vspace{0.5cm}
  }

  \setlength{\tabcolsep}{2pt}
  \begin{tabular}{@{}l rl @{}c@{} rl }
    \toprule
    Opt.\ problem~$\prob$   & \multicolumn{2}{c}{\PCOG-$\prob$-\probCV} &
    & \multicolumn{2}{c}{\PCOG-\prob-\probCE} \\\midrule

    \minVC & \own{\DPb-{complete}} & \VCCVcite
    && \own{\thetaCb-{complete}} & \VCCEcite\\

    \;\;~~\paramOrgNb & \own{\DPb-{complete}} & \VCCVparamcite
    &&\own{\thetaCb-{complete}} &  \VCCEparamcite\\

    \;\;~~\paramMaxOrg &  \own{\DPb-{complete}} & \VCCVparamcite
    && \own{\thetaCb-{complete}} & \VCCEOrgsizecite \\[0.5ex]

    \minDS & \own{\DPb-{complete}} & \DSCVcite
    && \own{\thetaCb-{complete}} & \DSCEcite\\
    
    \;\;~~\paramOrgNb & \own{\DPb-{complete}} & \DSCVparamcite &
           &\own{\thetaCb-{complete}} & \DSCEparamcite \\

    \;\;~~\paramMaxOrg &  \own{\DPb-{complete}} & \DSCVOrgsizecite
    && \own{\thetaCb-{complete}} & \DSCEOrgsizecite \\[0.5ex]

    \minST & \coNP-complete & \MSTCVcite
    && \PP$^*$ & \MSTCEcite\\

    \;\;~~\paramOrgNb & \own{\FPTb} & \MSTCVNbOrgcite
    && -- & --\\

    \;\;~~\paramMaxOrg &  \coNP-complete & \MSTCVOrgsizecite
    && -- & -- \\[0.5ex]

    \maxMatching & \coNP-complete & \MCVcite
    && \own{\coNPb-complete} & \MCEcite\\

    \;\;~~\paramOrgNb & \own{\FPTb} & \MCVNbOrgcite
    &&\own{\FPTb} &  \MCENbOrgcite\\

    \;\;~~\paramMaxOrg & \coNP-complete & \MCVcite
    && \own{\coNPb-complete} & \MCEcite \\
    \bottomrule
  \end{tabular}
\end{table}

\paragraph{Related Work.} %

\COG\ has been studied extensively across numerous graph optimization problems~\cite{claus1973cost,Bird76,Granot81,faigle1997complexity,deng1999algorithmic,potters1992traveling,hamers1999cost,Velzen04,Fang05,Fang08,hafezalkotob2016,Xiao21}.
We focus on prior work most relevant to our analysis. %

\citet{deng1999algorithmic} study three relevant problems: core verification (\probCV), core existence (\probCE), and the search variant of \probCE\ in the context of \COG.
They establish a foundational characterization result for \COG{s}:
a \stable\ allocation exists if and only if the underlying linear program has an integral optimal solution.
For minimum vertex cover, maximum matching, and maximum independent set games, they provide polynomial-time algorithms to determine core existence and find core allocations when possible.
However, these polynomial-time results do not transfer to the generalized \PCOG\ setting we study.
For the minimum vertex cover game, \citet{Fang08} analyze several structural properties including balancedness, extendibility, and core largeness.
They show that a \stable\ allocation exists if and only if every edge belongs to a maximum matching
and reconfirm that \probCE\ is polynomial-time solvable for minimum vertex cover game.

Van Velzen~\cite{Velzen04}  introduces several variants for the minimum dominating set game based on the domination distance~$k$ (the length of the shortest path from a dominating set vertex).
The standard variant, where $k=1$, corresponds to the restricted case of our \PCOG-\minDS\ where each agent owns exactly one vertex. We confirm \citeauthor{Velzen04}'s conjecture that determining core existence in his setting is NP-hard (see \cref{thm:DSdp:orgsize}).

\citet{claus1973cost} initiate the study of minimum-cost spanning tree games.
While \citet{Bird76} prove that a \stable\ allocation for this game always exists,
\citet{Granot81} show that such allocations can be computed in polynomial time from any minimum-cost spanning tree.
However, \citet{faigle1997complexity} establish that verifying core membership remains \coNP-complete despite these positive results.

Most directly related to our work,
\citet{Biro19} extend the maximum matching game to the partitioned setting (\PCOG) to model international kidney exchange.
They establish that \PCOG-\maxMatching\ is equivalent to \COG-\maxbMatching, i.e., maximum $b$-matching game, for core stability analysis.
Building on results from \citet{biro2018stable}, they show that \probCV\ is \coNP-complete and \probCE\ is \coNP-hard.
We complement the latter by proving \coNP-containment for \probCE.
\citet{Gourves12} study \PCOG-\maxbMatching\ for another fairness concept called \myemph{individual rationality} (i.e., no agent prefers to be alone) and show that determining whether an individually rational allocation exists is \NP-complete, even for bipartite graphs.

The \myemph{multi-located player model} represents another approach to generalizing \COG{s}, where agents may own multiple elements but the optimization goal considers only one element per agent.
This model has been studied for the Chinese postman problem~\cite{estevez2020chinese}, fixed tree games~\cite{miquel2006fixed}, and \minST~\cite{Le16}.

\PCOG\ concepts have been applied to model multi-organization collaboration in scheduling~\cite{Rzadca2007MultiScheduling,pascual2009introduction,cohen2011multi,cohen2014energy,durand2021efficiency}.
These studies typically focus on a solution concept that is related to individual rationality rather than core stability.
Specifically, the goal is to find a schedule such that no agent finishes her jobs later than if that agent were to use their local schedule on her own machines.
This schedule is generally either given, or computed using a local heuristic.

\paragraph{Paper Outline.}
In \cref{sec:prelim}, we introduce necessary concepts from cooperative game theory, 
define our four partitioned combinatorial optimization games,
and present key structural properties that connect core stability to the underlying optimization problems.
In \cref{sec:PVCG,sec:PDSG,sec:PMGPSTG}, we present our findings for the four games.
In each section, we follow a similar structure: We first address core verification and then core existence. 
We conclude in \cref{sec:conclude} with preliminary results on individual rationality and directions for future research.
\ifcr
\else
  Proofs and statements marked with (\appsymb) are deferred to the Appendix and can be found in the full version as well~\cite{fullpaper}.
\fi

\section{Preliminaries}\label{sec:prelim}\appendixsection{sec:prelim}

\toappendix{Here, we present a couple of further definitions.}

Given a non-negative integer~$t\in \mathds{N}$, let $[t]$ denote the set~$\{1,\ldots, t\}$.

\subsection{Cooperative Games and Core Stability}\label{subsec:coopgamesstable}
In this section, we formally define the relevant definitions from cooperative games. %
A \myemph{cooperative game with transferable utilities} (or simply \myemph{game}) %
is a pair~$(\players, \posutil)$ where $\players$ is a finite set of~\myemph{$\coln$} agents and
\myemph{$\posutil\colon 2^{\players}\to \mathds{R}_{\geq 0}$} a \myemph{value function} (aka.\ \myemph{characteristic function}) with $\posutil(\emptyset) = 0$.
A \myemph{coalition} is a subset of~$\players$.
An \myemph{\allocation} $\alloc = (\allocel 1, \dots, \allocel{\coln})\in \mathds{R}^{\coln}_{\geq 0}$ is called a \myemph{\preimputation} if $\sum_{i \in \players} \allocel i = \posutil(\players)$.
An \allocation\ is \myemph{\stable} if it is a \preimputation\ and $\sum_{i \in \pset}\allocel i \geq \posutil(\pset)$ holds for every $\pset \subseteq \players$; 
otherwise we say that a coalition~$\pset$ \myemph{is blocking}~$\alloc$ if $\sum_{i \in \pset}\allocel i < \posutil(\pset)$ holds.
The \myemph{core} of a game is a subset consisting of all \stable\ \allocation{s}.

Games with a value function are usually referred to as \myemph{maximization} games. 
In this work, however, we mostly consider \myemph{minimization} games where each coalition has a \costtxt\ rather than a value. %
In this case, we have a \myemph{\costtxt\ function $\util\colon 2^{\players}\to \mathds{R}_{\geq 0}$} with $\util(\emptyset) = 0$. %
The definitions of \allocation\ and \preimputation\ are the same as above.
An \allocation\ is \myemph{\stable} if it is a \preimputation\ and $\sum_{i \in \pset}\allocel i \leq \util(\pset)$ holds for every coalition~$\pset \subseteq \players$; otherwise $\pset$ \myemph{blocks} \alloc. %

We study the following two problems related to core stability:

\decprob{\probCVlong~(\probCV)}{An instance of a game~$(\players, \posutil)$ and an allocation  $\alloc \in \mathds{R}^{\coln}$.}{Is \alloc\ \stable?}

\decprob{\probCElong~(\probCE)}{An instance of a game~$(\players, \posutil)$.}{Does $(\players, \posutil)$ admit a \stable\ \allocation?}

\begin{remark}
  When the characteristic function $\posutil$ is given explicitly—that is, a value (or cost) for every coalition—the input length is exponential in $\coln$.
  Under this unrestricted encoding, \probCV\ is in \PP, whereas \probCE\ is in \NP.
  Our focus, however, is on combinatorial optimization games whose value functions admit compact encoding. For such games the input size is polynomial in~$\coln$, making the two problems more subtle and worthy of detailed complexity analysis.
\end{remark}

\subsection{(Partitioned) Combinatorial Optimization Games} \label{sec:pcogdef}
We consider \PCOGlong{s} (\PCOG{s}) that are defined on graphs. 
To this end, we recall the following concepts regarding induced subgraphs. 
Given an undirected graph~$G = (V,E)$ and two subsets~$V'\subseteq V$ and $E'\subseteq E$,
a subgraph \myemph{induced} by $V'$ (resp.\ $E'$), written as \myemph{$G[V']$} (resp.\ \myemph{$G[E']$}), is a subgraph $(V', E'')$ of $G$ where
$E''=\{e' \in E \mid e' \subseteq V'\}$ %
(resp.\ a subgraph~$(V'', E')$ of $G$ where $V'' \coloneqq \bigcup_{e' \in E'} e'$). %

\COGlong{} (\COG) is a class of games with compact representation on the value function, where we are provided with an undirected graph, and depending on the optimization problem, each agent owns either a vertex or an edge of the graph.
The value of a subset of agents is based on the objective function of the underlying optimization problem, such as vertex cover size, on the subgraph induced by the agents' vertices or edges.
\emph{\PCOG} extends \COG\ by allowing agents to  own a subset of vertices or edges instead of a single one.
We define the four \PCOG{s} separately. 

\begin{definition}[\ParVCgame]
  A \myemph{\ParVCgame\ (\ParVCgameshort)} is a minimization game which consists of a set~\players\ of agents and an edge-partitioned graph $G = (V, E = E_1 \cup \cdots \cup E_{\coln})$ such that every agent $i \in \players$ owns the edge set $E_i$.
  For every $\pset \subseteq \players$, the \costtxt\ of $\pset$ is defined as the cardinality of a minimum-cardinality vertex cover of~$G[\bigcup_{i \in \pset} E_i]$;~
  recall that a subset~$V' \subseteq V$ is a \myemph{vertex cover} if each edge~$e\in E$ is incident to at least one vertex in~$V'$, in which case we say that $e$ is covered by~$V'$.
\end{definition}

\begin{example}\label{Ex:VCov}
  Below are two \ParVCgameshort{s} on three agents $1,2,$ and $3$ with edge sets~$E_1, E_2$, and $E_3$, respectively.

  {\centering
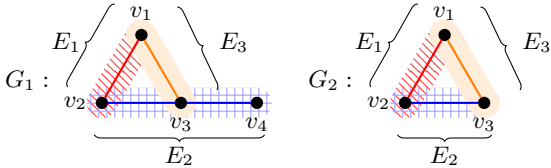
    \begin{tabular}{@{}c@{\quad}c}
      \scalebox{1}{
      \begin{tikzpicture}[node distance={10mm}, thick, main/.style = {draw, circle, scale=0.75, inner sep=1.8pt}]
        
        \foreach \s / \t/ \l/ \c/ \v in {v1/1/above/blue/v_1, v2/2/left/blue/v_2, v3/3/below/red/v_3} {
          \node[mainS, draw=black, fill =black, label=\l:$\v$] (\s) at (\t*360/3-30: .6 cm) {};
        }
        \node[mainS, draw=black, fill =black, label=below:$v_4$] (v4) at ($(v3)+(1,0)$) {};
        \foreach \s / \t/\c in {v1/v2/red, v2/v3/blue,v1/v3/orange, v3/v4/blue} {
          \draw[color=\c] (\s) -- (\t);
        }
        \node[] at (-1.5,0) {$G_1:$};
        \begin{pgfonlayer}{bg}
          \draw[firstagentA] \hedgeii{v2}{v1}{2mm};
          \draw[secondagentA] \hedgeiii{v4}{v3}{v2}{2mm};
          \draw[thirdagentA] \hedgeii{v1}{v3}{2mm};
        \end{pgfonlayer}
        \draw[decoration={brace,raise=18pt},decorate] ($(v2)+(0.1,-0.1)$) --node[midway,xshift=-25pt,yshift=10pt] (E1) {$E_1$} ($(v1)+(0.2,0.1)$);
        \draw[decoration={brace,raise=12pt},decorate] ($(v4)+(0.1,0)$) --node[midway, yshift=-20pt] (E2) {$E_2$} ($(v2)+(-0.1,0)$);
        \draw[decoration={brace,raise=18pt},decorate] ($(v1)+(-0.1,0.1)$) --node[midway, xshift=30pt,yshift=10pt] (E3) {$E_3$} ($(v3)+(-0.1,-0.1)$);
      \end{tikzpicture}}
      &
        \scalebox{1}{\begin{tikzpicture}[node distance={10mm}, thick, main/.style = {draw, circle, scale=0.75, inner sep=1.8pt}]
            
            \foreach \s / \t/ \l/ \c/ \v in {v1/1/above/blue/v_1, v2/2/left/blue/v_2, v3/3/below/red/v_3} {
              \node[vcvertex, label=\l:$\v$] (\s) at (\t*360/3-30: .6 cm) {};
            }
            \foreach \s / \t/\c in {v1/v2/firstagentedge, v2/v3/secondagentedge,v1/v3/thirdagentedge} {
              \draw[\c] (\s) -- (\t);
            }
            \node[] at (-1.5, 0) {$G_2:$};
            \begin{pgfonlayer}{bg}
              \draw[firstagentA] \hedgeii{v2}{v1}{2mm};
              \draw[secondagentA] \hedgeii{v3}{v2}{2mm};
              \draw[thirdagentA] \hedgeii{v1}{v3}{2mm};
            \end{pgfonlayer}
            \draw[decoration={brace,raise=18pt},decorate] ($(v2)+(0.1,-0.1)$) --node[midway,xshift=-25pt,yshift=10pt] (E1) {$E_1$} ($(v1)+(0.2,0.1)$);
            \draw[decoration={brace,raise=12pt},decorate] ($(v3)+(0.1,0)$) --node[midway, yshift=-20pt] (E2) {$E_2$} ($(v2)+(-0.1,0)$);
            \draw[decoration={brace,raise=18pt},decorate] ($(v1)+(-0.1,0.1)$) --node[midway, xshift=30pt,yshift=10pt] (E3) {$E_3$} ($(v3)+(-0.1,-0.1)$);
          \end{tikzpicture}}
    \end{tabular}
    \par}

  In $G_1$, for every~$i\in[3]$, it holds that $G[E_i]$ has a minimum vertex cover of size $1$, while the entire graph admits a minimum vertex cover of size $2$. The only \stable\ allocation is~$\alloc=(1,1,0)$.
  Clearly $\allocel 1 + \allocel 2 + \allocel 3 = 2 = \cost([3])$, where $\cost([3])$ is the size of a minimum vertex cover of $G_1$.
  The allocation is \stable, because $G[E_1\cup E_2]$ has a minimum vertex cover of size $2$, while $G[E_1\cup E_3]$ and $G[E_2\cup E_3]$ have a minimum vertex cover of size $1$ each. We can thus verify that every subset of agents $S$ satisfies $\sum_{i \in S}\allocel i \leq \cost(S)$; recall that $\cost(S)$ is the size of a minimum vertex cover of the graph restricted to the edges owned by the agents in $S$.

  In contrast, game~$G_2$ has an empty core because for each pair $i,j\in[3]$ with $i\neq j$, it holds that $G[E_i\cup E_j]$ has a minimum vertex cover of size $1$, while the entire graph has a minimum vertex cover of size~$2$.
  Therefore we need to satisfy $\allocel i+\allocel j\leq 1$, for all $i,j\in[3]$ such that $i\neq j$, which implies that $2(\allocel 1+\allocel 2+\allocel 3)\leq 3$. 
  This however contradicts $\allocel 1+\allocel 2+\allocel 3=~2$.

\end{example}

\begin{definition}[\ParDSgame]
  A \myemph{\ParDSgame\ (\ParDSgameshort)} is a minimization game which consists of 
  a set~\players\ of agents and a vertex-partitioned graph $G = (V = V_1 \cup \cdots \cup V_{\coln}, E)$ such that every agent $i \in \players$ owns the vertex set~$V_i$.
  For every $\pset \subseteq \players$, the \costtxt\ of~$\pset$ is defined as the cardinality of a minimum-cardinality dominating set of~$G[\bigcup_{i \in \pset} V_i]$;~
  recall that a vertex subset~$V' \subseteq V$ is a \myemph{dominating set} if each vertex~$v \in V \setminus V'$ is adjacent to at least one vertex in~$V'$; in this case we say that~$v$ is \myemph{dominated by} $V'$. 
\end{definition}

\begin{example}\label{Ex:DomSET}Below are two \ParDSgameshort{s} on three agents $1,2,$ and $3$ with vertex subsets $V_1, V_2$, and $V_3$, respectively.
  
  {\centering
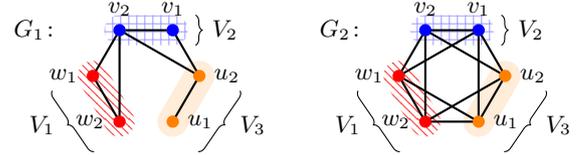
    \begin{tabular}{cc}
      {
      \begin{tikzpicture}[node distance={14mm}, thick, main/.style = {draw, circle, scale=0.75, inner sep=1.8pt}]
        
        \foreach \s / \t/ \l/ \c/ \v in {v1/1/above/blue/v_1,
          v2/2/above/blue/v_2,
          v4/3/left/red/w_1, v5/4/left/red/w_2, v7/5/right/orange/u_1, v8/6/right/orange/u_2} {
          \node[mainS, draw=\c, fill =\c, label=\l:$\v$] (\s) at (\t*360/6: .7 cm) {};
        }
        \foreach \s / \t in {v1/v2, v1/v8,v2/v8, v2/v4, v2/v5, v4/v5,v7/v8} {
          \draw (\s) -- (\t);
        }
        \node[left = 4ex of v2]  {$G_1\colon$};
        \begin{pgfonlayer}{bg}
          \draw[firstagentA] \hedgeii{v5}{v4}{2mm};
          \draw[secondagentA] \hedgeii{v2}{v1}{2mm};
          \draw[thirdagentA] \hedgeii{v8}{v7}{2mm};
        \end{pgfonlayer}
        \draw[decoration={brace},decorate] ($(v1)+(0.3,0.2)$) --node[right,xshift=3pt] (E1) {$V_2$} ($(v1)+(0.3,-0.2)$);
        \draw[decoration={brace,raise=15pt},decorate] ($(v5)+(0.1,-0.1)$) --node[midway, xshift=-24pt,yshift=-10pt] (E2) {$V_1$} ($(v4)+(-0.1,0.1)$);
        \draw[decoration={brace,raise=15pt},decorate] ($(v8)+(0.1,0.1)$) --node[midway, xshift=25pt,yshift=-10pt] (E3) {$V_3$} ($(v7)+(-0.1,-0.1)$);
      \end{tikzpicture}}
      &
        {
        \begin{tikzpicture}[node distance={14mm}, thick, main/.style = {draw, circle, scale=0.75, inner sep=1.8pt}]
          
          \foreach \s / \t/ \l/ \c/ \v in {v1/1/above/blue/v_1, v2/2/above/blue/v_2, v4/3/left/red/w_1, v5/4/left/red/w_2, v7/5/right/orange/u_1, v8/6/right/orange/u_2} {
            \node[mainS, draw=\c, fill =\c, label=\l:$\v$] (\s) at (\t*360/6: .7 cm) {};
          }
          \foreach \s / \t in {v1/v2, v1/v8,v2/v8, v2/v4, v2/v5, v4/v5,v7/v8, v1/v7, v4/v1, v5/v7, v5/v8, v7/v4} {
            \draw (\s) -- (\t);
          }
          \node[left = 4ex of v2]  {$G_2\colon$};
          \begin{pgfonlayer}{bg}
            \draw[firstagentA] \hedgeii{v5}{v4}{2mm};
            \draw[secondagentA] \hedgeii{v2}{v1}{2mm};
            \draw[thirdagentA] \hedgeii{v8}{v7}{2mm};
          \end{pgfonlayer}
          \draw[decoration={brace},decorate] ($(v1)+(0.3,0.2)$) --node[right,xshift=3pt] (E1) {$V_2$} ($(v1)+(0.3,-0.2)$);
          \draw[decoration={brace,raise=15pt},decorate] ($(v5)+(0.1,-0.1)$) --node[midway, xshift=-24pt,yshift=-10pt] (E2) {$V_1$} ($(v4)+(-0.1,0.1)$);
          \draw[decoration={brace,raise=15pt},decorate] ($(v8)+(0.1,0.1)$) --node[midway, xshift=25pt,yshift=-10pt] (E3) {$V_3$} ($(v7)+(-0.1,-0.1)$);
        \end{tikzpicture}} 
    \end{tabular}
    \par}

  It can be easily verified that a minimum dominating set of $G_1$ has size $2$, e.g., $\{u_2,v_2\}$ is a minimum dominating set.
  The only \stable\ allocation is~$\alloc=(1,0,1)$. %
  Instance~$G_2$, on the other hand, has an empty core. For all $i,j\in[3]$ with $i\neq j$, the induced subgraph~$G[V_i\cup V_j]$ has a minimum dominating set of size $1$, but the entire graph has a dominating set of size~$2$. %

\end{example}

\begin{definition}[\ParSTgame]
  A \myemph{\ParSTgame\ (\ParSTgameshort)} is a minimization game
  which consists of a set~\players\ of agents and
  a \emph{complete} vertex-partitioned graph~$G = (V = \{\supernode\} \cup V_1 \cup \cdots \cup V_{\coln}, E)$, 
  and a weight function $\weight \colon E \to \mathds{R}_{\geq 0}$
  such that every agent $i \in \players$ owns the vertex set~$V_i$, but no one owns the \supernodetext~\supernode.
  For every $\pset \subseteq \players$, the \costtxt\ of $\pset$ is defined as the total weight of a minimum-weight spanning tree of~$G[\{\supernode\} \cup\bigcup_{i \in \pset} V_i]$;~
  recall that a subgraph $T$ of $G$ is a \myemph{spanning tree} if $V(T) = V(G)$ and $T$ is a tree.
  The \myemph{weight} of $T$ is defined $\weight(T) \coloneqq \sum_{e \in E(T)}\weight(e)$.
\end{definition}

It is known that a \stable\ allocation for \STgameshort\ (i.e. the unpartitioned variant of \ParSTgameshort)
always exists and finding one can be done in polynomial time~\cite{Bird76,Granot81}.
We will see in \cref{cor:parSTalwaysY} that the same holds for \ParSTgameshort.
\medskip

\noindent
\begin{minipage}{0.75\linewidth}
  \begin{example}\label{Ex:MST}
    Consider the following instance, with two agents and three vertices, as seen in the graph on the right:
    One can verify that \allocation\ $\alloc = (2 + \varepsilon, 2 - \varepsilon)$ is \stable\ for every $\varepsilon \in [0,1]$, since the weight of a minimum-weight spanning tree for the entire graph is $4$, while for agent~1 it is $3$ and for agent~$2$ it is~$2$.
  \end{example}
\end{minipage}
\begin{minipage}{0.24\linewidth}
  {\centering
    \begin{tikzpicture}[node distance={14mm}, thick, main/.style = {draw, circle, scale=0.75, inner sep=1.8pt}]
      \node[mainS, draw=black, fill = black] (S) at (0.5,1.9) {};
      \node[above right = -5pt and 0pt of S] {$\supernode$};
      \node[firstagent, label=left:$v_1$] (A) at (-0,0) {};
      \node[firstagent, label=right:$v_2$] (B) at (0.5,0.87) {};
      \node[secondagent, label=right:$w_1$] (C) at (1,0) {};

      \foreach \s / \t / \ww / \b / \po in {S/A/2/-35/0.3,S/B/2/0/0.5, S/C/2/35/0.3, A/B/1/0/0.5, A/C/1/0/0.5, B/C/1/0/0.5} {
        \draw (\s) edge[bend left=\b]  node[pos=\po, fill=white, inner sep=1pt] {$\ww$}  (\t);
      }
      
      \begin{pgfonlayer}{bg}
        \draw[firstagentA] \hedgeii{A}{B}{2mm};
        \draw[secondagentA] \hedgei{C}{2mm};
        \draw[decoration={brace,raise=6pt,mirror},decorate] ($(B)+(-0.1,0.1)$) --node[midway, xshift=-13pt, yshift=10pt] (V1) {$V_1$} ($(A)+(-0.1,0.1)$);
        \draw[decoration={brace,raise=5pt},decorate] ($(C)+(0,0.2)$) --node[midway,xshift=6pt, yshift=14pt] (V2) {$V_2$} ($(C)+(0.2,0.1)$);
      \end{pgfonlayer}
    \end{tikzpicture} 
    \par}

\end{minipage}

\begin{definition}[\ParMgame]
  A \myemph{\ParMgame\ (\ParMgameshort)} is a maximization game which consists of
  a set~\players\ of agents and a vertex-partitioned graph $G = (V = V_1 \cup \cdots \cup V_{\coln}, E)$ such that every agent $i \in \players$ owns the vertex set~$V_i$.
  For every $\pset \subseteq \players$, the value of $\pset$ is defined as the cardinality of a maximum matching of~$G[\bigcup_{i \in \pset} V_i]$;~
  recall that an edge subset $E' \subseteq E$ is a matching if all edges in $E'$ are pairwise disjoint. %
\end{definition}

\begin{example}\label{Ex:PMG}
  For \ParMgameshort, we will use the same graphs from \cref{Ex:VCov}, but divide the vertices rather than the edges: 

  {\centering
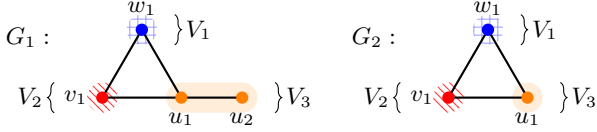
    \begin{tabular}{@{}c@{\quad}c}
      \scalebox{1}{
      \begin{tikzpicture}[node distance={10mm}, thick, main/.style = {draw, circle, scale=0.75, inner sep=1.8pt}]
        
        \foreach \s / \t/ \l/ \c/ \v in {v1/1/above/blue/w_1, v2/2/left/red/v_1, v3/3/below/orange/u_1} {
          \node[mainS, draw=\c, fill =\c, label=\l:$\v$] (\s) at (\t*360/3-30: .6 cm) {};
        }
        \node[mainS, draw=orange, fill =orange, label=below:$u_2$] (v4) at ($(v3)+(0.8,0)$) {};
        \foreach \s / \t/\c in {v1/v2/red, v2/v3/blue,v1/v3/orange, v3/v4/blue} {
          \draw (\s) -- (\t);
        }
        \node[] at (-1.5,0.5) {$G_1:$};
        \begin{pgfonlayer}{bg}
          \draw[firstagentA] \hedgei{v2}{2mm};
          \draw[secondagentA] \hedgei{v1}{2mm};
          \draw[thirdagentA] \hedgeii{v4}{v3}{2mm};
        \end{pgfonlayer}
        \draw[decoration={brace,raise=18pt},decorate] ($(v1)+(-0.2,0.2)$) --node[midway,xshift=27pt] (E1) {$V_1$} ($(v1)+(-0.2,-0.2)$);
        \draw[decoration={brace,raise=18pt},decorate] ($(v2)+(-0,-0.2)$) --node[midway, xshift=-27pt] (E2) {$V_2$} ($(v2)+(-0,0.2)$);
        \draw[decoration={brace,raise=18pt},decorate] ($(v4)+(-0.2,0.2)$) --node[midway,xshift=27pt] (E3) {$V_3$} ($(v4)+(-0.2,-0.2)$);
      \end{tikzpicture}}  &
                            \scalebox{1}{\begin{tikzpicture}[node distance={10mm}, thick, main/.style = {draw, circle, scale=0.75, inner sep=1.8pt}]
                                
                                \foreach \s / \t/ \l/ \c/ \v in {v1/1/above/blue/w_1, v2/2/left/red/v_1, v3/3/below/orange/u_1} {
                                  \node[mainS, draw=\c, fill =\c, label=\l:$\v$] (\s) at (\t*360/3-30: .6 cm) {};
                                }
                                \foreach \s / \t/\c in {v1/v2/red, v2/v3/blue,v1/v3/orange} {
                                  \draw (\s) -- (\t);
                                }
                                \node[] at (-1.5,0.5) {$G_2:$};
                                \begin{pgfonlayer}{bg}
                                  \draw[firstagentA] \hedgei{v2}{2mm};
                                  \draw[secondagentA] \hedgei{v1}{2mm};
                                  \draw[thirdagentA] \hedgei{v3}{2mm};
                                \end{pgfonlayer}
                                \draw[decoration={brace,raise=18pt},decorate] ($(v1)+(-0.2,0.2)$) --node[midway,xshift=27pt] (E1) {$V_1$} ($(v1)+(-0.2,-0.2)$);
                                \draw[decoration={brace,raise=18pt},decorate] ($(v2)+(-0,-0.2)$) --node[midway, xshift=-27pt] (E2) {$V_2$} ($(v2)+(-0,0.2)$);
                                \draw[decoration={brace,raise=18pt},decorate] ($(v3)+(-0.2,0.2)$) --node[midway,xshift=27pt] (E3) {$V_3$} ($(v3)+(-0.2,-0.2)$);
                              \end{tikzpicture}}
    \end{tabular}
    \par}
  Graph~$G_1$ has many \stable\ \allocation s, e.g., $\alloc=(0.5,0.5,1)$ is \stable: The maximum matching of the entire graph has size~$2$,
  while $G[V_1\cup V_2]$ has a maximum matching of size~$1$ and $G[V_3]$ as well.
  Instance~$G_2$ has an empty core. The reasoning is similar to that for \cref{Ex:VCov}. %
\end{example}

Note that the decision variant of whether a given graph has a vertex cover or a dominating set of size at most $k$, referred to as \DS\ and \VC, is \NP-complete~\cite{GJ79}.
In contrast, both a minimum-weight spanning tree~\cite{Prim} and a maximum-cardinality matching can be found in polynomial time~\cite{Edmonds65}.

We use \VCgameshort, \DSgameshort, \STgameshort, and \Mgameshort\ to refer to the unpartitioned classes of \ParVCgameshort, \ParDSgameshort, \ParSTgameshort, and \ParMgameshort\ respectively, i.e., when each agent owns a single unit of the graph.
Given a game class $X$, we define the problems~$X$-\probCE\ and $X$-\probCV\ by restricting the input game to the game class~$X$.
For example, the problem of deciding whether an instance of \ParVCgameshort\ admits a \stable\ \allocation\ is referred to as \probPVCCE.

\paragraph{Parameters.}
For the parameterized complexity study, we focus on two parameters: (1) the number of agents:~\myemph{$\nbOrg\coloneqq |\players|$} and
(2) the  maximum agent size \myemph{$\Orgsize$}, which is \myemph{$\Orgsize \coloneqq \max_{i \in \players}|V_i|$} for \ParDSgameshort, \ParSTgameshort, and \ParMgameshort, and \myemph{$\Orgsize \coloneqq \max_{i \in \players}|E_i|$} for \ParVCgameshort.
We refer to the textbook by Cygan et al.~\cite{CyFoKoLoMaPiPiSa2015} for more details on parameterized algorithmics.

\subsection{ Classes beyond \NP\ and \coNP: $\thetaC$, \DP, $\sigmaC$.}\label{subsec:complexity}
We define relevant complexity classes that lie beyond \NP\ and \coNP\ with the help of \NP-oracles.
An \myemph{\NP-oracle} is a special algorithm that can decide in \emph{constant} time whether a given instance is contained in a specific \NP\ problem~\cite[Chapter 3]{arora2009computational}.
The complexity class~\myemph{$\thetaC$} (aka.\ $\PP^{\NP[\log]}$ and $\PP^{\NP}_{\mid\mid}$) contains all problems~$\Pi$ which can be decided by a \emph{polynomial-time deterministic algorithm} with \emph{logarithmically} many \NP-oracle calls.
The class $\thetaC$ contains a special complexity class called \myemph{$\DP$}~\cite{PY84} (D standing for ``difference''); it contains all problems that are the differences of two \NP\ problems.
More precisely, a problem is contained in $\DP$ if an input instance is a yes-instance if and only if we can make two \NP-oracle queries with a yes answer to the first one and a no answer to the second one~\cite[Chapter 17]{papadimitrioubook}.
$\DP$ contains several problems that are related to computational social choice, in particular resource allocation~\cite{NNRR2014}. %
Note that it holds that
{$\PP\subseteq \NP \subseteq \DP \subseteq \thetaC  \subseteq \NPNP=\sigmaC$,}
where the last one contains all problems that can be decided by a \emph{polynomial-time non-deterministic algorithm} with \emph{polynomially} many \NP-oracle queries and is shown to be equivalent to $\sigmaC$~\cite[Chapter 5]{arora2009computational}.
It is widely assumed that all inclusion relations above are strict.
For more discussion, refer to for example the textbook by Papadimitriou~\cite{papadimitrioubook}.

%
%
%
%
%

%
%
%
%
%
%
%
%
%
%
%
%
%
%
%
%

\toappendixalter{\subsection{Ellipsoid Method}\label{sec:ellipsoid}
  Inspired by Bir\'{o} et al.~\cite{biro2018stable,Biro19}, we will design algorithms utilizing the ellipsoid method for \cref{thm:VCCEthetacont,thm:DSCEcont,lem:PMGcoNPcontain} to show containment in either \thetaC\ or \coNP.
  In general, the ellipsoid method finds a feasible solution to a linear program~(LP) or reports that none exists iteratively~\cite{GLS1988}. 
  Notably, this technique is applicable even if not all linear constraints are explicitly given~\cite{Schrijver99book}.
  The central part of the technique is a \myemph{separation oracle}, which, given a point, reports correctly whether this point is a feasible solution or returns a violated constraint. 
  Given such a separation oracle, the algorithm either returns a feasible solution of the LP or reports that none exist with polynomially many calls to the oracle by shrinking the solution space (the ellipsoid) in time $T\cdot\max\{L,n\}^{O(1)}$~\cite{Schrijver99book}, where $T$ is the running time of the separation oracle, $L$ the maximum encoding size of a constraint, and $n$ the dimension of the search space.
  More specifically, this algorithm runs through polynomially many iterations, polynomial in the number~$p$ of variables and the largest size~$q$ of a constraint.
  In our setting, $p$ is the number of agents and $q$ is polynomial bounded in the solution value of the associated optimization problem since the value function $\posutil$ is polynomial in the input graph.
  In each iteration, the algorithm calls the separation oracle and computes a new smaller ellipsoid space if the given point is not feasible.
  The computation of the new ellipsoid is also polynomial in $p+q$.
  For further information, see \cref{appsec:ellipsoid} in \ifcr\else the Appendix\fi.

}{
  \subsection{Ellipsoid Method}\label{appsec:ellipsoid}
  As we utilize the ellipsoid method throughout this paper, we give a brief description of the ellipsoid method. For a more detailed introduction we refer to Chapter $3$ in the book by \citet{GLS1988}, the appendix of the book by~\citet{chvatal83}, and the book by \citet{Schrijver99book}. 

  The ellipsoid method is a well-known method to solve linear programs in polynomial time. Linear programs consist of multiple linear constraints over a set of variables.
  We are given an upper-bound on the solution space in the form of an ellipsoid and a lower bound on the size of the solution space if it is non-empty.
  The ellipsoid method either finds a feasible solution for the linear program or determines that no solution can exist. In each step, the method checks whether the center of the ellipsoid violates a constraint.
  If not, the method returns this point as a solution,
  otherwise, using the violated constraint, it generates a new ellipsoid that must contain the solution space if it exists. However, the lower bound on the solution space can be circumvented by slightly modifying the ellipsoid method; see~\cite{chvatal83}. The size of the ellipsoid shrinks by a constant factor, and thus the algorithm terminates after polynomial many steps in the number of constraints and variables.
  Afterwards, it either returns a solution or reports ``no'' when the remaining ellipsoid is smaller than the aforementioned lower bound on the size of the solution space.

  The ellipsoid method can be applied even if we do not have an explicit encoding of the underlying linear program. This is described in detail in Chapter 14 of the book by \citet{Schrijver99book}. In this case, the ellipsoid method requires a separation oracle. 
  A separation oracle takes as an input a point from the search space and either verifies that the point is a valid solution (in the case of a linear program a feasible solution of the linear program) or it returns the direction in which a solution can be found (in the case of a linear program a violated constraint).
  These constraints and the separation oracle themselves may be hard to compute; however the number of steps is polynomially bounded in the dimension of the search space and the largest size among the inequality constraints. Note that the dimension of the search space is equal to the number of agents and the size of the inequality constraints is upper-bounded by the number of vertices in our problems. Hence the running time for our problems is $T\cdot n^{O(1)}$~\cite{Schrijver99book}, where $T$ is the running time of the separation oracle and $n$ is the number of variables.}{}
\subsection{Structural Observations}\label{sec:StructuralObs}
\appendixsection{sec:StructuralObs}
In this section, we present a few structural observations for our \PCOG{s}. %
First, we prove an auxiliary condition when a game has a non-empty core, with the help of another game.

\begin{restatable}[\appsymb]{proposition}{thmsimpletopart}
\label{thm:simpletopart}
Let $G_1 = (X, \posutil_1)$ and $G_2 = (Y, \posutil_2)$ be two maximization games,
$b \geq 0$ a non-negative constant,
and $f\colon X\rightarrow Y$ a \emph{surjective} function such that
$\posutil_2(\pset)=b\cdot \posutil_1(f^{-1}(\pset))$ holds for all coalitions~$\pset\subseteq Y$.
Let $\alloc$ be an allocation for~$G_1$.
The following holds:
If $\alloc$ is \stable\ for $G_1$, 
then in time polynomial in~$\max(|X|,|Y|)$
we can derive a \stable\ allocation for~$G_2$ from~$\alloc$.
An analogous statement holds for minimization games.
\end{restatable}
\appendixproofwithstatement{thm:simpletopart}{\thmsimpletopart*}{
  \begin{proof}
    We show the proof for maximization games; the proof for minimization games is analogous.
    Assume that $\alloc=(\allocel 1,\ldots,\allocel{|X|})$ is \stable\ for~$G_1$.
    We define an allocation~$\alloc'$ for $G_2$ and show that it \stable.
    For each agent~$j\in Y$, we set $\allocel{j}' \coloneqq b\cdot \sum_{i\in f^{-1}(j)}\allocel{i}$.
    First, we note that since $f$ is surjective,
    $\sum_{j\in S}\allocel{j}' =
    b\cdot\sum_{j\in S}\sum_{i\in f^{-1}(j)}\allocel{i} =  b\cdot \sum_{i\in f^{-1}(S)} \allocel{i}$ holds for all coalitions~$S\subseteq Y$.
    
    Hence
    $\allocel{i}'$ is a pre-imputation for~$G_2$
    because
    $\sum_{j\in Y}\allocel{j}' = b \cdot \sum_{i\in f^{-1}(Y)}\allocel{i} =
    b \cdot \sum_{i\in X}\allocel{i}
    =b\cdot \posutil_1(X)=\posutil_2(Y)$; recall that the second last equation holds since $f^{-1}(Y) = X$. 
    It remains to show that there is no blocking coalition.
    Consider an arbitrary coalition~$\pset\subset Y$. %
    By assumption, it holds that
    $\posutil_2(\pset)=b \cdot \posutil_1(f^{-1}(\pset))
    \leq b \cdot \sum_{j\in f^{-1}(\pset)}\allocel j=\sum_{i\in \pset}\allocel i'$;
    the second inequality holds since $\alloc$ is \stable, implying that $\pset$ is not blocking~$\alloc'$, as desired.
\end{proof}
}

The above result, together with the fact that every instance of \STgameshort\ admits a \stable\ \allocation\ which can be found in polynomial time~\cite{Bird76,Granot81}, implies the following: %

\begin{restatable}[\appsymb]{corollary}{corparSTalwaysY}
\label{cor:parSTalwaysY}
For every $X \in \{$\VCab, \DSab, \STab, \Mab$\}$, if an instance of \abcog-$X$ consisting of unpartitioned graph $G$ admits a \stable\ \allocation, then so does any instance of \parProb\abcog-$X$ which consists of a graph that is an edge- or vertex-partition of $G$.
In particular, every instance of \ParSTgameshort\ admits a \stable\ \allocation, which can be found in polynomial time.
\end{restatable}

\appendixproofwithstatement{cor:parSTalwaysY}{\corparSTalwaysY*}{
\begin{proof}
We show the proof for $X = \DS$; the proofs for other games are analogous.
Let $G = (V, E)$ be the original unpartitioned graph, and let $\hat{G} = (V = V_1 \cup \dots \cup V_{\coln}, E)$ be a vertex-partition of $G$.
Let $b = 1$.
For every $v \in V$, we set $f(v) = V_\ell$, where $\ell$ is the unique member of $[\coln]$ such that $v \in V_\ell$;
observe that $f^{-1}(V_\ell) = V_{\ell}$.
Clearly $\bigcup_{i \in \pset}V_i$ is equivalent to $\bigcup_{v \in f^{-1}(\pset)}\{v\}$, and thus the cardinality of a minimum dominating set of $\hat{G}[\bigcup_{i \in \pset}V_i]$ is equivalent to the cardinality of a minimum dominating set of $G[\bigcup_{v \in f^{-1}(\pset)}\{v\}]$, as required.

In particular, since every instance of \STgameshort\ admits a \stable\ \allocation\ by \citet{Bird76,Granot81}, every instance of \ParSTgameshort\ admits a \stable\ \allocation, which can be found in polynomial time.
\end{proof}
}

Finally, the next observation indicates when we can consider each connected component separately, which will be useful for our proofs.

\begin{restatable}[\appsymb]{independentobservation}{obssepgraphs}\label{obs:sepgraphs}
  Let $(\players, G=(V,E))$ be an instance of \ParVCgameshort, \ParDSgameshort, \ParSTgameshort, or \ParMgameshort\ such that~$G$ consists of two connected components~$G_1$ and $G_2$.
  Let $\players_{1}$ and $\players_2$ be the agents owning vertices (or edges) in~$G_1$ and $G_2$, respectively.
  If $\players_1\cap\players_2=\emptyset$, then an \allocation\ $\alloc$ is \stable\ if and only if \alloc\ restricted to~$\players_1$ and $\alloc$ restricted to~$\players_2$ are both \stable.

\end{restatable}

\appendixproofwithstatement{obs:sepgraphs}{\obssepgraphs*}{
\begin{proof}
We show the proof for maximization games; the proof for minimization games is analogous.
First assume that $\alloc$ is \stable\ on \players.
~Since $\alloc$ is a \preimputation, we have that $\posutil(\players) = \sum_{i \in \players} \allocel i$.
Because $G_1$ and $G_2$ are not connected, we also have that $\posutil(\players_1) + \posutil(\players_2) = \posutil(\players)$.
If $\posutil(\players_1) < \sum_{i \in \players_1} \allocel i$, then $\posutil(\players_2) > \sum_{i \in \players_2} \allocel i$ and $\players_2$ blocks~\alloc.
If $\posutil(\players_1) > \sum_{i \in \players_1} \allocel i$, then $\players_1$ blocks~\alloc.
Thus~$\alloc$ restricted on $\players_1$ must be a \preimputation.
Through analogous argument $\alloc$ restricted on $\players_2$ must be a \preimputation.
Moreover, if a coalition $\pset \subseteq \players_1$ or $\pset \subseteq \players_2$  blocks~\alloc\ restricted to $\players_1$ or $\players_2$, respectively, then clearly $\pset$ also blocks $\players$. Thus $\alloc$ restricted to $\players_1$ and~\alloc\ restricted to $\players_2$ must both be \stable.

Now assume that $\alloc$ restricted to $\players_1$ is \stable, and $\alloc$ restricted to $\players_2$ as well. 
We obtain that $\sum_{i \in \players} \allocel i  = \sum_{i \in \players_1} \allocel i + \sum_{i \in \players_2} \allocel i = \posutil(\players_1) + \posutil(\players_2) = \posutil(\players)$, and thus \alloc\ is a \preimputation.
Moreover, if a coalition $\pset$ blocks \alloc, we have that $\sum_{i \in \pset \cap \players_1}\allocel i + \sum_{i \in \pset \cap \players_2}\allocel i = \sum_{i \in \pset}\allocel i < \posutil(\pset) = \posutil(\pset \cap \players_1) + \posutil(\pset \cap \players_2)$, and thus we must have that $\sum_{i \in \pset \cap \players_1}\allocel i < \posutil(\pset \cap \players_1)$ or $\sum_{i \in \pset \cap \players_2}\allocel i < \posutil(\pset \cap \players_2)$, implying that $\alloc$ restricted to $\players_1$ or $\players_2$ is not \stable, a contradiction.
\end{proof}
}

\toappendix{

  We conclude this section with a result regarding super-additive games and \ir\ allocation. %
  First, let us recall the definitions.

  \begin{definition}[Individual rationality]
    Given a maximization game, an \allocation\ is \myemph{\ir} if it is a \preimputation\ and $\allocel i \geq \posutil(\{i\})$ holds for every agent~$i \in \players$.
    Given a minimization game, an \allocation\ is \myemph{\ir} if it is a \preimputation\ and and  $\allocel i \le \posutil(\{i\})$ holds for every agent~$i \in \players$.
    An \ir\ \preimputation\ is also referred to as an \myemph{imputation}. 
  \end{definition}
  
  \begin{definition}[Superadditive games]\label{def:superadditive}
    We say that a game is super-additive, if for every two subsets agents $X, Y \subseteq \players$ such that $X \cap Y =\emptyset$, it holds that
    (i) if the game is a maximization game, then $\posutil(X \cup Y) \geq \posutil(X) + \posutil(Y)$,
    (ii) if the game is a minimization game, then $\util(X \cup Y) \leq \util(X) + \util(Y)$. 
\end{definition}

\begin{restatable}[\appsymb]{independentobservation}{obssubaddir}
  \label{obs:supaddir}
Every super-additive game admits an \ir\ \allocation. 
\end{restatable}

\begin{proof}
We first show the proof for maximization games.
Let $\players=[\nbOrg]$ be the set of agents. Since the game is super-additive, it holds that $\sum_{i\in \players}\posutil(\{i\})\leq\posutil(\players)$.
Let $\beta=\frac{\posutil(\players)-\sum_{i\in \players}\posutil(\{i\})}{\nbOrg}\geq 0$. Then the \allocation\ $\alloc=(\posutil(\{1\})+\beta,\ldots,\posutil(\{\nbOrg\})+\beta)$ is \ir, as for each voter $i\in \players$ it holds that $\allocel i\geq\posutil(\{i\})$.

The proof for the minimization games is similar, except we have to take care that no agent receives negative cost.
Let $\players=[\nbOrg]$ be the set of agents. 
Let $r=\frac{\util(\players)}{\sum_{i \in \players}\util(\{i\})}$.
Since the game is super-additive, it holds that $\sum_{i\in \players}\util(\{i\})\geq\util(\players)$ and thus $0 < r \leq 1$.
Consider the \allocation\ $\alloc = (\util(\{1\})\cdot r, \dots, \util(\{\nbOrg\})\cdot r)$.
This is a \preimputation: $\sum_{i \in \players}\allocel i = \sum_{i \in \players}(\util(\{i\})\cdot r = r \sum_{i \in \players}(\util(\{i\}) = \frac{\util(\players)}{\sum_{i \in \players}\util(\{i\})}  \sum_{i \in \players}(\util(\{i\}) = \util(\players)$.
Since $0 < r \leq 1$, for every voter $i\in \players$ it holds that $0 \leq \allocel i\leq\util(\{i\})$, and \alloc\ is \ir.
\end{proof}

Since \ParVCgameshort, \ParDSgameshort, \ParSTgameshort, and \ParMgameshort\ are all super-additive, their instances always admit an \ir\ \allocation.

}

\newcommand{\Vars}[1]{\ensuremath{X_{#1}}}
\newcommand{\clanr}[1]{\ensuremath{m_{#1}}}
\newcommand{\varnr}[1]{\ensuremath{n_{#1}}}
\newcommand{\claver}[3]{\ensuremath{v^{#2}_{C^{#3}_{#1}}}}
\newcommand{\agentsym}{a}
\newcommand{\hagent}[1]{\ensuremath{\hat{\agentsym}_{#1}}}
\newcommand{\agent}[1]{\ensuremath{\agentsym_{#1}}}
\newcommand{\clagentM}[2]{\agent{C^{#2}_{#1}}}
\newcommand{\clagentL}[2]{\ensuremath{\agent{C^{#2}_{#1}}^{\leftarrow}}}
\newcommand{\clagentR}[2]{\ensuremath{\agent{C^{#2}_{#1}}^{\rightarrow}}}
\newcommand{\psetC}{\ensuremath{\pset_{\mathcal{C}}}}
\section{\ParVCgame}\label{sec:PVCG}
\appendixsection{sec:PVCG}
In this section, we consider partitioned vertex cover games.
We start with the core verification and show that \probPVCCV\ is \DP-complete.
This holds even if the number of agents or the agent size is a constant.
Afterwards we move to the core existence. %
We show that \probPVCCE\ is \thetaC-complete.
The hardness again holds even if the number of agents or the agent size is a constant.

\mypara{Core Verification.} 
To verify whether an allocation is \stable, one must first solve an $\NP$ problem—determining the minimum vertex cover size of the graph—and then address a complementary $\coNP$ problem, certifying that no induced subgraph admits a strictly smaller vertex cover.
Because these two conditions must hold simultaneously, $\probPVCCV$ belongs to the class $\DP$ (the conjunction of an $\NP$ and a $\coNP$ predicate) and is, in fact, $\DP$–complete.
The proof is similar to that for $\probPDSCV$ which we present in \cref{lem:verifDS}.

\newcommand{\chvertex}[2]{\ensuremath{v^{#2}_{#1}}}
\newcommand{\chagent}[1]{\ensuremath{\agent{#1}^*}}
\newcommand{\clavers}[1]{\claver{}{#1}{}}
\newcommand{\clagents}{\clagentM{}{}}

\begin{restatable}[\appsymb]{theorem}{thmVCverif}\label{thm:VCverif}
\probPVCCV\ is \DP-complete. The hardness remains even if $\nbOrg=1$ or $\Orgsize=~7$. 
\end{restatable}

\appendixproofwithstatement{thm:VCverif}{\thmVCverif*}{
\begin{proof}

The proof proceeds in  three parts.
First we show \DP-membership, and afterwards we present two reductions, first containing only one agent and the second satisfying that every agent has at most seven edges.

\mypara{\baseDP-membership.}
	We first show \DP-membership.
	By \citet{papadimitrioubook}, we need to provide two \NP-problems~$P$ and $Q$,
	and show that \probPVCCV~=$P\cap \overline{Q}$, in other words,
	every instance of \probPVCCV\ is a yes-instance if and only if it is a \emph{yes}-instance of~$P$ but a \emph{no}-instance of~$Q$.
	To this end, let $P$ and $Q$ be the following two \NP-problems. 
	\begin{description}
		\item[\text{\normalfont{Problem}}~$P$:] Given an instance~$I=(G = (V,E_1\cup\dots\cup E_{\coln}), \alloc)$, 
		does~$G$ admit a vertex cover of size at most~$\sum_{i \in \players}{\allocel i}$?
		\item[\text{\normalfont{Problem}}~$Q$:] Given an instance~$I=(G = (V,E_1\cup\dots\cup E_{\coln}), \alloc)$,
		is there a subset~$\pset \subseteq \players$ of agents such that the induced graph~$G[\cup_{i\in \pset}E_i]$ has a vertex cover with size strictly smaller than~$\sum_{i\in \pset} \allocel{i}$?    
	\end{description}
	It is straightforward to check that both problems are contained in \NP\ since the witness of either problem has polynomial size and can be verified in polynomial time as well.
	
	It remains to show that for each instance~$I=(G = (V,E_1\cup\dots\cup E_{\coln}),\alloc)$ of \probPVCCV,
	allocation~$\alloc$ is \stable\ if and only if $I$ is a yes-instance of~$P$ and a no-instance of~$Q$. 
	By definition, \allocation~$\alloc$ is \stable\ if and only if $\alloc$ is a \preimputation\ and no subset of agents is blocking~$\alloc$.
	The former is equivalent to~$I$ being a yes-instance of~$P$ and the latter is equivalent to $I$ being a no-instance of $Q$.
	
\mypara{\baseDP-hardness when $\nbOrg=1$.}
	We now show \DP-hardness for the restricted case where there is only one agent.
	We achieve this by providing a polynomial-time reduction from \textsc{SAT-UNSAT} problem.

	Let $(\varphi_1, \varphi_2)$ be an instance of \textsc{SAT-UNSAT} where $\varphi_1$ and $\varphi_2$ are over the variable sets~$\Vars{1}$ and $\Vars{2}$, respectively. 
	We create an instance~$I'=(G, \alloc)$ of \probPDSCV\ with only one agent.
	Graph~$G$ consists of three disjoint subgraphs: $G_{1}$, $G_2^{(1)}$, and $G_2^{(2)}$,  where $G_1$ corresponds to~$\varphi_1$ and~$G_2^{(1)}$ and $G_2^{(2)}$ to~$\varphi_2$.
	We note that each subgraph is a slight modification of the standard reduction from \textsc{3SAT} to \textsc{Vertex Cover}.
	
	\contrstitle{Graph~$G_1$.} We create the following vertices for~$G_1$:
	\begin{compactitem}[--]
		\item For every variable $\svar{} \in \Vars 1$, create a positive literal vertex~$v_{\svar{}}$ and a negative literal vertex~$v_{\snegvar{}}$.
		\item For every clause $C \in \varphi_1$, create a dummy vertex $v_{C,d}$ and for every literal $\ell \in C$, create a vertex $\clavers{\ell}$.
		\item Create a special vertex $v^*$. 
	\end{compactitem}
	We create the following edges for~$G_1$:
	\begin{compactitem}[--]
		\item The two vertices corresponding to a variable are adjacent:
		For every $\svar{}\in\Vars{1}$, add the edge $\{v_{\svar{}}, v_{\snegvar{}}\}$.
		\item For every clause $C \in \varphi_1$, the four vertices $\{\clavers{\ell} \mid \ell \in C\} \cup \{v_{C,d}\}$ form a clique. 
		\item Each vertex corresponding to a literal in a clause is adjacent to the vertex corresponding to that literal: 
		For every clause~$C \in \varphi_1$, every literal~$\ell \in C$, add edge~$\{\clavers{\ell}, v_{\ell}\}$.
		\item The special vertex $v^*$ is adjacent to all dummy vertices of clauses:
		For every clause~$C \in \varphi_1$, add edge~$\{v^*,v_{C,d}\}$.
	\end{compactitem}
	\contrstitle{Graphs~$G_2^{(1)}$ and $G_2^{(2)}$.} These two graphs are constructed in the same way as~$G_1$, except that $\varphi_2$ and $\Vars{2}$ are the underlying CNF formula and variable set. 
	
	For the sake of brevity, we define the number of clauses in the formulas as $\clanr 1 \coloneqq |\varphi_1|$ and $\clanr 2 \coloneqq |\varphi_2|$, and the number of variables in the formulas $\varnr{1}\coloneqq |\Vars{1}|$ and $\varnr{2}\coloneqq |\Vars{2}|$. Finally, we construct the \costtxt\ vector~$\alloc = ((\varnr{1}+3\clanr{1}) + 2(\varnr{2}+3\clanr{2}) + 2)$; note that we have only one agent, so there is only one \costtxt\ in the vector.
	Let $I$ denote the created instance~$I=(G_1\uplus G_2^{(1)} \uplus G_2^{(2)}, \alloc)$.
	The following claim builds a connection between satisfiable formula and the size of a minimum vertex cover. 
	
	\begin{claim}\label{clm:G1VCSATUNSAT}
		If $\varphi_1$ (resp.\ $\varphi_2$) is satisfiable, then every minimum vertex cover of $G_1$ (resp.\ $G_2^{(1)}$ and $G_2^{(2)}$) has $\varnr{1}+3\clanr{1}$ (resp.\ $\varnr{2}+3\clanr{2}$) vertices. 
		If $\varphi_1$ (resp.\ $\varphi_2$) is unsatisfiable, then every minimum vertex cover of $G_1$ (resp.\ $G_2^{(1)}$ and $G_2^{(2)}$) has~$\varnr{1}+3\clanr{1}+1$ (resp.\ $\varnr{2}+3\clanr{2} + 1$) vertices.
	\end{claim}
	\begin{proof}\renewcommand{\qed}{\hfill (end of the proof of~\cref{clm:G1VCSATUNSAT})~$\diamond$}
		We show the statements only for $G_1$; the reasoning for $G_2^{(1)}$ and $G_2^{(2)}$ is analogous.
		First, observe that for every variable~$\svar{} \in \Vars 1$, due to the edge between $v_{\svar{}}$ and $v_{\snegvar{}}$, one of the two corresponding vertices needs to be in the vertex cover. Since the vertices $\{\clavers{\ell} \mid \ell \in C\} \cup \{v_{C,d}\}$ form a clique, at least three of the vertices must be in any vertex cover. Therefore, every vertex cover of $G_1$ has to have size at least~$\varnr{1}+3\clanr{1}$. 
		
		For the first statement, assume that $\varphi_1$ is satisfiable, and let $\sigma\colon \Vars{1} \to \{\trueT,\falseF\}$ be a satisfying assignment for~$\varphi_1$. We construct a vertex cover~$V'$ of size~$\varnr{1}$ as follows: For each variable~$\svar{}\in \Vars{1}$, if $\sigma(x)=\trueT$, then add~$v_{\svar{}}$ to~$V'$; otherwise add~$v_{\snegvar{}}$ to~$V'$.
For each clause $C\in\varphi_1$, there must be at least one $\ell' \in C$ such that $\ell'$ is true under $\sigma$.
We add the vertices in $\{\clavers{\ell} \mid \ell \in C \setminus \ell'\} \cup \{v_{C,d}\}$ to~$V'$.
We have that $|V'| = \varnr{1}+3\clanr{1}$, and thus if it is a vertex cover, it is a minimum vertex cover.
It is straightforward to verify that $V'$ is a vertex cover: 
For each variable~$\svar{}\in \Vars{1}$, edge $\{v_{\svar{}}, v_{\snegvar{}}\}$  is covered. 
For each clause~$C\in \varphi_1$, the edges between vertices corresponding to this clause are covered, because we have all but one of the corresponding vertices in $V'$.
For each~$C\in \varphi_1$, $\ell \in C$, edge $\{v_\ell, \clavers{\ell}\}$ is covered by $v_\ell$ if $\ell$ is true under $\sigma$, otherwise it is covered by $\clavers{\ell}$.
Finally, each edge adjacent to $v^*$ is also covered, because every dummy vertex is in $V'$.
		
Now we show the second statement.
Graph $G_1$ has a vertex cover of size~$\varnr{1}+3\clanr{1}+1$, as $V' = \{v_{\svar{}} \mid \svar{} \in \Vars 1\} \cup \{\clavers{\ell} \mid C \in \varphi_1, \ell \in C\} \cup \{v^*\}$ forms a vertex cover of size $\varnr{1}+3\clanr{1}+1$:
For every $\svar{} \in \Vars 1$, the edge between $v_{\svar{}}$ and $v_{\snegvar{}}$ is covered by $v_{\svar{}}$.
For every $C \in \varphi_1$, the edges between the vertices in $\{\clavers{\ell} \mid \ell \in C \} \cup \{v_{C,d}\}$ are covered by $\{\clavers{\ell} \mid \ell \in C \}$, edge $\{v_{C,d}, v^*\}$ is covered by $v^*$, and for every $\ell \in C$, edge $\{\clavers \ell, v_\ell\}$ is covered by \clavers \ell.

Now, to show that this is indeed minimum, we show that if $G_1$ has a vertex cover of size~$\varnr{1}+3\clanr{1}$, then formula~$\varphi_1$ is satisfiable.
As we established previously, set~$V'$ must contain for every $\svar{} \in \Vars 1$ one of the vertices $v_{\svar{}}, v_{\snegvar{}}$ and for every $C \in \varphi_1$ three of the vertices $\{\clavers{\ell} \mid \ell \in C \} \cup \{v_{C,d}\}$.
Vertex $v^*$ cannot be contained in $V'$, as then the size would exceed $\varnr{1}+3\clanr{1}$. 
Let us construct an assignment  $\sigma\colon \Vars{1} \to \{\trueT,\falseF\}$ as follows: $\sigma(\svar{i}) = \trueT$ if $v_{\svar{}} \in V'$ and $\sigma(\snegvar{i}) = \falseF$ otherwise.
We claim that~$\sigma$ is a satisfying assignment.
Consider an arbitrary clause $C \in \varphi$.
We know that exactly three of the vertices $\{\clavers{\ell} \mid \ell \in C \} \cup \{v_{C,d}\}$ must be in $V'$.
Since $v^* \notin V'$, exactly one of $\{\clavers{\ell} \mid \ell \in C \}$ is not in $V'$.
Let $\clavers{\ell'} \notin V'$.
Since $V'$ is a vertex cover, we must have that $v_{\ell'} \in V'$, and thus $\ell'$ is true under $\sigma$, satisfying the clause $C$.
Thus $\sigma$ is a satisfying assignment.
	\end{proof}
	
	Using \cref{clm:G1VCSATUNSAT}, we show the correctness, i.e., instance~$(\varphi_1,\varphi_2)$ is a yes-instance of \textsc{SAT-UNSAT} if and only if $I'$ is a yes-instance of \probPVCCV.
	
	$(\Longrightarrow)$: Assume that $(\varphi_1,\varphi_2)$ is a yes-instance, i.e.,
	$\varphi_1$ is satisfiable but $\varphi_2$ is not.
	By \cref{clm:G1VCSATUNSAT}, graph $G_1$ has a minimum dominating set of size~$\varnr{1}+3\clanr{1}$ while ~$G_2^{(1)}$ (resp.\ $G_2^{(2)}$) has a minimum dominating set of size~$\varnr{2}+3\clanr{2}+1$.
	This means every minimum dominating set of the entire graph has size~$(\varnr{1}+3\clanr{1}) + 2(\varnr{2}+3\clanr{2}) + 2$.
	By definition, \allocation\ $\alloc=((\varnr{1}+3\clanr{1}) + 2(\varnr{2}+3\clanr{2}) + 2)$ is a \preimputation\ and also \stable, witnessing that~$I'$ is a yes-instance.

	$(\Longleftarrow)$: Assume $(\varphi_1,\varphi_2)$ is a no-instance. Then $\varphi_1$ is not satisfiable or $\varphi_2$ is satisfiable. 
	
	If $\varphi_1$ is not satisfiable, then by the second statement of \cref{clm:G1VCSATUNSAT}, the size of a minimum dominating set for~$G_1$ is $\varnr{1}+3\clanr{1}+1$.
	Again by  \cref{clm:G1VCSATUNSAT}, every minimum dominating set of the entire graph~$G_1\uplus G_2^{(1)} \uplus G_2^{(2)}$ has size either $\varnr{1}+3\clanr{1}+2(\varnr{2}+3\clanr{2})+3$ (if $\varphi_2$ is not satisfiable) or $\varnr{1}+3\clanr{1}+2(\varnr{2}+3\clanr{2})+1$ (if $\varphi_2$ is satisfiable).
	This implies that $\alloc$ is not a pre-imputation, and hence not \stable.
	
	If $\varphi_1$ is satisfiable, then $\varphi_2$ must be satisfiable to be a no-instance.
	By \cref{clm:G1VCSATUNSAT}, every minimum dominating set of the entire graph~$G_1\uplus G_2^{(1)} \uplus G_2^{(2)}$ has size $\varnr{1}+3\clanr{1}+2(\varnr{2}+3\clanr{2})$, again implying that $\alloc$ is not a pre-imputation, and hence not \stable.
	
As we conclude in both cases that $\alloc$ is not \stable, instance $I'$ is a no-instance as well.

\mypara{\baseDP-hardness when $\Orgsize=~7$.}
We now show \DP-hardness for the restricted case, where each agent owns at most $7$ edges. We assume that every literal appears at most twice, which we can ensure by applying the reduction from \citet{BKS-2bal3sat-2003} to $\varphi_1$ and $\varphi_2$. 
The reduction uses ideas similar to the previous proof, but we add a gadget that prevents a set of agents that does not contain all the variable agents from blocking.
	
	Let $(\varphi_1,\varphi_2)$ be an instance of \textsc{SAT-UNSAT} with variables sets $\Vars{1}$ and $\Vars{2}$, respectively. We create an instance $I'=(G=(V,E_1\cup\dots\cup E_{\coln}),\alloc)$ of \probPVCCV\ with $\Orgsize=7$. Graph $G$ consists of two disjoint subgraphs $G_1$ and $G_{2}$, where $G_1$ corresponds to $\varphi_1$ and $G_2$ to $\varphi_2$. We note that each subgraph is a slight modification of the standard reduction from \textsc{3SAT} to \textsc{Vertex Cover}.

	In what follows, let $\varnr{1} \coloneqq |\Vars{1}|$, $\varnr{2} \coloneqq |\Vars{2}|$, $\clanr 1 \coloneqq |\varphi_1|$, and $\clanr 2 \coloneqq |\varphi_2|$.
	
\def \xx {1}
\def \xy {0.8}
  \begin{figure}[t!]
    \centering
   
  \begin{tikzpicture}[black, scale=1,every node/.style={scale=0.9}]
  
\begin{pgfonlayer}{fg}
 \foreach \x / \y / \n / \nn  / \pos in {
	5/7.5/cj/\clavers{\svar{1}}/right,
	4/7.5/cj2/\clavers{\ell_{j}}/left,
	4.5/8/cj3/\clavers{\ell_{k}}/left,
      6/6/v1/v^1_1/left,
      6/5/v2/v^2_1/above left,
      6/4/v3/v^3_1/below left,
      5.5/7/vp/v_{\svar{1}}/left,
      6.5/7/vn/v_{\snegvar{1}}/right,
      4/6/vm1/v^1_{\varnr{1}}/left,
      4/5/vm2/v^2_{\varnr{1}}/left,
      4/4/vm3/v^3_{\varnr{1}}/left,
      8/6/vp1/v^1_{2}/right,
      8/5/vp2/v^2_{2}/right,
      8/4/vp3/v^3_{2}/right,
      6/3/d/v_d/left
    } {
      \node[vcvertex, label=\pos :$\nn$] (\n) at (\x*\xx,\y*\xy) {};
    }
 \end{pgfonlayer}
 
  \foreach \x / \y / \n  in {
  	6/7.5/c1,
  	6.5/7.5/c2,
  	7.2/7.5/c3} {
      \node[hiddenV] (\n) at (\x*\xx,\y*\xy) {};
    }
    
    \foreach \s/\t in {
    		v1/v2,v2/v3,
    		v2/vp3,v3/vp2,
    		v1/d} {
      \draw[firstagentedge] (\s) -- (\t);
      \begin{pgfonlayer}{bg}

      		\draw[firstagentA] \hedgeii{\s}{\t}{2mm};
     \end{pgfonlayer}
    }
     \foreach \s/\t in {
    		vm1/vm2,vm2/vm3,
    		vm2/v3,vm3/v2,
    		vp/cj,c1/vp,c2/vn,c3/vn,
    		v1/vp,v1/vn,vn/vp} {
      \draw[secondagentedge] (\s) -- (\t);
     \begin{pgfonlayer}{bg}

      		\draw[secondagentA] \hedgeii{\s}{\t}{2mm};
     \end{pgfonlayer}
    }
    \foreach \s/\t in {
    		vp1/vp2,vp2/vp3,
    		cj/cj2,cj2/cj3,cj/cj3,
    		v3/d} {
      \draw[thirdagentedge] (\s) -- (\t);
           \begin{pgfonlayer}{bg}

      		\draw[thirdagentA] \hedgeii{\s}{\t}{2mm};
     \end{pgfonlayer}
    }

\end{tikzpicture}

\caption{Illustration for the proof of \cref{thm:VCverif}. We assume that the variable $\svar{1}$ is in the clause $C = (\svar{1} \vee \ell_j \vee \ell_k)$. The different patterns around the edges indicate different agents.
\vspace{0.5cm} %
}\label{fig:verifVC}
\end{figure}	
	
	\contrstitle{Graph~$G_1$.} We create the following vertices for~$G_1$:
	\begin{compactitem}[--]
		\item  For every variable $\svar{i}\in\Vars 1$, we create a positive literal vertex~$v_{\svar{i}}$, a negative literal vertex $v_{\snegvar{i}}$, and three chain vertices $\chvertex i 1$, $\chvertex i 2$, and~$\chvertex i 3$.
		\item For every clause $C \in\varphi_1$ and every literal $\ell \in C$, we add the clause vertex $\clavers \ell$.
		\item Additionally, we create a special vertex $v^*$.
	\end{compactitem}
	We create the following edges for~$G_1$:

	\begin{compactitem}[--]
\item For every variable $\svar{i}\in\Vars 1$, let $\{v_{\svar{i}},$  $v_{\snegvar{i}},$ $\chvertex{i}{1}\}$ form a clique, and add the edges $\{\chvertex{i}{1}, \chvertex{i}{2}\}, \{\chvertex{i}{2}, \chvertex{i}{3}\}, \{\chvertex{i}{3}, \chvertex{i+1}{2}\}, \{\chvertex{i}{2}, \chvertex{i+1}{3}\}$. where $\varnr{1} + 1 = 1$.
\item For every clause $C = (\ell_1 \vee \ell_2 \vee \ell_3)\in \varphi_z$, let $\{\clavers{\ell_1}, \clavers{\ell_2}, \clavers{\ell_3}\}$ form a clique.
Also add the edges $\{\{v_{\svar{i}}, \clavers{}\} \mid C \in \varphi_z, \svar{i} \in C\} \cup \{\{v_{\snegvar{i}}, \clavers{}\} \mid C \in \varphi_z,  \snegvar{i} \in C\}$.
\item Add the edge $\{v_d, \chvertex{1}{3}\}$.

	\end{compactitem}  
Let us finally construct the agents:
\begin{compactitem}[--]
\item For every variable $\svar{i}\in\Vars 1$, add a vertex agent $\agent{\svar{i}}$. This agent owns the edges in the clique formed by $\{v_{\svar{i}}, v_{\snegvar{i}}, \chvertex{i}{1}\}$, and the edges $\{\{v_{\svar{i}}, \clavers{}\} \mid C \in \varphi_z, \svar{i} \in C\} \cup \{\{v_{\snegvar{i}}, \clavers{}\} \mid C \in \varphi_z,  \snegvar{i} \in C\}$, in total at most seven edges, as each literal appears in at most two clauses.
	\item For every variable $\svar{i}\in\Vars z$, add a chain agent $\chagent{i}$, who owns the edges $\{\chvertex{i}{1}, \chvertex{i}{2}\}, \{\chvertex{i}{2}, \chvertex{i}{3}\}, \{\chvertex{i}{3}, \chvertex{i+1}{2}\}, \{\chvertex{i}{2}, \chvertex{i+1}{3}\}$, in total four edges.
\item For every clause $C = (\ell_1 \vee \ell_2 \vee \ell_3)\in \varphi_1$, we add a a clause agent~\clagents, who owns the edges in the clique formed by $\{\clavers{\ell_1}, \clavers{\ell_2}, \clavers{\ell_3}\}$, in total three edges.
\item We add a dummy agent $\agent{d}$, who owns edge $\{v_d, \chvertex{1}{3}\}$.
\end{compactitem}
	\contrstitle{Graph~$G_2$.} This graph is constructed in the same way as~$G_1$, except that $\varphi_2$ and $\Vars{2}$ are the underlying CNF formula and variable set.
	Let~$\players_{1}$ be the set of agents constructed when constructing $G_{1}$  and~$\players_2$ the set of agents constructed when constructing $G_2$.

	We construct the \allocation\ \alloc\ as follows:
	\begin{compactitem}[--]
		\item For every $z \in [2]$:
		\begin{compactitem}[*]
			\item In graph $G_z$, for every $\svar{i} \in \Vars z$, we allocate \agent{\svar{i}} cost $2$ and allocate \chagent{i} cost $ 1$.
			\item In graph $G_z$, for every $C \in \varphi_z$, we allocate \clagents\ cost $2$.
		\end{compactitem}
		\item In graph $G_1$, we allocate $\agent{d}$ cost $0$.
		\item In graph $G_{2}$, we allocate $\agent{d}$ cost $1$. 
	\end{compactitem}

	We first show the following two claims that establish a connection between our construction and the SAT-formulas.
	\begin{claim}\label{orgsizeverifVCforw}
		If $\varphi_1$ (resp.\ $\varphi_2$) is satisfiable, then $G_1$ (resp.\ $G_2$) has a vertex cover of size $3\varnr 1 + 2\clanr 1$ (resp.\ $3\varnr 2 + 2\clanr 2$).
		
	\end{claim}
	
	\begin{proof}\renewcommand{\qedsymbol}{\hfill (end of the proof of~\cref{orgsizeverifVCforw})~$\diamond$}
		We show the statement for $G_1$, the reasoning for $G_2$ is analogous. 
		We construct a vertex cover $V'$ of $G_1$ of size $3\varnr{1} + 2\clanr{1}$.
		Let $\sigma \colon X_1\rightarrow\{\trueT,\falseF\}$ be a satisfying assignment of $\varphi_1$.
		For every variable $\svar{i} \in \Vars 1$, if $\sigma$ assigns $\svar{i}$ to true, we add $v_{\svar{i}}$ to $V'$, otherwise we add $v_{\snegvar{i}}$.
		Regardless, we add $\chvertex{i}{1}$ and $\chvertex{i}{3}$ to $V'$.
		For every clause $C = (\ell_1 \vee \ell_2 \vee \ell_3) \in \varphi_1$, there must be $w \in [3]$ such that~$\ell_w$ is true under $\sigma$. We add $\{\clavers{\ell_1}, \clavers{\ell_2}, \clavers{\ell_3}\} \setminus \{\clavers{\ell_w}\}$ to $V'$.
		Clearly $|V'| = 3\varnr{z} + 2\clanr{z}$.
		
		Clearly, for every $\svar{i} \in \Vars 1$, the edges in the clique formed by $v_{\svar{i}}, v_{\snegvar{i}}$, and $\chvertex{i}{1}$ are covered.
		For every $C = (\ell_1 \vee \ell_2 \vee \ell_3)\in \varphi_z$, the edges in the clique formed by $\{\clavers{\ell_1}, \clavers{\ell_2}, \clavers{\ell_3}\}$ are also covered.
		For every $w \in [3]$, edge $\{v_{\ell_w}, \clavers{\ell_w}\}$ is also covered: If $\ell_w$ is true under $\sigma$, then $v_{\ell_w} \in V'$; otherwise $\clavers{\ell_w} \in V'$.
		For every $\svar{i} \in \Vars 1$,
		edge $\{\chvertex{i}{1}, \chvertex{i}{2}\}$  is covered by $\chvertex{i}{1}$,
		edges $ \{\chvertex{i}{2}, \chvertex{i}{3}\}$ and $\{\chvertex{i}{3}, \chvertex{i + 1}{2}\}$ by~$\chvertex{i}{3}$,
		and edge $\{\chvertex{i}{2}, \chvertex{i+1}{3}\}$ by $\chvertex{i + 1}{3}$.
		Edge $\{v_d, \chvertex{1}{3}\}$ is covered by $\chvertex{1}{3}$.
Thus $V'$ is a vertex cover of $G_1$ of size  $3\varnr{1} + 2\clanr{1}$.
	\end{proof}

	\begin{claim}\label{orgsizeverifVCbackw}
		
		Let $\pset_X \subseteq \{\agent{\svar{i}} \mid \svar{i} \in \Vars 1\}$ (resp.~$\{\agent{\svar{i}} \mid \svar{i} \in \Vars 2\}$), $\pset^* \subseteq \{\chagent{i} \mid \svar{} \in \Vars 1\}$ (resp.  $\{\chagent{i} \mid \svar{} \in \Vars 2\}$) and let $\psetC \subseteq \{\clagentM{j}{z} \mid C^z_j \in \varphi_1\}$ (resp. $\{\clagentM{j}{z} \mid C^z_j \in \varphi_2\}$).
		If the subgraph of $G_1$ (resp. $G_2$) induced by the edges of $\pset_X \cup \pset^* \cup \psetC \cup \{\agent{d}\}$ admits a vertex cover of size at most $2|\pset_X| + |\pset^*| +2|\psetC|$, then $\varphi_1$ (resp. $\varphi_2$) admits a satisfying assignment. 
		
		In particular, if $G_1$ (resp. $G_2$) admits a vertex cover $V'$ of size $3\varnr 1 + 2\clanr 1$ (resp. $3\varnr 2 + 2\clanr 2$), then $\varphi_1$ (resp. $\varphi_2$) admits a satisfying assignment.
	\end{claim}

	\begin{proof}\renewcommand{\qedsymbol}{\hfill (end of the proof of~\cref{orgsizeverifVCbackw})~$\diamond$}
		We show the statement for $G_1$, the reasoning for $G_2$ is analogous. 
		Observe that for every clause $\clagents \in \psetC$, we need at least two vertices to cover the edges owned by \clagents.
		These vertices may also cover edges between literal vertices and clause vertices.
		However, for every variable agent $\agent{\svar{i}} \in \pset_X$, we need at least two further vertices from $\{v_{\svar{i}}, v_{\snegvar{i}}, \chvertex{i}{1}\}$ to cover her edges.
		The selected vertices may cover edge $\{\chvertex{i}{1}, \chvertex{i}{2}\}$.
		
		In total, we need at least $2|\psetC| + 2|\pset_X|$ many vertices to cover the aforementioned edges, leaving us with at most $|\pset^*|$ vertices to cover the remaining edges.
		These are at least the edges $E^* \coloneqq \{\chvertex{1}{3}, v_d\} \cup \bigcup_{\chagent{i} \in \pset^*} \{\{\chvertex{i}{2}, \chvertex{i}{3}\},$ $\{\chvertex{i}{3},$ $\chvertex{i + 1}{2}\}, \{\chvertex{i}{2}, \chvertex{i + 1}{3}\}\}$.
		For every $\chagent{i} \in \pset^*$, at least one of the vertices $\{\chvertex{i}{2}, \chvertex{i}{3}\}$ must be in $V'$ to cover edge $\{\chvertex{i}{2}, \chvertex{i}{3}\}$, meaning we need at least $|\pset^*|$ vertices to cover them all.
		Thus we must use precisely $|\pset^*|$ vertices to cover them—more precisely, for every $\chagent{i} \in \pset^*$, \emph{exactly} one of the vertices $\{\chvertex{i}{2}, \chvertex{i}{3}\}$ is in $V'$, and no further vertices, including $v_d$, from $G^C$ are in $V'$.

We will now show use this fact to show that every chain agent in $\players_{1}$ must be in $\pset^*$.
Assume, towards a contradiction, that is not the case.
Consider the subgraph $G^C \coloneqq G_{1}[E^*]$.
First assume that $\pset^* \neq \emptyset$.
Since not every chain agent in $\players_{1}$ is in~$\psetC$, there must be an agent $\chagent{i} \in \pset^*$ such that $\chagent{i + 1} \notin \pset^*$ (where $\varnr z + 1 = 1$).
By earlier reasoning, we have that either $\chvertex{i}{2}$ or $\chvertex{i}{3}$ is in $V'$, but not both.
Moreover, as $\chagent{i + 1} \notin \pset^*$, we have that $\chvertex{i + 1}{2}, \chvertex{i + 1}{3} \notin V'$.
However, this implies that either edge $\{\chvertex{i}{2}, \chvertex{i+1}{3}\}$ or $\{\chvertex{i}{3}, \chvertex{i + 1}{2}\}$ is uncovered, a contradiction.
Now assume that $\pset^* = \emptyset$.
The set $V'$ must cover edge $\{v^3_1, v_d\}$ that belongs to $a_d$, but we know that $v_d \notin V'$ and since $\chagent{1} \notin \pset^*$ since $\pset^*$ is empty, we have that $v^3_1 \notin V'$, a contradiction.

		Thus we must have that every chain agent in $\players_{1}$ is in $\pset^*$.
		Recall that as $v_d \notin V'$, we must have, by earlier reasoning, that $\chvertex{1}{3} \in V'$ and $\chvertex{1}{2} \notin V'$. 
		To cover edge $\{\chvertex{1}{2}, \chvertex{2}{3}\}$ we need that $\chvertex{1}{3} \in V'$ and $\chvertex{1}{2} \notin V'$. By repeating this argument we obtain that for every $\svar{i} \in \Vars z$, it holds that $\chvertex{i}{3} \in V'$ and $\chvertex{i}{2} \notin V'$.
		However, for every $\svar{i} \in \Vars 1$, we must still cover edge $\{\chvertex{i}{1}, \chvertex{i}{2}\}$.
		As $\chvertex{i}{2} \notin V'$, we must have that $\chvertex{i}{1} \in V'$.
		Thus the size of $V'$ is at least $2 \varnr 1$.
		Moreover, for every variable agent $a_{\svar{i}}$ in $\pset_X$, we still need at least one vertex from $\{v_{\svar{i}}, v_{\snegvar{i}}\}$ to cover this edge, and for every clause agent $\clagents \in \psetC$, where $C= (\ell_1 \vee \ell_2 \vee \ell_3)$, we need at least two agents from $\{\clavers{\ell_1}, \clavers{\ell_2}, \clavers{\ell_3}\}$ to cover the edges between these vertices.
		Thus $|V'| \geq 2 \varnr 1 + |\pset_X| + 2|\psetC|$.
		Moreover, we assume that $|V'| \leq |\pset^*|+ 2|\pset_X| + 2|\psetC|$. As $|\pset^*| = \varnr 1$, we obtain that
		
		\begin{align*}		
		 2\varnr 1 + |\pset_X| + 2|\psetC| \leq \varnr 1 + 2|\pset_X| + 2|\psetC| \quad& \Longrightarrow
		\quad  \varnr 1 \leq |\pset_X| \\ & \Longrightarrow\quad |\pset_X| = \varnr 1.
		 \end{align*}
		In other words, for every $\svar{i} \in \Vars 1$, we have that $\agent{\svar{i}} \in \pset_X$.
		
		As we still require  $ 2|\psetC|$ vertices to cover the edges belonging to the agents in $\psetC$ and $2\varnr 1$ vertices to cover the edges belonging to the chain agents, we can have at most $\varnr 1$-many vertices in $V'$ whose all incident edges—restricted to the subgraph induced by the edges owned by the agents in $\psetC \cup \pset_X \cup \pset^*$—belong to variable agents. 
		This means that for every $\svar{i} \in \Vars 1$, we must have \emph{exactly} one vertex from $\{v_{\svar{i}}, v_{\snegvar{i}}\}$ in $V'$.
		Moreover, for every $C = (\ell_1 \vee \ell_2 \vee \ell_3) \in \varphi_z$ such that $\clagents \notin \psetC$, we have that $\clavers{\ell_w} \notin V'$ for any $w \in [3]$.
		
Consider assignment $\sigma \colon \Vars{} \to \{\trueT, \falseF\}$ that assigns a variable $\svar{} \in \Vars 1$ to true if $v_{\svar{}} \in V'$ and false otherwise. 
Assignment $\sigma$ must be a satisfying assignment: Assume towards a contradiction that there is a clause $C = (\ell_1 \vee \ell_2 \vee \ell_3) \in \varphi_1$ that is not satisfied under $\sigma$.
This means that $v_{\ell_w} \notin V'$ for any $w \in [3]$.
If $\clagents \notin \psetC$, then edge $\{v_{\ell_w}, \clavers{\ell_w}\}$ is not covered for any $w \in [3]$, a contradiction.
Thus assume $\clagents \in \psetC$.
As we can have at most two vertices from $\{\clavers{\ell_w} \mid w \in [3]\}$ in $V'$, there must be $w \in [3]$ such that $\clavers{\ell_w} \notin V'$.
But then edge $\{v_{\ell_w}, \clavers{\ell_w}\}$ is not covered, a contradiction.
	\end{proof}
	
	We now show that $I'$ is a yes-instance of \textsc{SAT-UNSAT} if and only if $\alloc$ is \stable.
	
	$(\Longrightarrow)$ First assume $I'$ is a yes-instance of \textsc{SAT-UNSAT}.

	Let us start by showing the following claim that restricts the blocking coalitions we must consider:
	
	\begin{claim}\label{clm:VCdummyblock}
		No set of agents $\pset \subseteq \players_{1} \setminus \{\agent{d}\}$ (resp. $\players_{2} \setminus \{\agent{d}\}$) may block.
	\end{claim}
	\begin{proof}\renewcommand{\qedsymbol}{\hfill (end of the proof of~\cref{clm:VCdummyblock})~$\diamond$}
		We show the statement for $\players_1$, the reasoning for $\players_2$ is analogous. 
		Let $\pset_X$ be the set of variable agents in $\pset$, $\psetC$ be the set of clause agents in $\pset$, and $\pset^*$ the set of chain agents in $\pset$.
		We observe that by construction $\sum_{i \in \pset}\allocel i = 2|\pset_X| + 2|\psetC| + |\pset^*|$.
		
		Moreover, we claim that $\util(S) \geq 2|\pset_X| + 2|\psetC| + |\pset^*|$, i.e., the minimum vertex cover of $G_{z}[\bigcup_{i \in \pset}E_i]$ is at least size $2|\pset_X| + 2|\psetC| + |\pset^*|$.
		Let $V'$ be a minimum vertex cover of $G_{1}[\bigcup_{i \in \pset}E_i]$.
		For every clause agent $\clagents \in \psetC$, where $C = (\ell_1 \vee \ell_2 \vee \ell_3)$, there must be at least two vertices in $V'$  from $\{\clavers{\ell_w} \mid w \in [3]\}$ to cover the clique formed by these vertices.
		Similarly, for every $\agent{\svar{i}} \in \pset_X$, there must be at least two vertices in $V'$ to cover the clique $\{v_{\svar{i}}, v_{\snegvar{i}}, \chvertex{i}{1}\}$.
		For every chain agent $\chagent{i} \in \pset^*$, we need at least one vertex to cover edge $\{\chvertex{i}{2}, \chvertex{i}{3}\}$.
		As none of the above enumerated sets overlap, we must have that $|V'| \geq 2|\pset_X| + 2|\psetC| + |\pset^*|$.
		Thus $\sum_{i \in \pset}\allocel{i} = 2|\pset_X| + 2|\psetC| + |\pset^*| \leq \util(\pset)$, as required.
	\end{proof}
	Since the graphs $G_1$ and $G_2$ are not connected to each other and no agent owns edges on both of them, by \cref{obs:sepgraphs}, it is sufficient for us to show that $\alloc$ restricted to each of these graphs is \stable.
	We have four possible cases:
	\begin{description}
		\item[Case 1:]  A subset $\pset \subseteq \players_1$ blocks \alloc.
		By \cref{clm:VCdummyblock}, we have that $\agent{d} \in \pset$ and $\sum_{i \in \pset \setminus \{\agent{d}\}} \allocel i \leq \util(\pset \setminus \{\agent{d}\})$.
		Since $\agent{d}$  is allocated cost $0$, it holds that $\sum_{i \in \pset \setminus \{\agent{d}\}} \allocel i = \sum_{i \in \pset}\allocel i$.
		Additionally, since the size of a minimum vertex cover cannot decrease by adding edges, we have that $\util(\pset \setminus \{\agent{d}\}) \leq \util(\pset)$.
		Thus we obtain that $\sum_{i \in \pset} \allocel i = \sum_{i \in \pset \setminus \{\agent{d}\}} \allocel i \leq \util(\pset \setminus \{\agent{d}\}) \leq \util(\pset)$, a contradiction to $\pset$ blocking.
		\item[Case 2:] We have that \alloc\ restricted to $G_1$ is not a \preimputation.
		Observe that $\sum_{i \in \players_1}\allocel i = 3\varnr{1} + 2\clanr{1}$.
		As the previous case implies that $\sum_{i \in \players_1} \allocel i \leq \util(\players_1)$, we must have that $ 3\varnr{1} + 2\clanr{1} < \util(\players_1)$, i.e., a minimum vertex cover of $G_1$ is strictly larger than $3\varnr{1} + 2\clanr{1}$.
		However, as $\varphi_1$ admits a satisfying assignment, graph $G_1$ admits a vertex cover of size $3\varnr{1} +2 \clanr{1}$ by \cref{orgsizeverifVCforw}, a contradiction.
		\item[Case 3:]A subset $\pset \subseteq \players_{2}$ blocks \alloc.
		By \cref{clm:VCdummyblock}, we have that $\agent{d} \in \pset$.
		Let $\pset_X$ be the set of variable agents in $\pset$, $\psetC$ be the set of clause agents in $\pset$, and $\pset^*$ the set of chain agents in $\pset$.
		We have that $\sum_{i \in \pset}\allocel i =  2|\pset_X| + |\pset^*| + 2|\psetC| + 1$.
		Thus we must have that $G_{2}[\bigcup_{i \in \pset}E_i]$ admits a dominating set $V'$ of size at most $2|\pset_X| + |\pset^*| + 2|\psetC|$.
		By \cref{orgsizeverifVCbackw}, formula $\varphi_2$ is then satisfiable.
		This however contradicts that $I'$ is a yes-instance of \textsc{SAT-UNSAT}.
		\item[Case 4:] We have that \alloc\ restricted to $G_2$ is not a \preimputation.
		The previous case establishes that $\sum_{i \in \players_{2}}\allocel i \leq \util(\players_{2})$, as otherwise~$\players_{2}$ would block.
		Thus we must have that $ \util(\players_{2}) > \sum_{i \in \players_{2}}\allocel i = 3\varnr{2} + 2\clanr{2} + 1$, i.e., a minimum vertex cover of $G_{2}$ is strictly larger than $3\varnr{2} + 2\clanr{2} + 1$.
		Towards a contradiction, we construct a vertex cover $V'$ of~$G_{2}$ of size $3\varnr{2} + 2\clanr{2} + 1$.
		Let $V' \coloneqq \{\clavers{\ell_1}, \clavers{\ell_2} \mid C = (\ell_1 \vee \ell_2 \vee \ell_3) \in \varphi_2\} \cup \{v_{\svar{i}}, v_{\snegvar{i}} \mid \svar{i} \in \Vars 2\} \cup \{\chvertex{j}{2} \mid C \in \varphi_2\} \cup \{v_d\}$.
		Clearly $|V'| = 3\varnr{2} + 2\clanr{2} + 1$.
		It remains to show this is indeed a vertex cover.
		For every $C = (\ell_1 \vee \ell_2 \vee \ell_3) \in \varphi_2$, clearly $\clavers{\ell_1}$ and $\clavers{\ell_2}$ cover the edges belonging to \clagents.
		For every $\svar{i} \in \Vars 1$, the edges connecting $v_{\svar{i}}$ and $v_{\snegvar{i}}$ to the clause vertices are clearly covered by $v_{\svar{i}}$ and $v_{\snegvar{i}}$. The vertices $v_{\svar{i}}$ and $v_{\snegvar{i}}$ also cover the edges that connect them to $\chvertex{i}{1}$. Thus all the edges belonging to $\agent{\svar{i}}$ are covered.
		Edges $\{\chvertex{i}{1}, \chvertex{i}{2}\}, \{\chvertex{i}{2}, \chvertex{i}{3}\}$, and $\{\chvertex{i}{2}, \chvertex{i + 1}{3}\}$ are covered by~\chvertex{i}{2}, whereas edge $\{\chvertex{i}{3}, \chvertex{i+1}{2}\}$ is covered by \chvertex{i + 1}{2}.
		Thus all the edges belonging to \chagent{i}\ are covered.
		Finally, edge $\{v_d, \chvertex{i}{3}\}$ that belongs to $\agent{d}$ is covered by $v_d$.
	\end{description}
	This concludes the proof of forward direction.
	
	$(\Longleftarrow)$ Assume $\alloc$ is \stable.
	Because $G_1$ and $G_{2}$ are completely disjoint and no agent has edges on both graphs, we must have that $\alloc$ is \stable\ when restricted to either of these graphs by \cref{obs:sepgraphs}.
	
	We first show that $\varphi_1$ admits a satisfying assignment.
	Since $\sum_{i \in \players_{1}} \allocel i = 3\varnr{1} + 2\clanr{1}$, graph~$G_{1}$ must admit a vertex cover~$V'$ of size $3\varnr{1} + 2\clanr{1}$.
	By \cref{orgsizeverifVCbackw}, this implies that $\varphi_1$ admits a satisfying assignment.

	Next we show that $\varphi_2$ cannot admit a satisfying assignment.
	Since $\sum_{i \in \players_{2}} \allocel i = 3\varnr{2} + 2\clanr{2} + 1$, graph~$G_{2}$ cannot admit a vertex cover $V'$ of size $3\varnr{2} + 2\clanr{2}$.
	By the contra-positive of \cref{orgsizeverifVCforw}, the formula $\varphi_2$ cannot admit a satisfying assignment.
	Thus $I'$ is a yes-instance of \textsc{SAT-UNSAT}.	
\end{proof}
}

\mypara{Core Existence.}
We now move to consider the existence problem \probPVCCE.
We first use ellipsoid method inspired by \citet{Biro19} to show \thetaC-containment.

\begin{restatable}[\appsymb]{theorem}{thmVCCEcont}\label{thm:VCCEthetacont}
\probPVCCE\ is contained in~\thetaC.
\end{restatable}

\begin{proof}
 Our proof will utilize the fact that the following hinted version of \probPVCCE\ is in \coNP, where the input additionally has the cardinality of a minimum vertex cover:

\decprob{\probPVCCEhint~(\probPVCCEhintshort)}{An instance $G = (V, E = E_1 \cup \cdots \cup E_{\coln})$ and $\minvc$ which is the cardinality of a minimum vertex cover of $G$.}{Is $G$ a yes-instance of \probPVCCE?}

We show that this problem is contained in \coNP:

\begin{mainclaim}\label{clm:minvcprobpvcceconp}
\probPVCCEhintshort\ is contained in \coNP.
\end{mainclaim}

\begin{proof}\renewcommand{\qedsymbol}{\hfill (end of the proof of~\cref{clm:minvcprobpvcceconp})~$\diamond$}

  We show \coNP-containment using the ellipsoid method; see \cref{sec:ellipsoid} for more details.
We show a polynomial-time verifiable witness for a no-instance; i.e., given this witness, we can verify in polynomial time that $G$ does not admit a \stable\ \allocation.
Our proof proceeds in two parts: (1) Given a no-instance $(G, \minvc)$ of \probPVCCEhintshort, we show that a polynomial-space witness exists, and (2) we show that given a certificate constructed in (1), we can verify in polynomial time that $G$ does not admit a \stable\ \allocation.  

Note that our approach does \emph{not} show that the original \probPVCCE\ problem is in \coNP, because the polynomial-time verification of the witness in Part 2 relies on knowing the minimum vertex cover size of $G$, which is \NP-hard to find.

\mypara{Part 1.}
Let $(G, \minvc)$ be a no-instance of \probPVCCEhintshort.
We show that a certificate for no-instance must exist, although we do not yet show how to verify it.

Observe that an \allocation\ \alloc\ is \stable\ if and only if it is satisfies the linear constraints described in \cref{subsec:coopgamesstable}.
Thus we can formulate \probPVCCEhintshort\ as a linear program.
Let $Q$ be the separation oracle for solving this linear program. 
Recall that $Q$ does not necessarily have to run in polynomial time, but since we can brute force all the possible blocking coalitions, it must exist.
Since $G$ does not have a \stable\ allocation, given an allocation $\alloc$,
oracle $Q$ either returns a blocking coalition $\pset \subseteq \players$,
or answers that $\alloc$ is not a \preimputation, i.e., $\sum_{i \in \players}\allocel i \neq \minvc$, where $\minvc$ is the cardinality of a minimum vertex cover of~$G$.
Let $\hat{Q}$ be the set that contains a triple $(\alloc, \pset, V'_\pset)$ for every call to $Q$ over the course of solving \probPVCCEhintshort\ on $G$, where (1) $\alloc$ is the allocation that is queried for $Q$, (2) $\pset$ is a coalition that blocks $\alloc$, or $\pset = \players$ if $\alloc$ is not a \preimputation, and (3) $V'_\pset$ is a vertex cover of $G[\bigcup_{i \in \pset}E_i] $ such that $|V'| < \sum_{i \in \pset}\allocel i$, or $\emptyset$ if $\alloc$ is not a \preimputation.
Recall that since $G$ does not admit a \stable\ \allocation, oracle $Q$ never returns that \alloc\ is in the core.
Since the ellipsoid method runs in time $T \cdot \operatorname{poly}|V|$ where $T$ is the running time of $Q$, oracle $Q$ is called polynomially many times over the execution of the ellipsoid method, and thus the size of $\hat{Q}$ is polynomial on the input size.

\mypara{Part 2.} Assume we are given $\hat{Q}$ as a witness that $G$ does not admit a \stable\ \allocation.
We can verify that $\hat{Q}$ only contains violated constraints in polynomial time:
As every \preimputation\ of $G$ satisfies $\sum_{i \in \players}\allocel i = \minvc$, we can verify in polynomial time that for every $(\alloc, \pset, V'_\pset) \in \hat{Q}$ the conditions (1)-(3) are not satisfied.

Next, we solve \probPVCCEhintshort\ on $(G, \minvc)$ with $\hat{Q}$ as an oracle.
By construction, set $\hat{Q}$ contains a violated constraint for every $\alloc$ that is queried from $Q$ over the course of the execution of the ellipsoid method on $G$, and thus we can use $\hat{Q}$ as an oracle.
Since we query $\hat{Q}$ in polynomial time, the ellipsoid method runs in polynomial time.
As the ellipsoid method and~$\hat{Q}$ are both correct, the ellipsoid method returns no if and only if $G$ does not admit a \stable\ \allocation.
This verifies $\hat{Q}$ in polynomial time.
\end{proof}

We are now ready to show \thetaC-containment of \probPVCCE.
Let $G = (V, E = E_1 \cup \cdots \cup E_{\coln})$ be an instance of \probPVCCE.
We first use binary search on $\coln$ to find the cardinality~$\minvc$ of a minimum vertex cover of $G$.
This requires $\log(\coln)$ many calls to an \NP-oracle solving \VC.
Now we make one further \NP-oracle call to the \emph{co-problem} of \probPVCCEhintshort\ with the input $G$, which by \cref{clm:minvcprobpvcceconp} is contained in \NP.
If this call returns no, then~$G$ admits a \stable\ \allocation\ and we return yes, otherwise we return no. 

Clearly we return yes if and only if $G$ admits a \stable\ \allocation.
In total we require $\log(\coln)$ \NP-oracle calls, which is in $O(\log(|V|))$, as required.
\end{proof}

We complement the above result with \thetaC-hardness, thereby showing that \probPVCCE\ is \thetaC-complete.

\begin{restatable}[\appsymb]{theorem}{thmVCtheta}\label{thm:VCtheta}
\probPVCCE\ is \thetaC-hard even when $\nbOrg = 4$.
\end{restatable}

\begin{proof}

\def \xx {1}
\def \xy {0.7}
  \begin{figure}[t!]
    \centering
   
  \begin{tikzpicture}[black, scale=1,every node/.style={scale=0.9}]
  
\begin{pgfonlayer}{fg}
 \foreach \x / \y / \n / \nn  / \pos in {
	4.5/5/vp/v'/left,
	6/5/u1/u_1/below right,
	6/6/u2/u_2/above right} {
      \node[vcvertex, label=\pos :$\nn$] (\n) at (\x*\xx,\y*\xy) {};
    }
 \end{pgfonlayer}
 
 \node[draw=green!60!black, ellipse, inner sep=10pt, line width=2pt, dashed, minimum width=100pt] (G) at (3.3*\xx,5.3*\xy) {$G$};
 
  \foreach \x / \y / \n  in {3.7/4.7/v1,3.8/5.8/v2} {
      \node[hiddenV] (\n) at (\x*\xx,\y*\xy) {};
    }
    
    \foreach \s/\t in {vp/u1} {
      \draw[firstagentedge] (\s) -- (\t);
      \begin{pgfonlayer}{bg}

      		\draw[firstagentA] \hedgeii{\s}{\t}{2mm};
     \end{pgfonlayer}
    }
     \foreach \s/\t in {vp/u2} {
      \draw[secondagentedge] (\s) -- (\t);
     \begin{pgfonlayer}{bg}

      		\draw[secondagentA] \hedgeii{\s}{\t}{2mm};
     \end{pgfonlayer}
    }
    \foreach \s/\t in {u1/u2} {
      \draw[thirdagentedge] (\s) -- (\t);
           \begin{pgfonlayer}{bg}

      		\draw[thirdagentA] \hedgeii{\s}{\t}{2mm};
     \end{pgfonlayer}
    }
    
     \foreach \s/\t in {vp/v1,vp/v2} {
      \draw[opacity=0.5] (\s) -- (\t);
    }

\draw[decoration={brace},decorate] ($(u1)+(0,-0.2)$) --node[right,yshift=-10pt] (E1) {$E_1$} ($(vp)+(0,-0.2)$);
\draw[decoration={brace},decorate] ($(vp)+(-0.1,0.2)$) --node[right,yshift=10pt,xshift=-8pt] (E2) {$E_2$} ($(u2)+(-0.1,0.2)$);
\draw[decoration={brace},decorate] ($(u2)+(0.2,0)$) --node[right,xshift=10pt] (E3) {$E_3$} ($(u1)+(0.2,0)$);

\draw[decoration={brace},decorate] ($(G)+(-1.6,0.7)$) --node[right,yshift=8pt] (E4) {$E_4$} ($(G)+(1.6,0.7)$);
\end{tikzpicture}

\caption{Illustration for the proof of \cref{thm:VCtheta}.
}\label{fig:thm:VCtheta}
\vspace{0.5cm}
\end{figure}

We reduce from the $\thetaC$-hard \VCM problem~\cite{HSV05Kemeny}. 
	\decprob{\VCM}{A graph $G=(V,E)$ and a vertex $v'\in V$.}{Does $G$ admit a \emph{minimum} vertex cover $V'$ such that $v'\in V'$?}
	
	Let $I = (G=(V,E), v')$ be an instance of \VCM. 
	We construct an instance $I'$ of \probPVCCE\ with four agents $\players = [4]$. The reduction is illustrated in \cref{fig:thm:VCtheta}.
	Let the graph of the reduced instance be $\hat{G} \coloneqq (V \cup \{u_1, u_2\}, E_1 \cup E_2 \cup E_3 \cup E_4)$, where $E_1 \coloneqq \{v', u_1\}$, $E_{2} \coloneqq \{v', u_2\}$, $E_{3} \coloneqq \{u_1, u_2\}$, and $E_{4} \coloneqq E$.
	
	We claim that $I$ is a yes-instance of \VCM\ if and only $I'$ is a yes-instance of \probPVCCE.
	
	$(\Longrightarrow)$: Assume $I$ is a yes-instance of \VCM.
	Let \minvc\ be the cardinality of a minimum vertex cover of $G$ and let $V'$ be a vertex cover of $G$ of size $\minvc$ that contains $v'$.
	We claim that the allocation $\alloc = ( 0, 0, 1, \minvc )$ is in the core.
	
	First, observe that the minimum vertex cover of $\hat{G}$ is of size $\minvc + 1$.
	For example, the set $V' \cup \{u_1\}$ is such a vertex cover:
	All the edges in~$E$ are clearly covered, as $V'$ is a vertex cover of $G$.
	Edge $\{v', u_1\}$ is covered by both $v'$ and $u_1$, edge $\{v',u_2\}$ is covered by $v'$, and edge $\{u_1, u_2\}$ is covered by $u_1$.
	There cannot be a smaller vertex cover, as we need $\minvc$ vertices from $V$ to cover the edges in~$E$, leaving no vertices to cover edge $\{u_1, u_2\}$.
	Thus \alloc\ is a \preimputation.
	
	To show \alloc\ is in the core, observe that a minimum vertex cover of $\hat{G}[E_4]$ is of size \minvc, and thus $\{4\}$ does not block \alloc.
Every subset $\pset \subseteq [3]$ satisfies $\sum_{i \in \pset}\allocel i \leq 1 \leq \util(\pset)$, and thus cannot block~\alloc.
For every $\pset \subseteq [2], \pset \neq \emptyset$ we have that $\util(\pset \cup \{4\}) = \minvc = \sum_{i \in \pset \cup \{4\}} \allocel i$ and $\util(\pset \cup \{3,4\}) = \minvc + 1= \sum_{i \in \pset \cup \{3, 4\}} \allocel i$, as required.

	$(\Longleftarrow)$: Assume $I'$ is a yes-instance of \probPVCCE. Let $\alloc \in \mathds{R}^{4}$ be in the core.
	Assume, towards a contradiction, that~$G$ does not admit a minimum vertex cover containing $v'$.
	Then the size of a minimum vertex cover of $\hat{G}$ is necessary $\minvc + 2$, where $\minvc$ is the size of a minimum vertex cover of $G$.
	This is because we need $\minvc$ vertices from $V \setminus \{v'\}$ to cover the edges in $E$, and two vertices from $\{v', u_1, u_2\}$ to cover the edges $\{v', u_1\}, \{v', u_2\}, \{u_1, u_2\}$.
	We have that $\allocel 4 \leq \minvc$, as otherwise $\{4\}$ would block.
	Thus $\allocel 1 + \allocel 2 + \allocel 3 \geq \minvc + 2 - \minvc = 2$.
	
However, for each pair $i,j\in[3]$ with $i\neq j$, it holds that $G[E_i\cup E_j]$ has a minimum vertex cover of size $1$, which by reasoning of \cref{Ex:VCov} implies that $\allocel 1 + \allocel 2 + \allocel 3 \leq \frac{3}{2}$, a contradiction.
\end{proof}

\newcommand{\vcX}{\ensuremath{6\varnr{}}}
\newcommand{\vcXp}{\ensuremath{36\varnr{}^2}}
\newcommand{\vcXcost}{\ensuremath{\frac{1}{3\varnr{}}}}
\newcommand{\vcXcostmul}[1]{\ensuremath{\frac{#1}{3\varnr{}}}}
\newcommand{\satvcsizelong}{\ensuremath{2\varnr{}\cdot \vcX + \varnr{} \cdot \vcXp + 2 \clanr{} \cdot \vcX + \satscore}}
\newcommand{\satvcreq}{\ensuremath{12\varnr{}^2 + 36 \varnr{}^3 + 12\clanr{}\varnr{}}}
\newcommand{\satvcsize}{\ensuremath{\satvcreq + \satscore}}
\newcommand{\vcXpmnX}{30\varnr{}^2}
\newcommand{\vcXn}{\vcX^2}
\newcommand{\vcXmnp}{25\varnr{}^2}

The previous reduction has one large agent. Next, we show that \probPVCCE\ remains \thetaC-hard even when the agent size is bounded by a constant.
The main part of the reduction is inspired by the standard reduction from \textsc{SAT} to \VC, but we duplicate vertices and then add ``prize'' vertices to ensure that a minimum vertex cover of the graph both corresponds to a satisfying assignment that minimizes the number of true variables.
We also add some gadgets to ensure correctness.

\begin{restatable}[\appsymb]{theorem}{thmVCorgsize}\label{thm:VCorgsize}
\probPVCCE\ is \thetaC-hard even when $\Orgsize=~9$. 
\end{restatable}

\newcommand{\elag}[3]{\agent{y_{#1}}^{#2,#3}}
\newcommand{\varag}[3]{\agent{\svar{#1}}^{#2,#3}}
\newcommand{\clag}[3]{\agent{C_{#1}}^{#2,#3}}
\newcommand{\eag}[1]{\hagent{#1}}
\newcommand{\prizev}[1]{\ensuremath{v^*_{#1}}}
\newcommand{\Pver}[1]{\ensuremath{V_{#1}}}
\newcommand{\pver}[2]{\ensuremath{v^{#2}_{#1}}}
\newcommand{\Nver}[1]{\ensuremath{\bar{V}_{#1}}}
\newcommand{\nver}[2]{\ensuremath{\bar{v}^{#2}_{#1}}}
\newcommand{\Clavar}[2]{\ensuremath{U_{#1,#2}}}
\newcommand{\Clav}[2]{\Clavar{#1}{\ell_{#2}}}
\newcommand{\clavar}[3]{\ensuremath{u^{#3}_{#1,#2}}}
\newcommand{\clav}[3]{\clavar{#1}{\ell_{#2}}{#3}}
\newcommand{\dver}[1]{\ensuremath{w_{#1}}}
\newcommand{\Bcv}[2]{\ensuremath{D_{#1,#2}}}
\newcommand{\bcv}[3]{\ensuremath{d^{#2}_{#1,#3}}}
\newcommand{\Bccv}[2]{\ensuremath{D_{\varnr{} + #1,#2}}}
\newcommand{\bccv}[3]{\ensuremath{d^{#2}_{\varnr{} + #1,#3}}}
\newcommand{\lcv}[4]{\ensuremath{b^{#2,#3}_{#1,#4}}}
\newcommand{\ever}[1]{\ensuremath{s_{#1}}}
\newcommand{\multivertex}{multi-vertex}
\newcommand{\multivertices}{multi-vertices}
\newcommand{\multiedge}{multi-edge}

\newcommand{\Satel}{\ensuremath{Y}}
\newcommand{\satel}[1]{\ensuremath{y_{#1}}}
\newcommand{\sind}{\ensuremath{g}}

\newcommand{\Eag}[1]{\ensuremath{E_{#1}}}
\newcommand{\GindVC}[1]{\ensuremath{G[\bigcup_{\agent{} \in #1}\Eag{\agent{}}]}}

\newcommand{\Elag}[1]{\ensuremath{A_{#1}}}
\newcommand{\Varag}[1]{\Elag{\svar{#1}}}
\newcommand{\Clag}[1]{\Elag{C_{#1}}}

\newcommand{\psetcount}[1]{\ensuremath{\hat{k}(#1)}}

\newcommand{\satscore}{\ensuremath{\beta}}
\newcommand{\satscoreb}{\ensuremath{\beta'}}
\newcommand{\mind}{z}
\newcommand{\mindd}{w}
\newcommand{\hmind}{\hat{z}}
\newcommand{\hmindd}{\hat{w}}
\newcommand{\tind}{t}

\newcommand{\nonemptyvarcount}{\gamma}
\newcommand{\nonemptyfillvarcount}{\gamma'}
\newcommand{\allsatagents}{\ensuremath{V^{\varphi}}}

\appendixproofwithstatement{thm:VCorgsize}{\thmVCorgsize*}{
\begin{proof}

	We start by showing that the following problem is \thetaC-hard:
	\decprob{\minasgwosatmember}{
		A 2-SAT formula $\varphi$ over the set of variables \Vars{}, a variable $\svar{}' \in \Vars{}$.}{
		Does $\varphi$ admit a satisfying assignment $\sigma$ such that~$\svar{}'$ is true under $\sigma$ and there is no satisfying assignment $\sigma'$ such that~$\sigma'$ sets strictly fewer variables to true than $\sigma$?
	}
	
	\begin{claim}\label{clm:maxtwosat}
		\minasgwosatmember\ is \thetaC-hard even if every variable appears at most twice positive and once negative.
	\end{claim}
	\begin{proof}\renewcommand{\qedsymbol}{\hfill (end of the proof of~\cref{clm:maxtwosat})~$\diamond$}
		We reduce from \VCM. Let $(G = (V, E), v')$ be an instance of \VCM.
		Let us construct an assignment $\varphi$ as follows:
		We create  a variable $\svar{i}$ for every vertex $v_i \in V$.
		For every edge $\{v_i, v_j\} \in E$, we construct a clause $(\svar{i} \vee \svar{j})$.
		For every subset $S \subseteq V$, let $\sigma_S$ be an assignment that sets $\svar{i}$ to true for every $v_i \in V$, and $\svar{j}$ to false for every $v_j \in V \setminus S$.
		Clearly, we have that $S$ is a vertex cover if and only if $\sigma_S$ is a satisfying assignment.
		Moreover, the number of vertices in $S$ is precisely the number of variables set true by $\sigma_S$.
		Finally, we have that $v' \in S$ if and only if $\sigma_S$ sets $\svar{}'$ to true.
		Thus $I$ is a yes-instance of \VCM\ if and only if $(\varphi, \svar{}')$ is a yes-instance of \minasgwosatmember.
		
		Next we show that we can bound the number each variable appears by strategies of \citet{BKS-2bal3sat-2003}.
		Let $(\varphi, v')$ be an instance of \minasgwosatmember.
		Observe that, by the previous construction, we can assume that every variable only appears as a positive literal.
		Let $\delta$ be the maximum number of times a literal appears in $\varphi$.
		Let us construct a modified instance $(\varphi', v'')$ as follows:
		For every $\svar{i} \in \Vars{}, z \in [\delta]$, we construct a new variable $\svar{i}^z$.
		Order the clauses where each variable appears arbitrarily.
		For each clause $C = (\svar{i} \vee \svar{j})$, let $z_i$ tell us which appearance of $\svar{i}$ and $z_j$ which appearance of $\svar{j}$ this is.
		We add the clause $(\svar{i}^{z_i} \vee \svar{j}^{z_j})$ to $\varphi'$.
		Moreover, for every $\svar{i} \in \Vars{}, z \in [\delta]$, add the clause $(\snegvar{i}^z \vee \svar{i}^{z + 1})$, where $\delta + 1 = 1$.
		Observe that together these clauses are equivalent to $(\svar{i}^1 \Longrightarrow \svar{i}^2  \Longrightarrow \cdots  \Longrightarrow \svar{i}^\delta  \Longrightarrow \svar{i}^1)$ which means that $\svar{i}^1 \iff \svar{i}^2  \iff \cdots  \iff \svar{i}^\delta$, i.e., in any satisfying assignment, all the copies of $\svar{i}$ are assigned the same way.
		For the special vertex~$\svar{}'$, let the special vertex of the reduced instance be $\svar{}'^{1}$
		
It is easy to observe that every variable appears at most twice positive and once negative. 
Moreover, clearly every satisfying assignment of $\varphi$ has a corresponding satisfying assignment of $\varphi'$ and vice versa. 
Moreover, if a satisfying assignment $\sigma$ of $\varphi$ sets $\ell$ variables to true, clearly the corresponding satisfying assignment of $\sigma'$ sets~$\delta \ell$ variables true. 
Finally, if a a satisfying assignment sets $\svar{}'$ to true, then the special variable of the reduced instance $\svar{}'^1$ must also be true.
\end{proof}
	
	In what follows, we may call a set of \vcX-many vertices a \myemph{\multivertex}.
	A \myemph{\multiedge} between two \multivertices\ $U_1 = \{u^{\mind}_1 \mid \mind \in [\vcX]\}$, $U_2 = \{u^{\mind}_2 \mid \mind \in [\vcX]\}$ is the set of all possible edges between the sets $U_1$ and $U_2$, i.e., the set $\{\{u^{\mind}_1, u^{\mindd}_2\} \mid (\mind, \mindd) \in [\vcX] \times [\vcX]\}$.
	
	The following observation shows that \multivertices\ and \multiedge s behave similarly to normal vertices and edges:
	\begin{observation}\label{obs:multiedge}
		Let $G$ be a graph that contains \multivertices~$U_1$ and~$U_2$ and a \multiedge\ between them. Then for every vertex cover~$V'$ of $G$ it holds that $U_1 \subseteq V'$ or $U_2 \subseteq V'$.
	\end{observation}
	\begin{proof}\renewcommand{\qedsymbol}{\hfill (end of the proof of~\cref{obs:multiedge})~$\diamond$}
		Assume, towards a contradiction, that there is a vertex cover~$V'$ of $G$ such that there are $u^{\mind}_1 \in U_1$ and $u^{\mindd}_2 \in U_2$ such that $u^{\mind}_1,u^{\mindd}_2 \notin V'$.
		Then $V'$ does not cover edge $\{u^{\mind}_1,u^{\mindd}_2\}$, a contradiction.
	\end{proof}

	Let $I = (\varphi, \svar{}')$ be an instance of \minasgwosatmember\ where every variable appears at most twice positive and once negative. 
As shown in \cref{clm:maxtwosat}, this is \thetaC-hard.
Let $\varphi$ be over the set of variables $\Vars{} = \{\svar{i} \mid i \in [\varnr{}]\}$, and let $\clanr{} \coloneqq |\varphi|$.
	We may assume, without loss of generality, that $\svar{}' = \svar{1}$.
For notational convenience, let $\Satel = \varphi \cup \Vars{}$.
We let $\Satel = \{\satel 1, \dots, \satel{\varnr{} + \clanr{}}\}$, where for every $\svar{i} \in \varnr{}$, we name $\satel i = \svar i$, and for every $C_j \in \varphi$, we name $\satel{\varnr{} + j} = C_j$.
	
\def \xx {1}
\def \xy {0.8}
  \begin{figure}[t!]
    \centering
   
  \begin{tikzpicture}[black, scale=1,every node/.style={scale=0.9}]
  
\begin{pgfonlayer}{fg}
 \foreach \x / \y / \n / \nn  / \pos in {
 5.5/7/vp/\Pver{i}/{above right},
      6.5/7/vn/\Nver{i}/{below right},
      4.5/7.5/c1/\Clavar{i_1}{\svar{i}}/above,
      3.5/7.5/c1p/\Clavar{i_1}{\snegvar{i'}}/above,
      4.5/6.5/c2/\Clavar{i_2}{\svar{i}}/left,
      7.5/7.5/cn/\Clavar{\bar{i}}{\snegvar{i}}/right,
      4/4.7/B11/\Bcv{i}{1}/{below right},
      4.5/5.5/B12/\Bcv{i}{2}/{below right},
      3.5/5.5/Bx3/\Bcv{i - 1}{3}/{below left},
      7/4.7/B21/\Bcv{i + 1}{1}/{below right},
      7.5/5.5/B22/\Bcv{i + 1}{2}/{below right},
      6.5/5.5/B13/\Bcv{i}{3}/{below left}
      } {
      \node[bigvertex, label=\pos :$\nn$] (\n) at (\x*\xx,\y*\xy) {};
      \node[vcvertex] () at (\x*\xx,\y*\xy) {};
    }
 \end{pgfonlayer}

  \foreach \x / \y / \n  in {} {
      \node[hiddenV] (\n) at (\x*\xx,\y*\xy) {};
    }
    
    \foreach \s/\t in {vp/vn,vp/c1,vp/c2,vn/cn,c1/c1p,
    B11/B12,Bx3/B11,B21/B22,B13/B21} {
      \draw[multiedgeB] (\s) -- (\t);
       \draw[multiedgeA] (\s) -- (\t);

    }

\end{tikzpicture}

\caption{Illustration for the proof of \cref{thm:VCorgsize}: the \multiedge s and \multivertices. We assume that the variable $\svar{i}$ is in the clause $C_{i_1}$ and $C_{i_2}$, and variable $\snegvar{i}$ is in the clause $C_{\bar{i}}$. We also assume that $C_{i_1} = (\svar{i} \vee \snegvar{i'})$. 
\vspace{0.5cm} %
}\label{fig:VCorgsize}
\end{figure}	

\def \xx {1}
\def \xy {0.8}
  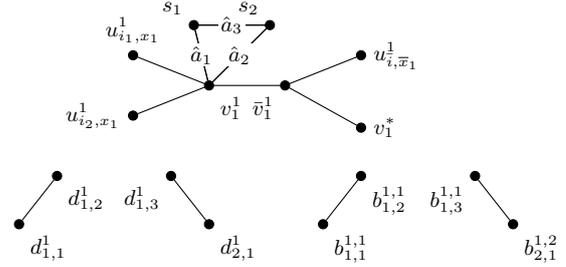
\begin{figure}[t!]
    \centering
   
  \begin{tikzpicture}[black, scale=1,every node/.style={scale=0.9}]	

\begin{pgfonlayer}{fg}
 \foreach \x / \y / \n / \nn  / \pos in {
 5.5/7/vp/\pver{1}{1}/{below right},
      6.5/7/vn/\nver{1}{1}/{below left},
      4.5/7.5/c1/\clavar{i_1}{\svar{1}}{1}/above,
      4.5/6.5/c2/\clavar{i_2}{\svar{1}}{1}/left,
      7.5/7.5/cn/\clavar{\bar{i}}{\snegvar{1}}{1}/right,
      7.5/6.3/prize/\prizev{1}/right,
      5.3/8/s1/\ever{1}/{above left},
      6.3/8/s2/\ever{2}/{above left},
      3/4.7/B11/\bcv{1}{1}{1}/{below right},
      3.5/5.5/B12/\bcv{1}{1}{2}/{below right},
      5.5/4.7/B21/\bcv{2}{1}{1}/{below right},
      5/5.5/B13/\bcv{1}{1}{3}/{below left},
      7/4.7/D11/\lcv{1}{1}{1}{1}/{below right},
      7.5/5.5/D12/\lcv{1}{1}{1}{2}/{below right},
      9.5/4.7/D21/\lcv{2}{1}{2}{1}/{below right},
      9/5.5/D13/\lcv{1}{1}{1}{3}/{below left}
      } {
      \node[vcvertex, label=\pos :$\nn$] (\n) at (\x*\xx,\y*\xy) {};
    }
 \end{pgfonlayer}

  \foreach \x / \y / \n  in {} {
      \node[hiddenV] (\n) at (\x*\xx,\y*\xy) {};
    }
    
    \foreach \s/\t in {vp/vn,vp/c1,vp/c2,vn/cn,
    B11/B12,B13/B21,D11/D12,D13/D21,
    s1/s2,s2/vp,s1/vp,
    prize/vn} {
       \draw[] (\s) -- (\t);

    }
    
        \foreach \s/\t/\label in {
    s1/s2/\eag{3},s2/vp/\eag{2},s1/vp/\eag{1}} {
       \draw[] (\s) -- (\t) node[midway,fill=white,inner sep=1pt] {$\label$};

    }

\end{tikzpicture}

\caption{Illustration for the proof of \cref{thm:VCorgsize}: the edges belonging to the agent $\varag{1}{1}{1}$ and the enforcer agents \eag{1}, \eag{2}, and \eag{3}. 
The edges belonging to the enforcer agent are labeled with the owner, and the unlabeled edges belong to $\varag{1}{1}{1}$.
We assume that the variable $\svar{1}$ is in the clause $C_{i_1}$ and $C_{i_2}$, and variable $\snegvar{1}$ is in the clause $C_{\bar{i}}$.
\vspace{0.5cm} %
}\label{fig:VCorgsize2}
\end{figure}		
	
	We construct an instance $I'$ of \probPVCCE\ with a graph~$G$ as follows:
	Let us first construct the vertices: The construction is depicted in \cref{fig:VCorgsize,fig:VCorgsize2}.
	\begin{compactitem}[--]
		\item For every variable $\svar{i} \in \Vars{}$, we construct:
		\begin{compactitem}[*]
			\item a prize vertex $\prizev{i}$,
			\item a positive literal \multivertex\ $\Pver{i} = \{\pver{i}{\mind} \mid \mind \in [\vcX]\}$ and a negative literal \multivertex\ $\Nver{i} = \{\nver{i}{\mind} \mid \mind \in [\vcX]\}$, and
			\item for every $(\mind, \mindd) \in [\vcX] \times [\vcX], \tind \in [3]$ a small chain vertex~$\lcv{i}{\mind}{\mindd}{\tind}$.
		\end{compactitem} 
		\item For every clause $C_j$ and every literal $\ell_i \in C_j$, we construct a clause \multivertex\ $\Clav{j}{i} = \{\clav{j}{i}{\mind} \mid \mind \in [\vcX]\}$.
		\item Moreover, for every $\satel \sind \in \Satel$, $\tind \in [3]$, we construct a big chain \multivertex\  $\Bcv{\sind}{\tind} = \{\bcv{\sind}{\mind}{\tind} \mid \mind \in [\vcX]\}$.
		\item Finally we construct two enforcer vertices $\ever{1}$ and $\ever{2}$.
	\end{compactitem}
	
	Let us now construct the edges:
	\begin{compactitem}[--]
		\item For every vertex $\svar{i} \in \Vars{}$, we construct:
		\begin{compactitem}[*]
			\item a \multiedge\ between the \multivertices\ \Pver{i} and \Nver{i},
			\item the edge $\{\nver{i}{1}, \prizev{i}\}$,
			\item for every $(\mind, \mindd) \in [\vcX] \times [\vcX]$, we add the edges $\{\lcv{i}{\mind}{\mindd}{1}, \lcv{i}{\mind}{\mindd}{2}\}$ and $\{\lcv{i}{\mind}{\mindd}{3}, \lcv{i}{\hmind}{\hmindd}{1}\}$, where $(\hmind,\hmindd)$ is $(\mind, \mindd + 1)$ if $(\mindd < \vcX)$, otherwise $(\mind + 1, 1)$, where $\vcX + 1 = 1$.
		\end{compactitem}
		\item For every clause $C_j = (\ell_i \vee \ell_{i'} )\in \varphi$, we add the \multiedge\ between $\Clav{j}{i}$ and $\Clav{j}{i'}$. For every $i'' \in \{i, i'\}$, we construct a \multiedge\ between $\Clav{j}{i''}$ and \Pver{i''}\ if $\ell_{i''}$ is a positive literal, otherwise between $\Clav{j}{i''}$ and \Nver{i''}.
		\item Moreover, for every $\satel \sind \in \Satel$, we construct a \multiedge\ between the \multivertices~\Bcv{\sind}{1} and~\Bcv{\sind}{2} and between the \multivertices~\Bcv{\sind}{3} and~\Bcv{\sind + 1}{1}, where $\varnr{} + \clanr{} + 1 = 1$.
		\item Finally, we add the edges $\{\ever{1}, \ever{2}\}, \{\ever{1}, \pver{1}{1}\}$, and $\{\ever{2}, \pver{1}{1}\}$.
	\end{compactitem}
	Let us now describe the set of agents $\players$:
	
	\begin{compactitem}[--]
		\item For every variable $\svar{i} \in \Vars{}$, let $C_{i_1}, C_{i_2}$ be the two clauses where~$\svar{i}$ appears (if they exist), and let $C_{\bar{i}}$ be the clause where~$\snegvar{i}$ appears.
		For every tuple $(\mind, \mindd) \in [\vcX] \times [\vcX]$, we construct a variable agent $\varag{i}{\mind}{\mindd}$ who owns the following edges:
		\begin{compactitem}[*]
			\item $\{\clavar{i_1}{\svar{i}}{\mind}, \pver{i}{\mindd}\}, \{\clavar{i_2}{\svar{i}}{\mind}, \pver{i}{\mindd}\}$, and $\{\clavar{\bar{i}}{\snegvar{i}}{\mindd}, \nver{i}{\mind}\}$, if $C_{i_1}, C_{i_2}$, and~$C_{\bar{i}}$ exist, respectively,
			\item $\{\pver{i}{\mindd}, \nver{i}{\mind}\}$,
			\item $\{\bcv{i}{\mind}{1}, \bcv{i}{\mindd}{2}\}$, and $\{\bcv{i}{\mind}{3}, \bcv{i + 1}{\mindd}{1}\}$,
			\item $\{\lcv{i}{\mind}{\mindd}{1}, \lcv{i}{\mind}{\mindd}{2}\}$ and $\{\lcv{i}{\mind}{\mindd}{3}, \lcv{i}{\hmind}{\hmindd}{1}\}$, where $\hmind,\hmindd$ is $\mind, \mindd + 1$ if $\mindd < \vcX$, otherwise $\mind + 1, 1$, where $\vcX + 1 = 1$, and
			\item if $\mind = \mindd = 1$, then $\{\nver{i}{1}, \prizev{i}\}$.
		\end{compactitem}
		\item For every clause $C_j = (\ell_i \vee \ell_{i'})\in \varphi$, for every tuple $(\mind, \mindd) \in [\vcX] \times [\vcX]$, we construct a clause agent $\clag{i}{\mind}{\mindd}$ who owns the following edges:
		\begin{compactitem}[*]
			\item $\{\clav{j}{\mind}{i}, \clav{j}{\mindd}{i'}\}$,
			\item $\{\bccv{j}{\mind}{1}, \bccv{j}{\mindd}{2}\}$ and $\{\bccv{j}{\mind}{3},\bccv{j + 1}{\mind}{1}\}$, where $\varnr{} + \clanr{} + 1 =~1$.
		\end{compactitem}
		\item We construct three enforcer agents $\eag{1}, \eag{2}$, and $\eag{3}$. Agent $\eag{1}$ owns edge $\{\ever{1}, \pver{1}{1}\}$, agent \eag{2}\ owns edge $\{\ever{2}, \pver{1}{1}\}$, and agent \eag{3}\ owns edge $\{\ever{1}, \ever{2}\}$.
	\end{compactitem}
	The variable agents own at most 9 edges, and the clause agents at most 3 edges. The enforcer agents own one edge each.
	Thus $\Orgsize =~9$.
	
	Given an agent $\agent{} \in \players$, we use \Eag{\agent{}} to denote the set of edges $\agent{}$ owns.
	For a variable or clause $\satel \sind \in \Satel$, let $\Elag{\satel{\sind}} \coloneqq \{\elag{\sind}{\mind}{\mindd} \mid (\mind, \mindd) \in [\vcX] \times [\vcX]\}$.
	Let $\allsatagents \coloneqq \bigcup_{\satel \sind \in \Satel} \Elag{\satel{\sind}} $; this is the set of all the agents except the enforcers.
	
	We start by showing the connection between our construction and the original SAT-formula:
	\begin{claim}\label{clm:vcsatconnect}
		\begin{compactenum}[(i)]
			\item If $\varphi$ admits a satisfying assignment $\sigma\colon \Vars{1} \to \{\trueT,\falseF\}$ that sets $\satscore$ variables to true, then $\GindVC{\allsatagents}$ admits a vertex cover $V'$ of size \satvcsize. Moreover, if $\sigma(\svar{1})=\trueT$, then $V'$ contains $\varag{1}{1}{1}$.\label{clm:vcsatconnect2}
			\item If $\GindVC{\allsatagents}$ admits a minimal vertex cover $V'$ of size \satvcsize, then $\varphi$ admits a satisfying assignment $\sigma$ that sets at most $\satscore$ variables to true. Moreover, if~$V'$ contains $\varag{1}{1}{1}$, then $\sigma(\svar{1}) = \trueT$.\label{clm:vcsatconnect1}
		\end{compactenum}
	\end{claim}
	\begin{proof}\renewcommand{\qedsymbol}{\hfill (end of the proof of~\cref{clm:vcsatconnect})~$\diamond$}
		We start by showing Statement~\eqref{clm:vcsatconnect2}.
		We construct a vertex cover $V'$ as follows:
		For every $\svar{i} \in \Vars{}$ such that $\sigma(\svar{i}) = \trueT$, we add the vertices in $\Pver{i} \cup \{\nver{i}{1}\}$ to $V'$.
		For every  $\svar{i} \in \Vars{}$ such that $\sigma(\svar{i}) = \falseF$, we add the vertices in $\Nver{i}$ to $V'$.
		For every $\svar{i} \in \Vars{}$ we add the vertices in $\{\lcv{i}{\mind}{\mindd}{1} \mid (\mind, \mindd) \in [\vcX] \times [\vcX]\}$ to $V'$.
		For every clause $C_j \in \varphi$, let $\ell_i \in C_j$ be a literal of $C_j$ that is false under $\sigma$, or if no literal of $C_j$ is false under $\sigma$, then an arbitrary literal of $C_j$. 
		Observe that since $\sigma$ is a satisfying assignment, at most one literal in $C_j$ is false under $\sigma$.
		We add the vertices in $\Clav{j}{i}$ to $V'$.
		For every $\satel \sind \in \Satel$, we add $\Bcv \sind 1$ to $V'$.
		Since $\sigma$ sets $\satscore$ variables to true, we get that the size of the vertex cover $V'$ is
		\begin{align*}
			&\satscore(\vcX + 1) + (\varnr{} - \satscore)\vcX + \varnr{} \cdot \vcXp + \clanr{} \cdot \vcX + (\varnr{} + \clanr{})\vcX\\
			=&\satvcsize,
		\end{align*}
		as required.
		
		It remains to show that $V'$ is indeed a vertex cover.
		Clearly for every $\svar{i} \in \Vars{}$, the edges that belong to the \multiedge\ between~$\Pver{i}$ and $\Nver{i}$ are covered.
		Edge $\{\nver{i}{1},\prizev{i}\}$ is also covered, as $\nver{i}{1} \in V'$ regardless of whether $\sigma$ assigns $\svar{i}$ to true or false.
		For every $C_j \in \varphi$ such that $\svar{i} \in C_j$, if $\svar{i}$ is true under $\sigma$, then $\Pver{i} \subseteq V'$, otherwise $\Clavar{j}{\svar{i}} \subseteq V'$.
		Thus the edges in the \multiedge\ between $\Pver{i}$ and $\Clavar{j}{\svar{i}}$ are covered.
		Through similar argumentation all the edges in the \multiedge\ between $\Nver{i}$ and $\Clavar{j}{\snegvar{i}}$ are covered for any $C_j \in \varphi$ that contains $\snegvar{i}$.
		For every $(\mind, \mindd) \in [\vcX] \times [\vcX]$, edge $\{\lcv{i}{\mind}{\mindd}{1}, \lcv{i}{\mind}{\mindd}{2}\}$ is covered by \lcv{i}{\mind}{\mindd}{1}, and edge $\{\lcv{i}{\mind}{\mindd}{3}, \lcv{i}{\hmind}{\hmindd}{1}\}$ by  \lcv{i}{\hmind}{\hmindd}{1}, where $(\hmind,\hmindd)$ is $(\mind, \mindd + 1)$ if $\mindd < \vcX$, otherwise $(\mind + 1, 1)$, where $\vcX + 1 = 1$.	
		For every $C_j = (\ell_i \vee \ell_{i'}) \in \varphi$, the vertices in the \multiedge\ between \Clav{j}{i} and \Clav{j}{i'} are covered by either the vertices in \Clav{j}{i} or \Clav{j}{i'}.	
		For every $\satel \sind \in \Satel$, the \multiedge\ between $\Bcv{\sind}{1}$ and $\Bcv{\sind}{2}$ is covered by the vertices in $\Bcv{\sind}{1}$, and the \multiedge\ between $\Bcv{\sind}{3}$ and $\Bcv{\sind + 1}{1}$ is covered by $\Bcv{\sind + 1}{1}$, where $\varnr{} + \clanr{} + 1 = 1$.	
		Thus $V'$ is indeed a vertex cover.
		Moreover, if $\svar{1}$ is true under $\sigma$, then clearly $\pver{1}{1} \in V'$.\\
		
		Next we show Statement~\eqref{clm:vcsatconnect1}.
		By \cref{obs:multiedge}, for every $\svar{i} \in \Vars{}$ we need at least $\vcX$ vertices to cover the \multiedge\ between $\Pver{i}$ and $\Nver{i}$, and we must have that $\Pver{i} \subseteq V'$ or $\Nver{i} \subseteq V'$.
		Also observe that if $\Pver{i} \subseteq V'$, then we need an additional vertex to cover edge $\{\nver{i}{1}, \prizev{i}\}$.
		We also need at least $\vcXp$ further vertices to cover the edges $\{\lcv{i}{\mind}{\mindd}{1},\lcv{i}{\mind}{\mindd}{2}\}$ for every $(\mind, \mindd) \in [\vcX] \times [\vcX]$.
		Similarly we need at least $\vcX$ vertices for every $C_j = (\ell_i \vee \ell_{i'})$ to cover the \multiedge\ between $\Clav{j}{i}$ and $\Clav{j}{i'}$.
		For every $\satel \sind \in \Satel$, we need $\vcX$ vertices to cover the \multiedge\ between \Bcv{\sind}{1} and \Bcv{\sind}{2}.
		This is in total $\varnr{} \cdot \vcX + \varnr{} \cdot \vcXp + \clanr{} \cdot \vcX + (\varnr{} + \clanr{})\vcX + \alpha = \satvcreq + \alpha$ vertices, where $\alpha$ is the number of variables $\svar{i} \in \Vars{}$ such that $\Pver{i} \subseteq V'$.
		Since $|V'| = \satvcsize$, we must have that $\alpha \leq \satscore$.
		Moreover, we can have at most $\satscore - \alpha \leq \varnr{}$ vertices in $V'$ that are not listed here.
		
		Let us construct an assignment $\sigma$ as follows:
		For every $\svar{i} \in \Vars{}$, if $\Pver{i} \subseteq V'$, then $\svar{i}$ is true under $\sigma$, otherwise it is false.
		Clearly this sets $\alpha \leq \satscore$ variables to true, as required.
		Assume, towards a contradiction, that $\sigma$ is not a satisfying assignment.
		Then there is a clause $C_j = (\ell_i \vee \ell_{i'})$ such that neither of its literals is true under $\sigma$.
		Assume, without loss of generality, that $\Clav{j}{i'} \subseteq V'$.
		Also, assume that $\ell_i$ is the true literal $\svar{i}$; the proof for the case where $\ell_i$ is a false literal is analogous.
		Then we have that $\Nver{i} \subseteq V'$.
		By earlier reasoning we have at most $\varnr{}$ vertices left to cover the \multiedge\ between $\Pver{i}$ and $\Clav{j}{i}$; however this requires by \cref{obs:multiedge} at least $\vcX > \varnr{}$ vertices, a contradiction.
		Thus $\sigma$ must be a satisfying assignment.
		
		Finally, we must show that if $V'$ contains $\pver{1}{1}$, then $\sigma$ sets $\svar{i}$ to true.
		Assume, towards a contradiction, that this is not the case.
		Then $\{\pver{1}{1}\} \cup \Nver{i} \subseteq V'$.
		We show that $V' \setminus \{\pver{1}{1}\}$ is also a vertex cover, contradicting the minimality of $V'$.
		The edges in the \multiedge\ between~$\Pver{1}$ and~$\Nver{1}$ are covered by $\Nver{1}$, so they remain covered by $V'\setminus \{\pver{1}{1}\}$.
		Consider $C_j \in \varphi$ such that $\svar{1} \in C_j$.
		By \cref{obs:multiedge}, it holds that $\Pver{1} \subseteq V'$ or $\Clavar{j}{\svar{1}} \subseteq V'$.
		Since $\Nver{i} \subseteq V'$, there are at most $\varnr{}$ vertices in $\Pver{i} \cap V'$, meaning that $\Clavar{j}{\svar{i}} \subseteq V'$. But then $V' \setminus \{\pver{1}{1}\}$ covers the \multiedge\ between \Clavar{j}{\svar{i}} and \Pver{1}.
		Thus $V' \setminus \{\pver{1}{1}\}$ is also a vertex cover, as required.
	\end{proof}
	
	Now we move on to prove that the two instances are equivalent.
	
	$(\Longrightarrow)$: Assume $I$ is a yes-instance of \minasgwosatmember.
	Let $\sigma \colon \Vars{} \to \{\trueT, \falseF\}$ be a satisfying assignment such that $\sigma(\svar{1}) = \trueT$ and there is no satisfying assignment that sets fewer variables to true.
	Let $\satscore$ be the number of variables that $\sigma$ sets to true.
	
	We construct an allocation \alloc\ as follows:
	\begin{compactitem}[--]
		\item For every $\svar{i} \in \Vars{}, (\mind, \mindd) \in [\vcX] \times [\vcX]$, 
		if $\mind = \mindd = 1$, then we allocate $\varag{i}{\mind}{\mindd}$ the \costtxt\ 
		$\vcXcost + 1 + \frac{\satscore}{\varnr{}}$, 
		otherwise we allocate $\vcXcost + 1$. 
		\item For every $C_j \in \varphi, (\mind, \mindd) \in [\vcX] \times [\vcX]$, the we allocate agent~$\clag{i}{\mind}{\mindd}$ the \costtxt\ $\vcXcost$.
		\item The \costtxt\ of $\eag{1}$ and $\eag{2}$ is 0, and the \costtxt\ of $\eag{3}$ is $1$.
	\end{compactitem}
	Let us first show that $\alloc$ is a \preimputation, i.e., a minimum vertex cover of $G$ has the cardinality $\varnr{}(\vcXp(\vcXcost + 1) + \frac{\satscore}{\varnr{}}) + m \cdot \vcXp \cdot \vcXcost + 1 = \satvcsize + 1$.
	By \cref{clm:vcsatconnect}\eqref{clm:vcsatconnect2}, we have that $\GindVC{\allsatagents}$ admits a vertex cover~$V''$ of size $\satvcsize$, and $\pver{1}{1} \in V'$.
	Thus the edges $\{\pver{1}{1}, \ever{1}\}$ and $\{\pver{1}{1}, \ever{2}\}$ are covered by~$V''$. By adding one further vertex from $\{\ever{1}, \ever{2}\}$, we obtain a vertex cover of size $\satvcsize + 1$.
	
	Now assume towards a contradiction that $G$ admits a vertex cover~$V'$ of size at most $\satvcsize$.
	We need at least one vertex from $\{\ever{1}, \ever{2}\}$ in~$V'$.
	Thus $|V' \setminus \{\ever{1}, \ever{2}\}| \geq \satvcsize - 1$, and $V' \setminus \{\ever{1}, \ever{2}\}$ is a vertex cover of $\GindVC{\allsatagents}$ of size $\satvcsize - 1$.
	By  \cref{clm:vcsatconnect}\eqref{clm:vcsatconnect1}, it holds that $\varphi$ admits a satisfying assignment with $\satscore - 1$ true variables, a contradiction.
	
	Next we show that \alloc\ is core stable.
	Assume, towards a contradiction, that $\alloc$ admits an inclusion-wise minimal blocking coalition~$\pset$.
	First we observe that $\eag{1}, \eag{2}, \eag{3} \notin \pset$:
	In what follows, let $\gamma$ be the size of a minimum vertex cover of $\GindVC{\pset}$ and $\gamma'$ be the size of a minimum vertex cover of $\GindVC{\pset \setminus \{\eag{1}, \eag{2}, \eag{3}\}}$.
	If $\eag{3} \in \pset$, then a minimum vertex cover of $\GindVC{\pset}$ must contain $\ever{1}$ or~$\ever{2}$.
	Because $\ever{1}$ and $\ever{2}$ do not cover any edges belonging to the agents in $\pset \setminus \{\eag{1}, \eag{2}, \eag{3}\}$, we must have that $\gamma' \leq \gamma - 1$.
	Thus since $\pset$ blocks~\alloc, we have that $\util(\pset \setminus \{\eag{1}, \eag{2}, \eag{3}\}) =  \gamma' \leq  \gamma - 1  = \util(\pset) - 1
	< \sum_{\agent{} \in \pset} \allocel{\agent{}} - 1 =
	\sum_{\agent{} \in \pset \setminus \{\eag{1}, \eag{2}, \eag{3}\}}\allocel{\agent{}} + \sum_{\agent{} \in \pset \cap \{\eag{1}, \eag{2}, \eag{3}\}} \allocel{\agent{}} - 1 
	= \sum_{\agent{} \in \pset \setminus \{\eag{1}, \eag{2}, \eag{3}\}}\allocel{\agent{}} + 1 - 1 
	=  \sum_{\agent{} \in \pset \setminus \{\eag{1}, \eag{2}, \eag{3}\}}\allocel{\agent{}}$.
	Thus $\pset \setminus \{\eag{1}, \eag{2}, \eag{3}\}$ is also a blocking coalition, and it is a subset of $\pset$, a contradiction to $\pset$ being inclusion-wise minimal.
	
	Now assume that $\eag{1} \in \pset$ or $\eag{2} \in \pset$ and $\eag{3} \notin \pset$. 
	If $\eag{1} \in \pset$, then the degree of $\ever{1}$ in $\GindVC{\pset}$ is $1$. Similarly, if $\eag{2} \in \pset$, then the degree of $\ever{2}$ is $1$ in $\GindVC{\pset}$.
	Thus there is a minimum vertex cover of $\GindVC{\pset}$ that does not contain them. 
	Since the \costtxt\ of both of these agents is $0$, we can remove them from $\pset$ and obtain a smaller blocking coalition.
	
	We have now shown that $\pset$ must consist of only variable and clause agents.
	For every $\satel \sind \in \Satel$, let $\psetcount{\satel{\sind}} \coloneqq |\agent{\satel{\sind}}\cap \pset|$, i.e., the number of agents corresponding to this variable or clause that are in~$\pset$.
	
	Now we can make the following observation about the \costtxt\ allocated to the agents in $\pset$:
	\begin{observation}\label{obs:minallocutil}
		It holds that $\sum_{\agent{} \in \pset}\allocel{\agent{}} \leq  \displaystyle\sum_{\svar{i} \in \Vars{}}\psetcount{\svar{i}}( \vcXcost + 1) + \nonemptyvarcount\frac{\satscore}{\varnr{}} + \displaystyle\sum_{C_j \in \varphi} \vcXcostmul{\psetcount{C_j} }$,
		where $\nonemptyvarcount$ is the number of variables that have at least one corresponding agent in~\pset.
	\end{observation}
	\begin{proof}\renewcommand{\qedsymbol}{\hfill (end of the proof of~\cref{obs:minallocutil})~$\diamond$}
		Observe that for every $\svar{i} \in \Vars{}$ that has at least one corresponding agent in $\pset$, it holds that $\sum_{\agent{} \in \Varag{i}} \allocel{\agent{}} \leq \psetcount{\svar{i}}( \vcXcost + 1) + \frac{\satscore}{\varnr{}}$.
		For every $\svar{i} \in \Vars{}$ that does not have a corresponding agent in $\pset$, we trivially have that $\sum_{\agent{} \in \Varag{i}} \allocel{\agent{}} = 0 = 0 (\vcXcost + 1) =\psetcount{\svar{i}}(\vcXcost + 1)$.
		For every $C_j \in \varphi$, we have that $\sum_{\agent{} \in \Clag{j}} \allocel{\agent{}}= \vcXcostmul{\psetcount{C_j} }$.
		The statement follows.
	\end{proof}
	
	Next we observe some requirements of a minimum vertex cover of $\GindVC{\pset}$:
	\begin{claim}\label{clm:minvertexcoverprops}
		Every minimum vertex cover $V'$ of $\GindVC{\pset}$ must contain vertices from the following pairwise disjoint vertex sets:
		\begin{compactenum}[(i)]
			\item for every $\svar{i} \in \Vars{}$, at least $\lceil \frac{\psetcount{\svar{i}}}{\vcX} \rceil$ vertices from $\Pver{i} \cup \Nver{i}$,\label{clm:minvcsizes:X}
			\item for every $C_j = (\ell_i \vee \ell_{i'}) \in \varphi$, at least $\lceil \frac{\psetcount{C_j}}{\vcX} \rceil$ vertices from $ \Clav{j}{i} \cup\Clav{j}{i'}$,\label{clm:minvcsizes:C}
			\item for every $\svar{i} \in \Vars{}$,
			if $\Varag{i} \subseteq \pset$, then
			at least $\psetcount{\svar{i}}$ vertices from $\{\lcv{i}{\mind}{\mindd}{\tind} \mid (\mind, \mindd) \in [\vcX] \times [\vcX], \tind \in [3]\}$,
			otherwise if $\Varag{i} \cap \pset \neq \emptyset$, then at least $\psetcount{\svar{i}} + 1$ vertices from $\{\lcv{i}{\mind}{\mindd}{\tind} \mid (\mind, \mindd) \in [\vcX] \times [\vcX], \tind \in [3]\}$, and\label{clm:minvcsizes:lc}
			\item \label{clm:minvcsizes:bc}
for every $\satel \sind \in \Satel$, at least $\lceil \frac{\max(\psetcount{\satel{\sind}}, \psetcount{\satel{\sind + 1}})}{\vcX}  \rceil$ vertices from $\Bcv{\sind}{3} \cup \Bcv{\sind+1}{1} \cup \Bcv{\sind+1}{2}$, where $\varnr{} + \clanr{} + 1 = 1$.
		\end{compactenum}
	\end{claim}
	\begin{proof}\renewcommand{\qedsymbol}{\hfill (end of the proof of~\cref{clm:minvertexcoverprops})~$\diamond$}
		We first show Statement~\eqref{clm:minvcsizes:X}. The edges from \multiedge\ between $\Pver{i}$ and $\Nver{i}$ that are owned by agents in \pset\ must be covered.
		A vertex from $\Pver{i}$ or $\Nver{i}$ is incident to precisely \vcX\ edges from this \multiedge, and can thus cover at most \vcX\ edges, meaning that we need $\lceil \frac{\psetcount{\svar{i}}}{\vcX} \rceil$ vertices from $(\Pver{i} \cup \Nver{i})$.
		The proof of Statement~\eqref{clm:minvcsizes:C} is analogous: The edges from \multiedge\ between $\Clav{j}{i}$ and $\Clav{j}{i'}$ that are in $\GindVC{\pset}$ must be covered, and minimum vertex cover of $\GindVC{\pset}$ must contain at least $\lceil \frac{\psetcount{C_j}}{\vcX} \rceil$ vertices from $(\Clav{j}{i} \cup \Clav{j}{i'})$.

		Now we show Statement~\eqref{clm:minvcsizes:lc}.
		For every $(\mind, \mindd) \in [\vcX] \times [\vcX]$ such that $\varag{i}{\mind}{\mindd} \in \pset$, we need at least one vertex from  edge $\{\lcv{i}{\mind}{\mindd}{1}, \lcv{i}{\mind}{\mindd}{2}\}$ in~$V'$. 
		In total we need at least $\psetcount{\svar{i}}$ vertices.
		Moreover, if there is at least one $(\mind, \mindd) \in [\vcX] \times [\vcX]$ such that $\varag{i}{\mind}{\mindd} \notin \pset$ and $\Varag{i} \cap \pset \neq \emptyset$, then there must be at least one $(\hmind, \hmindd) \in [\vcX] \times [\vcX]$ such that $\varag{i}{\hmind}{\hmindd} \in \pset$ and edge $\{\lcv{i}{\hmind}{\hmindd}{3}, \lcv{i}{\mind}{\mindd}{1}\}$ exists in $G$.
		Then at least one of these vertices in $\{\lcv{i}{\hmind}{\hmindd}{3}, \lcv{i}{\mind}{\mindd}{1}\}$ must be in a minimum vertex cover of $\GindVC{\pset}$, and this vertex is not counted in the earlier $\psetcount{\svar{i}}$-many vertices.
		
		Finally, we show Statement~\eqref{clm:minvcsizes:bc}.
		The edges from the \multiedge\ between $\Bcv{\sind}{3}$ and $\Bcv{\sind + 1}{1}$ must be covered.
		These edges belong to the agents in \Elag{\satel{\sind}}, and thus there are $\psetcount{\satel{\sind}}$ edges that need to be covered.
		A vertex from  $\Bcv{\sind}{3} \cup \Bcv{\sind + 1}{1}$ is incident to precisely~\vcX\ edges from this \multiedge, and thus we need at least $\lceil \frac{\psetcount{\satel{\sind}}}{\vcX} \rceil$ vertices from $\Bcv{\sind}{3} \cup \Bcv{\sind + 1}{1}$ in $V'$.
		Similarly, the edges from the \multiedge\ between $\Bcv{\sind + 1}{1}$ and $\Bcv{\sind + 1}{2}$ must be covered.
		These edges belong to the agents in \Elag{\satel{\sind + 1}}, and thus we need at least $\lceil \frac{\psetcount{\satel{\sind}}}{\vcX} \rceil$ vertices from $\Bcv{\sind + 1}{1} \cup \Bcv{\sind + 1}{2}$ in $V'$.
		Thus we need at least $\lceil \frac{\max(\psetcount{\satel{\sind}}, \psetcount{\satel{\sind + 1}})}{\vcX}  \rceil$ vertices from $\Bcv{\sind}{3} \cup \Bcv{\sind+1}{1} \cup \Bcv{\sind+1}{2}$ in~$V'$.
	\end{proof}
	
	By adding together all the sets from the previous theorem we obtain the following two useful observations:
	\begin{observation}\label{clm:minvcsizes}
		
		It holds that
		\begin{align*}
			\util(\pset)
			&\geq \displaystyle\sum_{\svar{i} \in \Vars{}}\psetcount{\svar{i}} \cdot (\frac{1}{\vcX} + 1)
			+ \nonemptyfillvarcount
			+ \displaystyle\sum_{C_j \in \varphi}\frac{\psetcount{C_j}}{\vcX}
			\\
			& \qquad
			+ \displaystyle\sum_{\satel \sind \in \Satel} \frac{\max(\psetcount{\satel{\sind}}, \psetcount{\satel{\sind + 1}})}{\vcX}  \\
			& \geq \displaystyle\sum_{\svar{i} \in \Vars{}}\psetcount{\svar{i}} \cdot (\vcXcost + 1) +  \nonemptyfillvarcount +\displaystyle\sum_{C_j \in \varphi}\vcXcostmul{\psetcount{C_j}} 
			,
		\end{align*}
		where $\satel{\varnr{} + \clanr{} + 1} = \satel{1}$ and $\nonemptyfillvarcount$ is the number of variables in $\Vars{}$ who have at least one and at most $\varnr{} - 1$ corresponding agents in $\pset$.
	\end{observation}
	\begin{proof}\renewcommand{\qedsymbol}{\hfill (end of the proof of~\cref{clm:minvcsizes})~$\diamond$}
		
		By combining \cref{clm:minvertexcoverprops}(\ref{clm:minvcsizes:X}--\ref{clm:minvcsizes:bc}), we obtain the first inequality. Observe that we can remove the ceiling functions, because they can only decrease the \costtxt\ on the right hand side of the bound.
		The second inequality is obtained by lower-bounding the \costtxt\ of every $\max$-function with its first argument.
	\end{proof}

	\begin{claim}\label{clm:minvcvertexbound}
		For every minimum vertex cover $V'$ of $\GindVC{\pset}$, the following holds:
		\begin{compactenum}[(i)]
			\item For every $\svar{i} \in \Vars{}$, we have that $ |(\Pver{i} \cup \Nver{i}) \cap V'| < \vcX + \satscore$.\label{clm:minvcvertexbound:X}
			\item For every $C_j = (\ell_1 \vee \ell_2) \in \varphi$, we have that $|(\Clav{j}{i} \cup \Clav{j}{i'}) \cap V'| < \vcX + \satscore$.\label{clm:minvcvertexbound:C}
		\end{compactenum}
	\end{claim}
	\begin{proof}\renewcommand{\qedsymbol}{\hfill (end of the proof of~\cref{clm:minvcvertexbound})~$\diamond$}
		We first show that for every $\svar{i} \in \Vars{}$, $|(\Pver{i} \cup \Nver{i}) \cap V'| \leq \lceil \frac{\psetcount{\svar{i}}}{\vcX}\rceil + \satscore$.
		Assume, towards a contradiction, that $|(\Pver{i} \cup \Nver{i}) \cap V'| > \lceil \frac{\psetcount{\svar{i}}}{\vcX} \rceil+ \satscore$.
		By combining the statements of \cref{clm:minvertexcoverprops} and utilizing that $\lceil x \rceil \geq x$ and $\max(x,y) \geq x$ for every $x,y > 0$, we obtain that $|V'| \geq \displaystyle\sum_{\svar{i} \in \Vars{}}\psetcount{\svar{i}}\cdot (\vcXcost + 1) + \displaystyle\sum_{C_j \in \varphi}\vcXcostmul{\psetcount{C_j}}  + \satscore$.
		From \cref{obs:minallocutil} it follows that $\sum_{\agent{} \in \pset}\allocel{\agent{}} \leq  \displaystyle\sum_{\svar{i} \in \Vars{}}\psetcount{\svar{i}}\cdot ( \vcXcost + 1) + \satscore + \displaystyle\sum_{C_j \in \varphi} \vcXcostmul{\psetcount{C_j}} $.
		However, this implies that $\util(\pset) = |V'| \geq \displaystyle\sum_{\svar{i} \in \Vars{}}\psetcount{\svar{i}}\cdot (\vcXcost + 1) + \displaystyle\sum_{C_j \in \varphi}\vcXcostmul{\psetcount{C_j}}  + \satscore \geq \sum_{\agent{} \in \pset}\allocel{\agent{}}$, a contradiction to $\pset$ blocking.
		Thus $|(\Pver{i} \cup \Nver{i}) \cap V'| \leq \lceil \frac{\psetcount{\svar{i}}}{\vcX} \rceil + \satscore$. Since $\psetcount{\svar{i}} \leq \vcXp$, the statement follows.

		The proof for the second statement is analogous.
	\end{proof}
	
We can now show that there must be at least one variable that has all of its corresponding agents in \pset.
	\begin{claim}\label{clm:fullagentexist}
		There is $\svar{\hat{i}} \in \Vars{}$ such that $\Varag{\hat{i}} \subseteq \pset$.
	\end{claim}
	\begin{proof}\renewcommand{\qedsymbol}{\hfill (end of the proof of~\cref{clm:fullagentexist})~$\diamond$}
		Assume, towards a contradiction, that no variable has all of its corresponding agents in $\pset$.
		Let $\nonemptyfillvarcount$ be the number of variables such that neither all nor none  of the agents corresponding to this variable are in $\pset$, and $\nonemptyvarcount$ is the number of variables that have at least one agent in $\pset$.
		We must have that $\nonemptyfillvarcount = \nonemptyvarcount$.
		By \cref{obs:minallocutil,clm:minvcsizes}, we obtain that
		
		\begin{align*}
			&\util(\pset) - \sum_{\agent{} \in \pset} \allocel{\agent{}}\\
			&\geq  \displaystyle\sum_{\svar{i} \in \Vars{}}\psetcount{\svar{i}}\cdot (\vcXcost + 1) + \nonemptyvarcount  +\displaystyle\sum_{C_j \in \varphi}\vcXcostmul{\psetcount{C_j}} 
			\\
			&\qquad - \displaystyle\sum_{\svar{i} \in \Vars{}}\psetcount{\svar{i}}\cdot( \vcXcost + 1) -\nonemptyvarcount\frac{\satscore}{\varnr{}} - \displaystyle\sum_{C_j \in \varphi} \vcXcostmul{\psetcount{C_j}} ,\\
			&\geq \nonemptyvarcount - \nonemptyvarcount\frac{\satscore}{\varnr{}} \geq 0,
		\end{align*}
		where the last inequality follows from the fact that $\satscore \leq \varnr{}$ by definition.
		Thus $\util(\pset) \geq \sum_{\agent{} \in \pset} \allocel{\agent{}}$, which contradicts \pset\ blocking.
	\end{proof}
	
	Now we are ready to show that for every clause and variable, most of the agents corresponding to it must be in the blocking coalition:
	\begin{claim}\label{clm:agentalmostfull}
		For every $\satel{\sind} \in \Satel$, it holds that $\psetcount{\satel{\sind}} > \vcXp - \vcX\satscore$.
	\end{claim}
	\begin{proof}\renewcommand{\qedsymbol}{\hfill (end of the proof of~\cref{clm:agentalmostfull})~$\diamond$}
		Let $\nonemptyfillvarcount$ be the number of variables such that neither all nor none  of the agents corresponding to this variable are in $\pset$, and $\nonemptyvarcount$ is the number of variables that have at least one agent in $\pset$.
		Since $\pset$ is a blocking coalition, by \cref{obs:minallocutil,clm:minvcsizes}, and the fact that $0 \leq \nonemptyfillvarcount \leq \nonemptyvarcount \leq \varnr{}$, we obtain that
		
		\begin{align*}
			0 >& \util(\pset) - \sum_{\agent{} \in \pset} \allocel{\agent{}}\\
			\geq& \displaystyle\sum_{\svar{i} \in \Vars{}}\psetcount{\svar{i}}\cdot(\frac{1}{\vcX} + 1)
			+ \nonemptyfillvarcount
			+ \displaystyle\sum_{C_j \in \varphi}\frac{\psetcount{C_j}}{\vcX}
			\\
			& \qquad
			+ \displaystyle\sum_{\satel \sind \in \Satel} \frac{\max(\psetcount{\satel{\sind}}, \psetcount{\satel{\sind + 1}})}{\vcX}  \\
			&\qquad - \displaystyle\sum_{\svar{i} \in \Vars{}}\psetcount{\svar{i}}\cdot( \vcXcost + 1) -\nonemptyvarcount\frac{\satscore}{\varnr{}} - \displaystyle\sum_{C_j \in \varphi}  \vcXcostmul{\psetcount{C_j}}\\
			\geq& -\displaystyle\sum_{\svar{i} \in \Vars{}}\frac{\psetcount{\svar{i}}}{\vcX} 
			-\displaystyle\sum_{C_j \in \varphi}\frac{\psetcount{C_j}}{\vcX} 
			\\ &\qquad + \displaystyle\sum_{\satel \sind \in \Satel} \frac{\max(\psetcount{\satel{\sind}}, \psetcount{\satel{\sind + 1}})}{\vcX} - \satscore \\
			\geq&\displaystyle\sum_{\satel \sind \in \Satel} \frac{\max(\psetcount{\satel{\sind}}, \psetcount{\satel{\sind + 1}}) - \psetcount{\satel{\sind}}}{\vcX}  - \satscore\\
		=&\displaystyle\sum_{\substack{\satel \sind \in \Satel\\ \psetcount{\satel{\sind + 1}} > \psetcount{\satel \sind}}} \frac{ \psetcount{\satel{\sind + 1}} - \psetcount{\satel{\sind}}}{\vcX}  - \satscore,
		\end{align*}
		where $\satel{\varnr{} + \clanr{} + 1} = \satel{1}$. 
		
		Let $\satel{\min} \coloneqq \argmin_{\satel \sind \in \Satel}\psetcount{\satel{\sind}}$ and let $\satel{\max} \coloneqq \argmax_{\satel \sind \in \Satel}\psetcount{\satel{\sind}}$.
Observe that we must have that $\displaystyle\sum_{\substack{\satel \sind \in \Satel\\ \psetcount{\satel{\sind + 1}} > \psetcount{\satel \sind}}} \frac{ \psetcount{\satel{\sind + 1}} - \psetcount{\satel{\sind}}}{\vcX} \geq \frac{\psetcount{\satel{\max}} - \psetcount{\satel{\min}}}{\vcX}$.
		By \cref{clm:fullagentexist}, there is $\svar{\hat{i}} \in \Vars{}$ such that $\psetcount{\svar{\hat{i}}} = \vcXp$, and thus $\psetcount{y_{\max}} = \vcXp$.
		We obtain that
		\begin{align*}
			0 > \frac{\vcXp - \psetcount{\satel{\min}}}{\vcX} - \satscore \iff \psetcount{y_{\min}} >  \vcXp - \vcX \satscore.
		\end{align*}
		Thus it holds that for every $\satel \sind \in \Satel$ that 
		$\psetcount{\satel{\sind}} >  \vcXp - \vcX \satscore$, as required.
	\end{proof}

	Let $V'_{\pset}$ be a minimum vertex cover of $\GindVC{\pset}$.
	We proceed to build an assignment $\sigma' \colon \Vars{} \to \{\trueT, \falseF\}$ from $\pset$ as follows: For every $\svar{i} \in \Vars{}$, we set $\sigma'(\svar{i}) = \trueT$ if and only if $|\Pver{i} \cap V'_{\pset}| \geq |\Nver{i} \cap V'_{\pset}|$.
Clearly this is an assignment.
	
We first show that $\sigma'$ must be a satisfying assignment.
	Assume, towards a contradiction, that $\sigma'$ is not a satisfying assignment.
	Then there must be a clause $C_j = (\ell_i \vee \ell_{i'})\in \varphi$ such that neither of its literals is true under $\sigma'$.
	We must have that  $|\Clav{j}{i} \cap V'_{\pset}| \leq |\Clav{j}{i'} \cap V'_{\pset}|$ or $|\Clav{j}{i} \cap V'_{\pset}| \geq |\Clav{j}{i'} \cap V'_{\pset}|$.
	Assume, without loss of generality, that $|\Clav{j}{i} \cap V'_{\pset}| \leq |\Clav{j}{i'} \cap V'_{\pset}|$.
	By \cref{clm:minvcvertexbound}\eqref{clm:minvcvertexbound:C}, we have that $|(\Clav{j}{i} \cup \Clav{j}{i'}) \cap V'_{\pset}| < \vcX + \satscore$ and thus $|\Clav{j}{i} \cap V'_{\pset}| < \frac{\vcX + \satscore}{2}$.
	Also, assume that~$\ell_i$ is the true literal $\svar{i}$; the proof for the case where~$\ell_i$ is the false literal will be analogous.
	Since $C_j$ is not satisfied under $\sigma'$, we must have that $|\Pver{i} \cap V'_{\pset}| \leq |\Nver{i} \cap V'_{\pset}|$.
	By \cref{clm:minvcvertexbound}\eqref{clm:minvcvertexbound:X}, we have that $|(\Pver{i} \cup \Nver{i}) \cap V'_{\pset}| < \vcX + \satscore$ and thus $|\Pver{i} \cap V'_{\pset}| < \frac{\vcX + \satscore}{2}$.
	
	Recall that the \multiedge\ between $\Pver{i}$ and $\Clavar{j}{\svar{i}}$ consists of \vcXp\ edges, which are all owned by different agents in \Varag{i}.
	Since $V'_{\pset}$ is a vertex cover of $\GindVC{\pset}$, each of these edges must
\begin{compactenum}[(i)]
\item be covered by $V'_{\pset}$ or  \label{prop:covered}
\item not be owned by an agent in \pset.\label{prop:notowned}
\end{compactenum}
Let us first count the number of edges in \eqref{prop:covered}.
The edges that are not covered by $V'_{\pset}$ are those between the vertices in $\Pver{i} \setminus V'$ and $\Clavar{j}{\svar{i}} \setminus V'_{\pset}$.
	Since there is an edge between every pair of vertices in the two aforementioned sets, we obtain that there are at most
\begin{align*}
vcXp - (\vcX - \frac{\vcX + \satscore}{2})^2 &\geq \vcXp - (\frac{2\cdot\vcX - \vcX - \varnr{}}{2})^2 \\&= \vcXp - \frac{\vcXmnp}{4}
\end{align*}
 edges that are covered by $V'_{\pset}$.

Let us now count the number of edges in \eqref{prop:notowned}.
	Since $\psetcount{\svar{i}} > \vcXp - \vcX\satscore $ by \cref{clm:agentalmostfull}, there are strictly fewer than $\vcX\satscore \leq \varnr{} \cdot \vcX$ edges that are not owned by agents in \pset.
	
We conclude that the number of edges in \eqref{prop:covered} and \eqref{prop:notowned} is at most 
$\vcXp - \frac{\vcXmnp}{4} + \varnr{} \cdot \vcX < \vcXp$.
This contradicts  $V'_{\pset}$ being a vertex cover of $\GindVC{\pset}$.

	We now have that $\sigma'$ must be a satisfying assignment.
	Let us use this fact to compute a lower bound for the size of $V'_{\pset}$.
	Let $\satscoreb$ be the number of variables $\sigma'$ sets to true.
	Recall that no satisfying assignment can set fewer than $\satscore$ variables to true.
	Because $\sigma'$ is a satisfying assignment, we have that $\satscoreb \geq \satscore$.
	\begin{compactitem}[--]
		\item For every $\svar{i} \in \Vars{}$ such that $\sigma'(\svar{i}) = \trueT$, we have two cases:
		\begin{description}
			\item[Case 1:] $\Varag{i} \subseteq \pset$. Every edge in the \multiedge\ between \Pver{i} and~\Nver{i} and edge $\{\nver{i}{1}, \prizev{i}\}$ are owned by agents in \pset.
			By \cref{obs:multiedge}, we must have that $\Pver{i} \subseteq V'_{\pset}$ or $\Nver{i} \subseteq V'_{\pset}$.
			Since~$\sigma'$ sets $\svar{i}$ to true, we have that $|\Pver{i} \cap V'_{\pset}| \geq |\Nver{i} \cap V'_{\pset}|$. and therefore if $\Nver{i} \subseteq V'_{\pset}$, then $\Pver{i} \subseteq V'_{\pset}$.
			Thus $\Pver{i} \subseteq V'_{\pset}$ in either case.
			The vertices in $\Pver{i}$ cannot cover edge $\{\nver{i}{1}, \prizev{i}\}$, so we need at least one vertex from this edge in $V'_{\pset}$.
			Thus we need at least $\vcX + 1 = \lceil\frac{\psetcount{\svar{i}}}{\vcX}\rceil + 1$ vertices from $\Pver{i} \cup \Nver{i} \cup \{\prizev{i}\}$ in $V'_{\pset}$.
			By \cref{clm:minvertexcoverprops}\eqref{clm:minvcsizes:lc}, we need at least further $\psetcount{\svar{i}}$ vertices from $\{\lcv{i}{\mind}{\mindd}{\tind} \mid (\mind, \mindd) \in [\vcX] \times [\vcX], \tind \in [3]\}$.
			\item[Case 2:]  $\Varag{i} \nsubseteq \pset$ in $V'$. 
			By \cref{clm:agentalmostfull}, there is at least one agent corresponding to $\svar{i}$ in $\pset$.
			Thus by \cref{clm:minvertexcoverprops}\eqref{clm:minvcsizes:lc}, the set $V'_{\pset}$ must contain at least $\psetcount{\svar{i}} + 1$ vertices from $\{\lcv{i}{\mind}{\mindd}{\tind} \mid (\mind, \mindd) \in [\vcX] \times [\vcX], \tind \in [3]\}$.
			By \cref{clm:minvertexcoverprops}\eqref{clm:minvcsizes:X}, we need further at least $\lceil \frac{\psetcount{\svar{i}}}{\vcX} \rceil$ vertices from $\Pver{i} \cup \Nver{i}$.

		\end{description}
		In both cases we need $ \lceil\frac{\psetcount{\svar{i}}}{\vcX}\rceil + \psetcount{\svar{i}} + 1$ vertices from $\Pver{i} \cup \Nver{i} \cup \{\prizev{i}\} \cup \{\lcv{i}{\mind}{\mindd}{\tind} \mid (\mind, \mindd) \in [\vcX] \times [\vcX], \tind \in [3]\}$ in $V'_{\pset}$.
		\item For every variable $\svar{i} \in \Vars{}$ such that $\sigma'(\svar{i}) = \falseF$, we need $\lceil \frac{\psetcount{\svar{i}}}{\vcX} \rceil$ vertices from $\Pver{i} \cup \Nver{i}$ by \cref{clm:minvertexcoverprops}\eqref{clm:minvcsizes:X} and at least $\psetcount{\svar{i}}$ vertices from $\{\lcv{i}{\mind}{\mindd}{\tind} \mid (\mind, \mindd) \in [\vcX] \times [\vcX], \tind \in [3]\}$ by \cref{clm:minvertexcoverprops}\eqref{clm:minvcsizes:lc}.
		\item For every clause $C_j = (\ell_i \vee \ell_{i'})\in \varphi$, we obtain from \cref{clm:minvertexcoverprops}\eqref{clm:minvcsizes:C} that we need at least $\lceil\frac{\psetcount{C_j}}{\vcX}\rceil $ vertices from $ \Clav{j}{i} \cup\Clav{j}{i'}$.
		\item For every $\satel \sind \in \Satel$, by \cref{clm:minvertexcoverprops}\eqref{clm:minvcsizes:bc}, we need at least  $\lceil\frac{\psetcount{\satel{\sind}}}{\vcX}\rceil $ from $\Bcv{\sind}{3} \cup \Bcv{\sind+1}{1} \cup \Bcv{\sind+1}{2}$, where $\varnr{} + \clanr{} + 1 = 1$.
	\end{compactitem}
	By combining the above, we have that 
\begin{align*}
\util(\pset) &= |V'_{\pset}| \\
& \geq \sum_{\substack{\svar{i} \in \Vars{} \\ \sigma'[\svar{i}] = \trueT}}(\lceil\frac{\psetcount{\svar{i}}}{\vcX}\rceil + \psetcount{\svar{i}} + 1) \\
&\qquad + \sum_{\substack{\svar{i} \in \Vars{} \\ \sigma'[\svar{i}] = \falseF}}(\lceil\frac{\psetcount{\svar{i}}}{\vcX}\rceil + \psetcount{\svar{i}}) + \\
&\qquad \sum_{C_j \in \varphi}\lceil\frac{\psetcount{C_j}}{\vcX}\rceil + \sum_{\satel \sind \in \Satel} \lceil\frac{\psetcount{\satel{\sind}}}{\vcX}\rceil\\
&\geq \sum_{\svar{i} \in \Vars{}}\psetcount{\svar{i}}\cdot(\vcXcost + 1)+ \sum_{C_j \in \varphi}\vcXcostmul{\psetcount{C_j}} + \satscoreb.
\end{align*}
	By \cref{obs:minallocutil} and the fact that $\satscore \leq \satscoreb$, we obtain
	$\sum_{\agent{} \in \pset}\allocel{\agent{}} \leq  \displaystyle\sum_{\svar{i} \in \Vars{}}\psetcount{\svar{i}}\cdot( \vcXcost + 1) + \satscore + \displaystyle\sum_{C_j \in \varphi} \vcXcostmul{\psetcount{C_j} } \leq \util(\pset)$, a contradiction to \pset\ blocking.
	This shows that $\alloc$ cannot admit a blocking coalition, and must be core stable.
	
	$(\Longleftarrow)$: Assume that the constructed instance $I'$ admits a core stable allocation \alloc. 
	Let $\minvc$ be the size of a minimum vertex cover of  $\GindVC{\allsatagents}$.
	Assume first that $\GindVC{\allsatagents}$ does not admit a minimum vertex cover containing $\varag{1}{1}{1}$.
	Then the size of a minimum vertex cover of $G$ is necessarily $\minvc + 2$.
	This is because we need $\minvc$ vertices from $\allsatagents \setminus \{\varag{1}{1}{1}\}$ to cover the edges belonging to the agents in \allsatagents,
	and two vertices from $\{\pver{1}{1}, \ever{1}, \ever{2}\}$ to cover the triangle between them.
	We have that $\sum_{\agent{} \in \players \setminus \{\eag{1}, \eag{2}, \eag{3}\}}\allocel{\agent{}} \leq \minvc$, as otherwise $\players \setminus \{\eag{1}, \eag{2}, \eag{3}\}$ would block.
	Thus $\allocel{\eag{1}} + \allocel{\eag{2}} + \allocel{\eag{3}} \geq 2$.
However, for each pair $i,j\in[3]$ with $i\neq j$, it holds that $G$ restricted to the edges of $\eag{i}$ and $\eag{j}$ has a minimum vertex cover of size $1$, which by reasoning of \cref{Ex:VCov} implies that $\eag{1} + \eag{2} + \eag{3 }\leq \frac{3}{2}$, a contradiction.
	Thus $\GindVC{\allsatagents}$ must admit a minimum vertex cover containing~$\pver{1}{1}$.
	
	By \cref{clm:vcsatconnect}\eqref{clm:vcsatconnect1}, since $\GindVC{\allsatagents}$ admits a minimum vertex cover of size $\minvc$ that contains $\pver{1}{1}$, we have that $\varphi$ admits a satisfying assignment that sets at most $\minvc - \satvcreq$ variables to true, including $\svar{1}$.
	If $\varphi$ admits a satisfying assignment which sets strictly fewer than $\minvc - \satvcreq$ variables to true, then by \cref{clm:vcsatconnect}\eqref{clm:vcsatconnect2}, graph $\GindVC{\allsatagents}$ admits a minimum vertex cover that is strictly smaller than $\minvc$, a contradiction to $\minvc$ being the size of a minimum vertex cover of $\GindVC{\allsatagents}$. 
Thus~$\varphi$ must admit a satisfying assignment $\sigma$ such that $\svar{1}$ is true under $\sigma$ and there is no satisfying assignment that sets fewer variables to true.
	This concludes the proof.
\end{proof}
}

\section{\ParDSgame}\label{sec:PDSG}
\appendixsection{sec:PDSG}

We now move on to study partitioned dominating set games.
Their computational complexities behave similarly to that for partitioned vertex cover games.

\mypara{Core Verification.} As already discussed for the case of minimum vertex cover, verifying  whether a given allocation is \stable\ reduces to addressing an \NP- and a \coNP-problem.
Indeed, this problem is \DP-complete.

\begin{restatable}[\appsymb]{theorem}{lemverifDS}\label{lem:verifDS}
	\probPDSCV\ is \DP-complete.
	The hardness remains even if $\nbOrg = 1$ or $\Orgsize = 4$.
\end{restatable}

\appendixproofwithstatementandsketch{lem:verifDS}{\lemverifDS*}{
\begin{proof}[Proof sketch]
The proof proceeds in three steps.
First we show \DP-membership, and afterwards we present a \DP-hardness reduction with only one agent.
We defer the proof for $\Orgsize = 4$ to \cref{proof:lem:verifDS} \ifcr\else in the appendix\fi.

\mypara{\baseDP-membership.}
	By \citet{papadimitrioubook}, we need to provide two \NP-complete problems~$P$ and $Q$,
	and show that \probPDSCV~=$P\cap \overline{Q}$, in other words,
	every instance of \probPDSCV\ is a yes-instance if and only if it is a \emph{yes}-instance of~$P$ but a \emph{no}-instance of~$Q$.
	To this end, let $P$ and $Q$ be the following two \NP-problems. 
	\begin{description}
		\item[\text{\normalfont{Problem}}~$P$:] Given an instance~$I=(G = (V_1\cup\dots\cup V_{\coln},E), \alloc)$, 
		does~$G$ admit a dominating set of size at most~~$\sum_{i \in \players}{\allocel i}$?
		\item[\text{\normalfont{Problem}}~$Q$:] Given an instance~$I=(G = (V_1\cup\dots\cup V_{\coln},E), \alloc)$,
		is there a subset~$\pset \subseteq \players$ such that the induced graph~$G[\cup_{i\in \pset}V_i]$ has a dominating set with size strictly smaller than~$\sum_{i\in \pset} \allocel{i}$?    
	\end{description}
	It is straightforward to check that both problems are contained in \NP\ since the witness of either problem has polynomial size and can be verified in polynomial time as well.
	
	It remains to show that for each instance~$I=(G = (V_1\cup\dots\cup V_{\coln},E),\alloc)$ of \probPDSCV,
	allocation~$\alloc$ is \stable\ if and only if $I$ is a yes-instance of~$P$ and a no-instance of~$Q$. 
	By definition, allocation~$\alloc$ is \stable\ if and only if $\alloc$ is a \preimputation\ and no subset of agents is blocking~$\alloc$.
The former is equivalent to~$I$ being a yes-instance of~$P$ and the latter is equivalent to $I$ being a no-instance of $Q$.

\mypara{\baseDP-hardness when $\nbOrg=1$.}	
To show \DP-hardness for the restricted case where there is only one agent,
we reduce from the following \DP-complete \textsc{SAT-UNSAT} problem~\cite{papadimitrioubook}.
	\decprob{SAT-UNSAT}{Two CNF-formulas $\varphi_1$ and $\varphi_2$ over the sets of variables~\Vars 1\ and \Vars 2, respectively, where each clause has exactly three literals.}{Is $\varphi_1$ satisfiable and $\varphi_2$ unsatisfiable?} 
	
	Let $(\varphi_1, \varphi_2)$ be an instance of \textsc{SAT-UNSAT} where $\varphi_1$ and $\varphi_2$ are over the variable sets~$\Vars{1}$ and $\Vars{2}$, respectively.
	We create an instance~$I'=(G, \alloc)$ of \probPDSCV\ with only one agent.
	Graph~$G$ consists of three disjoint subgraphs: $G_{1}$, $G_2^{(1)}$, and $G_2^{(2)}$,  where $G_1$ corresponds to~$\varphi_1$ and $G_2^{(1)}$ and $G_2^{(2)}$ to~$\varphi_2$.
	We note that each subgraph is a slight modification of the standard reduction from \textsc{3SAT} to \textsc{Dominating Set}.
	
	\contrstitle{Graph~$G_1$.} We create the following vertices for~$G_1$:
	\begin{compactitem}[--]
		\item For every variable $\svar{} \in \Vars 1$, create a positive literal vertex~$v_{\svar{}}$,
		a negative literal vertex~$v_{\snegvar{}}$, and a dummy vertex~$d_{\svar{}}$.
		\item For every clause $C \in \varphi_1$, create a clause vertex~$v_C$.
		\item Create a special vertex $v^*$. 
	\end{compactitem}
	We create the following edges for~$G_1$:
	\begin{compactitem}[--]
		\item The three vertices corresponding to the same variable form a triangle:
		For every $\svar{} \in \Vars 1$, add the edges~$\{v_{\svar{}}, v_{\snegvar{}}\}$, $\{v_{\svar{}}, d_{\svar{}}\}$, and $\{v_{\snegvar{}}, d_{\svar{}}\}$.
		\item Each clause vertex is adjacent to the vertices that correspond to the literals in the clause:
		For every clause~$C \in \varphi_1$ and every literal~$\ell \in C$, add edge~$\{v_C, v_{\ell}\}$.
		\item The special vertex is adjacent to all clause and literal vertices, except the dummies:
		For every $v_y \in \{v_{\svar{}}, v_{\snegvar{}} \mid \svar{} \in \Vars 1\} \cup \{v_C \mid C \in \varphi_1\}$, we add edge $\{v^*, v_y\}$.
	\end{compactitem}
	\contrstitle{Graphs~$G_2^{(1)}$ and $G_2^{(2)}$.} These two graphs are constructed in the same way as~$G_1$, except that $\varphi_2$ and $\Vars{2}$ are the underlying CNF formula and variable set.

For the sake of brevity, define $\varnr 1 \coloneqq |\Vars 1|$ and $\varnr 2 \coloneqq |\Vars 2|$. Finally, we construct the value vector~$\alloc = (\varnr 1 + 2\varnr 2 + 2)$; note that we have only one agent, so there is only one value in the vector.
Let $I$ denote the created instance~$I=(G_1\uplus G_2^{(1)} \uplus G_2^{(2)}, \alloc)$.

	The following claim builds a connection between satisfiable formula and the size of a minimum dominating set. 
	\begin{restatable}[\appsymb]{mainclaim}{clmGoneDOMSETSATUNSAT}
\label{clm:G1DOMSETSATUNSAT}
		If $\varphi_1$ (resp.\ $\varphi_2$) is satisfiable, then graph $G_1$ (resp.\ $G_2^{(1)}$ and $G_2^{(2)}$ each) has a minimum dominating set of size $\varnr{1}$ (resp.\ $\varnr{2}$). 
		If $\varphi_1$ (resp.\ $\varphi_2$) is unsatisfiable, then every minimum dominating set of $G_1$ (resp.\ $G_2^{(1)}$ and $G_2^{(2)}$) has~$\varnr{1}+1$ (resp.\ $\varnr{2} + 1$) vertices.

	\end{restatable}
	\appendixproofwithstatement{clm:G1DOMSETSATUNSAT}{\clmGoneDOMSETSATUNSAT*}{
	\begin{proof}\renewcommand{\qedsymbol}{\hfill (end of the proof of~\cref{clm:G1DOMSETSATUNSAT})~$\diamond$}
		We show the statements only for $G_1$; the reasoning for $G_2^{(1)}$ and $G_2^{(2)}$ is analogous.
		First, observe that for every variable~$\svar{} \in \Vars 1$, the dummy~$d_{\svar{}}$ is only adjacent to $v_{\svar{}}$ and $v_{\snegvar{}}$.
		Since there are exactly~$\varnr{1}$ such variables, every dominating set of $G_1$ has to have size at least~$\varnr{1}$.
		
		For the first statement, assume that $\varphi_1$ is satisfiable, and let $\sigma\colon \Vars{1} \to \{\trueT,\falseF\}$ be a satisfying assignment for~$\varphi_1$.
		We construct a dominating set~$V'$ of size~$\varnr{1}$ as follows, which by our previous reasoning a minimum-size dominating set:
		For each variable~$\svar{}\in \Vars{1}$, if $\sigma(x)=\trueT$, then add~$v_{\svar{}}$ to~$V'$; otherwise add~$v_{\snegvar{}}$ to~$V'$.
		Clearly $|V'|=\varnr{1}$.
		It is also straightforward to verify that $V'$ is a dominating set:
		For each variable~$\svar{}\in \Vars{1}$, all three corresponding vertices are dominated.
		For each clause~$C\in \varphi_1$, at least one literal in it is set to~$\trueT$, so by our construction the corresponding literal vertex is added to~$V'$ and dominates~$v_C$.
		The special vertex~$v^*$ is dominated by any vertex in~$V'$.
		
		Now we show the second statement.
		We first show that $G_1$ has a dominating set of size~$\varnr{1}+1$ since all dummies and the special vertex yield a dominating set. 
		Now, to show that this is indeed minimum, we show that if $G'$ has a dominating set of size~$\varnr{1}$, then the formula~$\varphi_1$ is satisfiable.
		Assume that $V'$ is a dominating set for~$G_1$ and of size~$\varnr{1}$.
		As we established previously, the set~$V'$ must contain for every $\svar{} \in \Vars 1$ one of the vertices $v_{\svar{}}, v_{\snegvar{}}, d_{\svar{}}$.
		Since $|V'| = \varnr{1}$, it must contain \emph{precisely} one of these vertices for every $\svar{} \in \Vars 1$, and no other vertices.
		Construct the following assignment~$\sigma\colon \Vars{1} \to \{\trueT, \falseF\}$ with $\sigma(\svar{}) = \trueT$ if $v_{\svar{}}\in V'$, and $\sigma(\svar{})=\falseF$ otherwise.
		We claim that $\sigma$ satisfies~$\varphi_{1}$.
		Consider an arbitrary clause~$C\in \varphi_{1}$.
		Since $V'$ is a dominating set and contains only vertices that correspond to the variables,
		$v_C$ must be dominated by a literal vertex~$v_{\ell}\in V'$ such that $\ell\in C$.
		Clearly we have that $\ell\in \{\svar{}, \snegvar{}\}$ for some variable~$\svar{}$.
		If $\ell=\svar{}$, then by construction, variable~$\svar{}$ is set true, and hence satisfies~$C$. 
		Otherwise $\ell=\snegvar{}$, meaning that $v_{\svar{}}\notin V'$.
		Then, variable~$\svar{}$ is set false, and hence also satisfies~$C$.
	\end{proof}
	}
	
	Using \cref{clm:G1DOMSETSATUNSAT}, we can show the correctness, i.e., instance~$(\varphi_1,\varphi_2)$ is a yes-instance of \textsc{SAT-UNSAT} if and only if $I$ is a yes-instance of \probPDSCV.
	
	$(\Longrightarrow)$: Assume that $(\varphi_1,\varphi_2)$ is a yes-instance, i.e.,
	$\varphi_1$ is satisfiable but $\varphi_2$ is not.
	By \cref{clm:G1DOMSETSATUNSAT}, graph $G_1$ has a minimum dominating set of size~$\varnr{1}$ while every minimum dominating set of~$G_2^{(1)}$ (resp.\ $G_2^{(2)}$) has size~$\varnr{2}+1$.
	This means every minimum dominating set of the entire graph has size~$\varnr{1}+2(\varnr{2}+1)$.
	By definition, \allocation\ $\alloc=(\varnr{1}+2\varnr{2}+2)$ is a \preimputation\ and also \stable, witnessing that $I'$ is a yes-instance.
	
	$(\Longleftarrow)$: Assume $(\varphi_1,\varphi_2)$ is a no-instance.
	Then, formula~$\varphi_1$ is not satisfiable or $\varphi_2$ is satisfiable.
	If $\varphi_1$ is not satisfiable, then by the first statement of \cref{clm:G1DOMSETSATUNSAT}, the size of a minimum dominating set for~$G_1$ is $\varnr{1}+1$.
	Again by  \cref{clm:G1DOMSETSATUNSAT}, every minimum dominating set of the entire graph~$G_1\uplus G_2^{(1)} \uplus G_2^{(2)}$ has size either $\varnr{1}+2\varnr{2}+3$ (if $\varphi_2$ is not satisfiable) or $\varnr{1}+2\varnr{2}+1$ (if $\varphi_2$ is satisfiable).
	This implies that $\alloc$ is not a \preimputation, and hence not \stable.
	
	If $\varphi_1$ is satisfiable, then $\varphi_2$ must be satisfiable to be a no-instance.
	By \cref{clm:G1DOMSETSATUNSAT}, every minimum dominating set of the entire graph~$G_1\uplus G_2^{(1)} \uplus G_2^{(2)}$ has size $\varnr{1}+2\varnr{2}$, again implying that $\alloc$ is not a \preimputation, and hence not \stable.
	
	Since in each case, we conclude that $\alloc$ is not \stable, instance~$(\varphi_1, \varphi_2)$ is a no-instance as well.
\end{proof}

}{
\begin{proof}

\mypara{\baseDP-hardness when $\Orgsize = 4$.}
We now show \DP-hardness when each agent owns at most $4$ vertices.
	We achieve this by providing a polynomial-time reduction from the \textsc{SAT-UNSAT} problem. 
	
	Let $(\varphi_1,\varphi_2)$ be an instance of \textsc{SAT-UNSAT} with variables sets $\Vars{1}$ and $\Vars{2}$, respectively. We create an instance $I'=(G=(V_1\cup\dots\cup V_{\coln},E),\alloc)$ of \probPDSCV\ with $\Orgsize = 4$. Graph~$G$ consists of two disjoint subgraphs: $G_1$ and $G_{2}$, where $G_1$ corresponds to $\varphi_1$ and $G_2$ to $\varphi_2$. We note that each subgraph is a slight modification of the standard reduction from \textsc{3SAT} to \textsc{Dominating Set}.
The construction is illustrated in \cref{fig:verifDS}.
	
\def \xx {1}
\def \xy {0.8}
  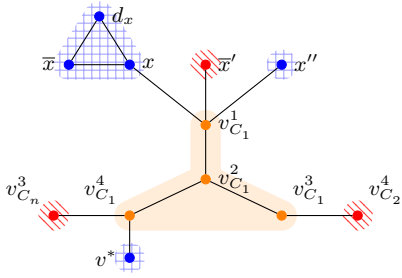
\begin{figure}[t!]
    \centering
   
  \begin{tikzpicture}[black, scale=1,every node/.style={scale=0.9}]
  
\begin{pgfonlayer}{fg}
 \foreach \x / \y / \n / \nn  / \pos / \c in {
	5/5/c11/\claver{1}{1}{}/right/orange,
	5/4.1/c12/\claver{1}{2}{}/right/orange,
	6/3.5/c13/\claver{1}{3}{}/{above right}/orange,
	4/3.5/c14/\claver{1}{4}{}/{above left}/orange,
	7/3.5/c24/\claver{2}{4}{}/{above right}/red,
	3/3.5/cn3/\claver{n}{3}{}/{above left}/red,
	4/6/x/\svar{}/right/blue,
	5/6/xp/\snegvar{}'/right/red,
	6/6/xpp/\svar{}''/right/blue,
	3.2/6/xn/\snegvar{}/left/blue,
	3.6/6.8/dx/d_{\svar{}}/right/blue,
	4/2.8/sp/v^*/left/blue} {
      \node[vcvertex, label=\pos :$\nn$, draw=\c, fill=\c] (\n) at (\x*\xx,\y*\xy) {};
    }
 \end{pgfonlayer}
 
  \foreach \x / \y / \n  in {} {
      \node[hiddenV] (\n) at (\x*\xx,\y*\xy) {};
    }
    
    \foreach \s/\t in {c11/c12,c12/c13,c12/c14,cn3/c14,c13/c24,
    c11/x,c11/xp,c11/xpp,x/xn,x/dx,dx/xn,sp/c14} {
      \draw[] (\s) -- (\t);
    }
    
    \begin{pgfonlayer}{bg}
     \draw[thirdagentA] \hedgeii{c11}{c12}{2mm};
     \draw[thirdagentA] \hedgeiii{c12}{c13}{c14}{2mm};
     \draw[firstagentA] \hedgei{cn3}{2mm};
     \draw[firstagentA] \hedgei{c24}{2mm};
     \draw[firstagentA] \hedgei{xp}{2mm};
     \draw[secondagentA] \hedgei{xpp}{2mm};
      \draw[secondagentA] \hedgei{sp}{2mm};
     \draw[secondagentA] \hedgeiii{x}{xn}{dx}{2mm};
    \end{pgfonlayer}

\end{tikzpicture}

\caption{
Illustration for the proof of \cref{clm:G1DOMSETSATUNSAT}.
We assume that $C_1 = (\svar{} \vee \snegvar{}' \vee \svar{}'')$. The different patterns around the vertices indicate different agents.
\vspace{0.5cm} %
}\label{fig:verifDS}
\end{figure}	

	\contrstitle{Graph~$G_1$.} Assume that the clauses are numbered $C_1,\dots,C_{\clanr{1}}$. We create the following agents and their vertices for~$G_1$:
	
	\begin{compactitem}[--]
		\item  For every variable $\svar{}\in\Vars 1$, we create a positive literal vertex~$v_{\svar{}}$, a negative literal vertex $v_{\snegvar{}}$, and a dummy~$d_{\svar{}}$.
		\item For every clause $C_j\in\varphi_1$, we create the clause vertices $\{\claver{j}{z}{} \mid z \in [4]\}$. 
		\item We create a special vertex $v^*$.
	\end{compactitem}
	We create the following edges for~$G_1$:
	
	\begin{compactitem}[--]
		\item The three vertices corresponding to the same variable form a triangle:
		For every $\svar{} \in \Vars 1$, add the edges~$\{v_{\svar{}}, v_{\snegvar{}}\}$, $\{v_{\svar{}}, d_{\svar{}}\}$, and $\{v_{\snegvar{}}, d_{\svar{}}\}$.
		\item  For every clause $C_j\in\varphi_1$, add the edges $\{\claver{j}{1}{},$ $\claver{j}{2}{}\},$ $\{\claver{j}{2}{}, \claver{j}{3}{}\},$ $\{\claver{j}{2}{}, \claver{j}{4}{}\}, \{\claver{j}{3}{}, \claver{j+ 1}{4}{}\}$, where $\clanr{1} + 1 = 1$.
		\item For every clause $C_j\in\varphi_1$, add edge $\{v_\ell, \claver{j}{1}{}\}$ for every literal~$\ell$ in the clause $C$.
		\item  For the clause $C_1\in\varphi_1$, add edge $\{\claver{1}{4}{},v^*\}$.
	\end{compactitem}
	
\noindent Finally, we create the following agents for $G_1$:
	\begin{compactitem}[--]
		\item  For every variable $\svar{}\in\Vars 1$, create a variable agent \agent{\svar{}}, who owns vertices $v_{\svar{}}$, $v_{\snegvar{}}$, and~$d_{\svar{}}$.
		\item For every clause $C_j\in\varphi_1$, we add a clause agent~\agent{C_j}  who owns vertices $\{\claver{j}{z}{} \mid z \in [4]\}$. 
		\item Create a special agent \agent{*} who owns vertex~$v^*$.
	\end{compactitem}
	
	\contrstitle{Graph~$G_2$.} This graph is constructed in the same way as~$G_1$, except that $\varphi_2$ and $\Vars{2}$ are the underlying CNF formula and variable set.
	
	Let $\players_{1}$ be the  set of agents constructed when constructing $G_{1}$  and~$\players_2$ the set of agents constructed when constructing $G_2$.
	
	We construct the \allocation\ \alloc\ as follows:
	\begin{compactitem}[--]
		\item For every $z \in [2]$:
		\begin{compactitem}[*]
			\item In the set $\players_z$, for every $\svar{} \in \Vars z$, we allocate $\agent{\svar{}}$ cost $1$.
			\item In the set $\players_z$, for every $C_j \in \varphi_z$, we allocate $\clagentM{j}{}$ cost $1$.
		\end{compactitem}
		\item In the set $\players_1$, we allocate $\agent{*}$ cost $0$.
		\item In the set $\players_{2}$, we set allocate $\agent{*}$ cost $1$. 
	\end{compactitem}
	
	For the sake of brevity, we define the number of clauses in the formula as $\clanr 1 \coloneqq |\varphi_1|$ and $\clanr 2 \coloneqq |\varphi_2|$, and we define $\varnr{1}\coloneqq |\Vars{1}|$ and $\varnr{2}\coloneqq |\Vars{2}|$.
	
	We first show the following two claims that establish a connection between our construction and the SAT-formulas.
	
	\begin{claim}\label{orgsizeverifDSforw}
		If $\varphi_1$ (resp.\ $\varphi_2$) is satisfiable, then $G_1$ (resp.~$G_2$) has a minimum dominating set of size $\varnr 1 + \clanr 1$ (resp.\ $\varnr 2 + \clanr 2$).
	\end{claim}
	
	\begin{proof}\renewcommand{\qedsymbol}{\hfill (end of the proof of~\cref{orgsizeverifDSforw})~$\diamond$}
		We show the statement for $G_1$, the reasoning for $G_2$ is analogous. First, observe that for every variable $\svar{}\in\Vars{1}$ the dummy~$d_{\svar{}}$ is only adjacent to $v_{\svar{}}$ and $v_{\snegvar{}}$. Therefore, we need to add one of these three vertices to every dominating set. Similarly, for every clause $C_j\in\varphi_1$, vertex $\claver{j}{2}{}$ can be dominated only by vertices in $\{\claver{j}{z}{} \mid z \in [4]\}$, and thus at least one of them needs to be in every dominating set.
Therefore, any dominating set has cardinality at least $\varnr{1} + \clanr{1}$.
		
Assume that $\varphi_1$ is satisfiable, and let $\sigma\colon \Vars{1}\rightarrow\{\trueT,\falseF\}$ be a satisfying assignment of $\varphi_1$.
We construct a dominating set~$V'$ of~$G_1$ of size $\varnr{1} + \clanr{1}$.
For every variable $\svar{} \in \Vars 1$, if $\sigma$ assigns $\svar{}$ to true, we add $v_{\svar{}}$ to~$V'$, otherwise we add $v_{\snegvar{}}$.
For every clause $C_j\in \varphi_1$, we add the agent~$\claver{j}{4}{}$ to~$V'$.
Clearly $|V'| = \varnr{1} + \clanr{1}$.
		
For every $\svar{} \in \Vars z$, the agents $v_{\svar{}}, v_{\snegvar{}}$, and $d_{\svar{}}$ are dominated.
Also, for every $C_j\in \varphi_1$, vertex $\claver{j}{2}{}$ is dominated by~$\claver{j}{4}{}$, vertex $\claver{j}{3}{}$ by~$\claver{j+1}{4}{}$ (where $\clanr{z} + 1 = 1$), and $\claver{j}{4}{}\in V'$.
Moreover, vertex $v^*$ is dominated by $\claver{1}{4}{}$.
		
It remains to show that for every $C_j\in \varphi_1$, vertex $\claver{j}{1}{}$ is dominated.
Since $\sigma$ is a satisfying assignment, there must be a literal $\ell$ such that $\{\claver{j}{1}{}, v_{\ell}\} \in E(G_1)$ and $v_{\ell} \in V'$.
Thus $V'$ is a dominating set of $G_1$ of size  $\varnr{1} + \clanr{1}$.
	\end{proof}
	
	\begin{claim}\label{orgsizeverifDSbackw}
		Let $\pset_X \subseteq \{\agent{\svar{}} \mid \svar{} \in \Vars 1\}$ (resp.~$\{\agent{\svar{}} \mid \svar{} \in \Vars 2\}$) and let $\psetC \subseteq \{\agent{C_j} \mid C_j\in \varphi_1\}$ (resp.~$\{\agent{C_j} \mid C_j\in \varphi_2\}$). If the subgraph of $G_1$ (resp.~$G_2$) induced by the vertices of $\pset_X \cup \psetC \cup \{\agent{*}\}$ admits a dominating set of size at most $|\pset_X| + |\psetC|$, then $\varphi_1$ (resp.~$\varphi_2$) admits a satisfying assignment.
		In particular, if $G_1$ (resp.~$G_2$) admits a dominating set of size $\varnr 1 + \clanr 1$ (resp.\ $\varnr 2 + \clanr 2$), then $\varphi_1$ (resp.\ $\varphi_2$) admits a satisfying assignment.
	\end{claim}

	\begin{proof}\renewcommand{\qed}{\hfill (end of the proof of~\cref{orgsizeverifDSbackw})~$\diamond$}
		We show the statement for $G_1$, the reasoning for $G_2$ is analogous. 
		Observe that for every $\agent{\svar{}} \in \pset_X$, vertex $d_{\svar{}}$ must be dominated by one of the vertices from $\{d_{\svar{}}, v_{\svar{}}, v_{\snegvar{}}\}$.
		Vertices~$v_{\svar{}}$ and $v_{\snegvar{}}$ may also dominate vertices in $\{\claver{j}{1}{} \mid \agent{C_j} \in \psetC\}$, but no other vertices belonging to the clause agents.
		Thus we can use at most $|\psetC|$ many vertices to dominate the vertices in $\{\claver{j}{z}{} \mid \agent{C_j} \in \psetC, z \in \{2,3,4\}\} \cup \{v^*\}$.
		We will now show that every clause agent in $\players_{1}$ must be in $\psetC$.
		Assume towards a contradiction that this is not the case.
		Consider the subgraph $G^C \coloneqq G_{1}[\{\claver{j}{z}{} \mid C_j \in \psetC, z \in \{2,3,4\} \cup \{v^*\}]$.
Since not every clause agent in $\players_{1}$ is in $\psetC$, every connected component of this subgraph must be either (1) a path of length 3$\ell$ for some $\ell \in \mathds{N}$ or (2) the component that contains $v^*$.
As $G_{1}[\{\claver{j}{z}{} \mid z \in \{2,3,4\}]$ is connected for every  $\agent{C_j} \in \psetC$, every clause agent in \psetC\ owns vertices on exactly one connected component.

Every connected component in (1) must contain vertices from~$k$ clause agents for some $k \in \mathds{N}$. Moreover, a minimum dominating set on such a component is of size~$k$, because one vertex can be used to dominate at most two other vertices on a path.
Since the components in (1) require at least one vertex for every clause agent in it, the component in (2) must also be dominated by the number of vertices that corresponds to the number of clause agents in it.
If $(2)$ contains only $\{v^*\}$, we must dominate it with zero vertices, which is impossible.
Therefore the component in (2) must contain at least the vertices in $\{\claver{j}{2}{},\claver{j}{3}{},\claver{j}{4}{}, v^*\}$.
		Without~$v^*$, this component contains $3k$ vertices, where $k$ is the number of clause agents whose vertices are a part of the connected component. As previously, a minimum dominating set on such a component is of size $k$, because one vertex can be used to dominate at most two other vertices.
		Moreover, this set is unique: It consists of the agents $\claver{j}{2}{}$ for every clause $C_j$ whose corresponding vertices are a part of this component.
		However, as this set does not contain $\claver{1}{4}{}$, it is not a minimum dominating set of the whole component.
		Thus we need at least $k + 1$ vertices to cover the whole component.
		But this implies that we need $|\psetC| + 1$ many vertices to dominate all the vertices in $\{\claver{j}{z}{} \mid \agent{C_j} \in \psetC, z \in \{2,3,4\}\} \cup \{v^*\}$ and thus $|\psetC| + |\pset_X| + 1$ many vertices to dominate $\bigcup_{i \in \psetC \cup \pset_X \cup \{\agent{*}\}}V_i$, a contradiction.
		
		Thus we must have that every clause agent in $\players_{1}$ is in $\psetC$.
		Recall that we can use at most $\clanr{1}$ many vertices to dominate the vertices in $\{\claver{j}{z}{}\mid \agent{C_j}\in \varphi_1, z \in \{2,3,4\}\} \cup \{v^*\}$.
		Therefore for every $C_j \in \varphi_1$, we need exactly one vertex from $\{\claver{j}{2}{}, \claver{j}{3}{},\claver{j}{4}{}\}$ in the dominating set, as no other vertex dominates $\claver{j}{2}{}$.
		Because $v^*$ must be dominated, we must have that $\claver{1}{4}{}$ is in the dominating set~$V'$.
		Because $\claver{1}{3}{}$ must be dominated, we must have that $\claver{2}{4}{} \in V'$.
		By repeating this argument we obtain that for every $C_j \in \varphi_1$, $\claver{j}{4}{}\in V'$ and $\claver{j}{2}{}, \claver{j}{3}{} \notin V'$;
		moreover, vertex $\claver{j}{1}{}$ is not dominated by~$\claver{j}{2}{}$ and must be dominated by some vertex belonging to a variable agent.
		
		Consider assignment $\sigma$ that assigns a variable $\svar{} \in \Vars 1$ to true if $v_{\svar{}} \in V'$ and false otherwise. Because we can have at most one vertex from every variable agent in $V'$—otherwise $V'$ would be too large—we have that $v_{\snegvar{}} \in V'$ implies that $\sigma$ assigns $\svar{}$ to false.
Assignment $\sigma$ must be a satisfying assignment: As we established previously, for every $C_j \in \varphi_1$ there must be a literal $\ell \in C_j$ such that $v_{\ell} \in V'$, meaning that the literal $\ell$ is true under $\sigma$.
	\end{proof}
	Using \cref{orgsizeverifDSforw,orgsizeverifDSbackw}, we can show correctness, i.e., instance~$(\varphi_1,\varphi_2)$ is a yes-instance of \textsc{SAT-UNSAT} if and only if $I'$ is \stable.
	
	$(\Longrightarrow)$ First assume $(\varphi_1,\varphi_2)$ is a yes-instance, i.e., formula~$\varphi_1$ is satisfiable but $\varphi_2$ not.
	
	Let us start by showing the following claim that restricts the blocking coalitions we must consider:
	\begin{claim}\label{clm:DSdummyblock}
		No set of agents $\pset \subseteq \players_{1} \setminus \{\agent{*}\}$ (resp. $\players_{2} \setminus \{\agent{*}\}$) may block.
	\end{claim}
	
	\begin{proof}\renewcommand{\qed}{\hfill (end of the proof of~\cref{clm:DSdummyblock})~$\diamond$}
		We show the statement for $\players_1$, the reasoning for $\players_2$ is analogous. 
		Let $\pset \subseteq \players_{1} \setminus \{\agent{*}\}$.
		Let $\pset_X$ be the set of variable agents in $\pset$ and $\psetC$ be the set of clause agents in $\pset$.
		We observe that by construction $\sum_{i \in \pset}\allocel i = |\pset_X| + |\psetC|$.
		Let $V'$ be a minimum dominating set of $G_{1}[\bigcup_{i \in \pset}V_i]$.
		For every variable agent $\agent{\svar{}} \in \pset_X$, there must be a vertex that dominates the dummy vertex $d_{\svar{}}$. By construction, this must be one of the three vertices owned by $\agent{\svar{}}$, as no other vertices can dominate $d_{\svar{}}$.
		For every clause agent $\agent{C_j} \in \psetC$, vertex~$\claver{j}{2}{}$ must be dominated by one of the vertices in $\{\claver{j}{z}{} \mid z \in [4]\}$.
		As none of the above enumerated sets overlap, we must have that $|V'| \geq |\pset_X| + |\psetC|$.
		Thus $\sum_{i \in \pset}\allocel i = |\pset_X| + |\psetC| \geq \util(\pset)$, as required.
	\end{proof}
	Since graphs $G_1$ and $G_2$ are not connected to each other, by \cref{obs:sepgraphs} it is sufficient for us to show that $\alloc$ restricted to each of these graphs is \stable.
We have four possible cases to consider:

	\begin{description}
		\item[Case 1:] A subset $\pset \subseteq \players_1$ blocks \alloc\ restricted to $\players_1$.
		By \cref{clm:DSdummyblock}, we have that $\agent{d} \in \pset$ and $\sum_{i \in \pset \setminus \{\agent{d}\}} \allocel i \leq \util(\pset \setminus \{\agent{d}\})$.
		Since $\allocel {\agent{*}} = 0$, it holds that $\sum_{i \in \pset \setminus \{\agent{d}\}} \allocel i = \sum_{i \in \pset}\allocel i$.
		Additionally, since $v*$ is adjacent only to $\claver{1}{4}{}$, we have that $\util(\pset \setminus \{\agent{*}\}) \leq \util(\pset)$.
		Thus we obtain that $\sum_{i \in \pset} \allocel i = \sum_{i \in \pset \setminus \{\agent{d}\}} \allocel i \leq \util(\pset \setminus \{\agent{*}\}) \leq \util(\pset)$, a contradiction to $\pset$ blocking.
		\item[Case 2:] We have that \alloc\ restricted to $\players_2$ is not a \preimputation.
		Observe that $\sum_{i \in \players_1}\allocel i = \varnr{1} + \clanr{1}$.
		As the previous case implies that $\sum_{i \in \players_1} \allocel i \leq \util(\players_1)$, we must have that $\varnr{1} + \clanr{1} < \util(\players_1)$, i.e., a minimum dominating set of $G_1$ is strictly larger than $\varnr{1} + \clanr{1}$.
		However, as $\varphi_1$ admits a satisfying assignment, graph $G_1$ admits a dominating set of size $\varnr{1} + \clanr{1}$ by \cref{orgsizeverifDSforw}, a contradiction.
		\item[Case 3:] A subset $\pset \subseteq \players_{2}$ blocks \alloc\ restricted to $\players_2$.
		By \cref{clm:DSdummyblock}, we have that $\agent{*} \in \pset$.
		Let $\pset_X$ be the set of variable agents in $\pset$ and $\psetC$ be the set of clause agents in $\pset$.
		We have that $\sum_{i \in \pset}\allocel i =  |\pset_X| + |\psetC| + 1$.
		Thus we must have that $G_{2}[\bigcup_{i \in \pset}V_i]$ admits a dominating set $V'$ of size at most $|\pset_X| + |\psetC|$.
		By \cref{orgsizeverifDSbackw}, the formula $\varphi_2$ is then satisfiable.
		This however contradicts that $(\varphi_1,\varphi_2)$ is a yes-instance of \textsc{SAT-UNSAT}.
		\item[Case 4:]  We have that \alloc\ restricted to $\players_2$ is not a \preimputation.
		The previous case establishes that $\sum_{i \in \players_{2}}\allocel i \leq \util(\players_{2})$, as otherwise~$\players_{2}$ would block.
		Thus we must have that $ \util(\players_{2}) > \sum_{i \in \players_{2}}\allocel i = \varnr{2} + \clanr{2} + 1$, i.e., a minimum dominating set of~$G_{2}$ is strictly larger than $\varnr{2} + \clanr{2} + 1$.
		Towards a contradiction, we construct a dominating set $V'$ of~$G_{2}$ of size $\varnr{2} + \clanr{2} + 1$.
		Let $V' \coloneqq \{v_{\svar{}} \mid \svar{} \in \Vars 2\} \cup \{\claver{j}{2}{} \mid C_j \in \varphi_2\} \cup \{v^*\}$.
		Clearly $|V'| = \varnr{2} + \clanr{2} + 1$.
		For every variable $\svar{} \in \Vars 2$, the vertices $v_{\snegvar{}}$ and $d_{\svar{}}$ are dominated by $v_{\svar{}}$.
		For every clause $C \in \varphi_2$, the vertices $\claver{j}{1}{}, \claver{j}{3}{},$ and $\claver{j}{4}{}$ are dominated by $\claver{j}{4}{}$.
Vertex $v^*$ is in $V'$.
		Thus every vertex is either in $V'$ or dominated by a vertex in~$V'$.
	\end{description}
	This concludes the proof of forward direction.
	
	$(\Longleftarrow)$ Assume $I'$ is a yes-instance of \probPDSCV, i.e., \allocation~$\alloc$ is \stable.
	Because $G_1$ and $G_{2}$ are not connected, we must have that $\alloc$ is \stable\ when restricted to both of these graphs.
	
	We first show that $\varphi_1$ admits a satisfying assignment.
	Since $\sum_{i \in \players_{1}} \allocel i = \varnr{1} + \clanr{1}$, graph~$G_{1}$ must admit a dominating set~$V'$ of size $\varnr{1} + \clanr{1}$.
	By \cref{orgsizeverifDSbackw}, this implies that $\varphi_1$ admits a satisfying assignment.

	Next we show that $\varphi_2$ cannot admit a satisfying assignment.
	Since $\sum_{i \in \players_{2}} \allocel i = \varnr{2} + \clanr{2} + 1$, graph~$G_{2}$ cannot admit a dominating set $V'$ of size $\varnr{2} + \clanr{2}$.
	By the contra-positive of \cref{orgsizeverifDSforw}, formula $\varphi_2$ cannot admit a satisfying assignment.
	Thus $(\varphi_1,\varphi_2)$ is a yes-instance of \textsc{SAT-UNSAT}.
\end{proof}
}

\mypara{Core Existence.}
Through a technique analogous to the one for  \cref{thm:VCCEthetacont}, we also obtain \thetaC-containment for dominating set.

\begin{restatable}[\appsymb]{theorem}{thmDSCEcont}\label{thm:DSCEcont}
	\probPDSCE\ is contained in \thetaC.
\end{restatable}
\appendixproofwithstatement{thm:DSCEcont}{\thmDSCEcont*}{
\begin{proof}

The proof is highly analogous to the proof of \cref{thm:VCCEthetacont}.
Instead of showing \coNP-containment of \probPVCCEhintshort, we show the \coNP-containment of the corresponding problem for \probPDSCE:
\decprob{\probPDSCEhint}{An instance $G = (V = V_1 \cup \dots V_{\coln}, E)$ and $\minvc$ which is the cardinality of a minimum dominating set of $G$.}{Is $G$ a yes-instance of \probPDSCE?}
The proof is analogous to \cref{clm:minvcprobpvcceconp}: Our witness $\hat{Q}$ contains triples $(\alloc, \pset, V'_\pset)$ for every call to oracle $Q$ over the course of solving \probPDSCE\ on $G$, where (1) $\alloc$ is the allocation that is queried for $Q$, (2) $\pset$ is a coalition that blocks $\alloc$, or $\pset = \players$ if $\alloc$ is not a \preimputation, and (3) $V'_\pset$ is a \emph{dominating set} of $G[\bigcup_{i \in \pset}V_i] $ such that $|V'| < \sum_{i \in \pset}\allocel i$, or $\emptyset$ if $\alloc$ is not a \preimputation.
We can verify this witness as in the proof of \cref{clm:minvcprobpvcceconp}.

The rest of the proof proceeds as the proof of \cref{thm:VCCEthetacont}: We find the cardinality of a minimum dominating set of $G$ through $\log(|V|)$ calls to \DS, and then conclude by one call to the co-problem of  \probPDSCEhint.
\end{proof}
}

We complement the $\thetaC$-containment result above by showing $\thetaC$-hardness.
The proof is similar to the one for \cref{thm:VCtheta}. %

\begin{restatable}[\appsymb]{theorem}{thmDStheta}
\label{thm:DStheta}
\probPDSCE\ is \thetaC-hard even when $\nbOrg = 4$.
\end{restatable}

\appendixproofwithstatement{thm:DStheta}{\thmDStheta*}{

\begin{proof}
We first show the \thetaC-hardness of the following problem.
	\decprob{Dominating Set Membership}{A graph $G=(V,E)$ and a vertex $v'\in V$.}{Does $G$ admit a \emph{minimum} dominating set $V'$ such that $v'\in V'$?}
	\begin{claim}\label{clm:DSmembership}
		\textsc{Dominating Set Membership} is $\thetaC$-hard.
	\end{claim}
	\begin{proof}\renewcommand{\qed}{\hfill (end of the proof of~\cref{clm:DSmembership})~$\diamond$}
		To show hardness, we reduce from \VCM. 
We utilize standard reduction from \textsc{Vertex Cover} to \textsc{Dominating Set}. Let $I=(G = (V,E),v')$ be an instance of \VCM. We now create the following instance $I'=(G' = (U,E'),u')$ for \textsc{Dominating Set Membership}:
Let us first construct the vertices.
\begin{compactitem}[--]
\item For each edge $e\in E$, we add a vertex $u_e$. Vertices added in this way are referred to as edge-vertices.
\item For each vertex $v\in V$, we add a vertex $u_v$. Vertices added in this way are referred to as vertex-vertices. 
\end{compactitem}

Next we construct the edges.
\begin{compactitem}[--]
\item For each $v, w \in V, v \neq w$, we add the edge $\{u_v, u_w\}$
\item For each $e \in E, v \in e$, we add the edge $\{u_v, u_e\}$.
\end{compactitem}
Finally, we set $u' \coloneqq u_{v'}$.

We start by noting that there exists a minimum dominating set of~$G'$ consisting only of vertex-vertices: Given a dominating set $U'$ of $U$ containing an edge vertex $u_e, e \in e$, we can construct a dominating set $U''$ with $|U''| \leq |U'|$ by replacing $u_e$ with $u_v$ such that $v \in e$, since $u_v$ dominates $u_e$ and every vertex adjacent to $u_e$.

Given a vertex cover $V'$ of $G$, we can verify that $U' \coloneqq \{u_v \mid v \in V'\}$ is a dominating set of $G'$ such that $|U'| = |V'|$, and $v' \in V'$ if and only if $u_{v'} \in U'$.
Similarly, given a dominating set $U'' \subseteq U$ such that $U''$ only contains vertex-vertices, we can verify that $V'' \coloneqq \{v \in V \mid u_v \in U''\}$ is a vertex cover of $G$ such that $|U''| = |V''|$ and $v' \in V''$ if and only if $u_{v'} \in U''$.
Therefore it follows directly that $v'$ is in a minimum vertex cover in $G$ if and only if $u_{v'}$ is in a minimum dominating set of $G'$, thereby showing the statement. 
	\end{proof}
	
	We can now reduce from \textsc{Dominating Set Membership} to \probPDSCE. Let $I=(G = (V,E),v')$ be an instance of \textsc{Dominating Set Membership}.
We will now generate an instance $I' = ([4], G' = (V_1 \cup V_2 \cup V_3 \cup V_4, E'))$ of \probPDSCE\ that contains four agents. We describe the instance $I'$ by describing the sub-graph induced by each agent's vertices first, and then describing the edges connecting these sub-graphs.
The reduction is described in \cref{fig:DStheta}.

\def \xx {1}
\def \xy {1}

\newcommand{\drawtriangle}[5]{
\begin{pgfonlayer}{fg}
 \foreach \x / \y / \n / \nn  / \pos / \c in {
 	#1/#2/v#41/v_{#4,1}/left/#3,
 	{#1+1}/#2/v#42/v_{#4,2}/right/#3,
 	{#1 +0.5}/{#2 - 0.7}/v#43/v_{#4,3}/right/#3} {
      \node[vcvertex, label=\pos :$\nn$, draw=\c, fill=\c] (\n) at (\x*\xx,\y*\xy) {};
    }

  \draw[decoration={brace,raise=15pt},decorate] ($(v#42)+(0.1,0.1)$) --node[midway, xshift=25pt,yshift=-10pt] (E3) {$E_{#4}$} ($(v#43)+(-0.1,-0.1)$);

 \end{pgfonlayer}
 
    \foreach \s/\t in {v#41/v#42,v#41/v#43,v#42/v#43} {
      \draw[] (\s) -- (\t);
    }
    
       \begin{pgfonlayer}{bg}
     \draw[#5] \hedgeiii{v#41}{v#42}{v#43}{2mm};
    \end{pgfonlayer}
}  

\newcommand{\drawtrianglep}[5]{
\begin{pgfonlayer}{fg}
 \foreach \x / \y / \n / \nn  / \pos / \c in {
 	#1/#2/v#41/v_{#4,1}/left/#3,
 	{#1+1}/#2/v#42/v_{#4,2}/right/#3,
 	{#1 +0.5}/{#2 - 0.7}/v#43/v_{#4,3}/left/#3} {
      \node[vcvertex, label=\pos :$\nn$, draw=\c, fill=\c] (\n) at (\x*\xx,\y*\xy) {};
    }

 \draw[decoration={brace,raise=15pt},decorate] ($(v#43)+(-0.1,-0.1)$) --node[midway, xshift=-25pt,yshift=-10pt] (E3) {$E_{#4}$} ($(v#41)+(-0.1,+0.1)$);
 
    \foreach \s/\t in {v#41/v#42,v#41/v#43,v#42/v#43} {
      \draw[] (\s) -- (\t);
    }

  \end{pgfonlayer}
    
       \begin{pgfonlayer}{bg}
     \draw[#5] \hedgeiii{v#41}{v#42}{v#43}{2mm};
    \end{pgfonlayer}
}  

  \begin{figure}[t!]

\centering
  \begin{tikzpicture}[black, scale=1,every node/.style={scale=0.9}]
  
   \node[draw=green!60!black, ellipse, inner sep=10pt, line width=2pt, dashed, minimum width=70pt] (G) at (3.2*\xx,5.3*\xy) {$G$};

\draw[decoration={brace},decorate] ($(G)+(-1.1,0.7)$) --node[right,yshift=8pt] (E4) {$E_4$} ($(G)+(1.1,0.7)$);

\node[vcvertex, label=left :$v'$] (vp) at (4*\xx,5.3*\xy) {};
  
\drawtriangle{5}{5}{red}{1}{firstagentA}

\drawtrianglep{2.5}{3}{blue}{2}{secondagentA} 

\drawtriangle{5.5}{3}{orange}{3}{thirdagentA}
 
  \foreach \x / \y / \n  in {} {
      \node[hiddenV] (\n) at (\x*\xx,\y*\xy) {};
    }
    
    \foreach \s/\t/\b in {v12/v21/-20,v12/v22/20,v12/v23/20,
    v13/v31/20,v13/v32/20,v13/v33/20,
    v21/v11/20,v21/v13/20,
    v23/v31/-20,v23/v32/-20,v23/v33/-20,
    v31/v11/30,v31/v12/30,
    v32/v21/20,v32/v22/20,
    vp/v11/40,vp/v12/40,vp/v13/40} {
      \draw[] (\s) edge[bend left=\b] (\t);
    }

\end{tikzpicture}

\caption{
Illustration for the proof of \cref{thm:DStheta}.  %
}\label{fig:DStheta}\vspace{0.5cm}
\end{figure}
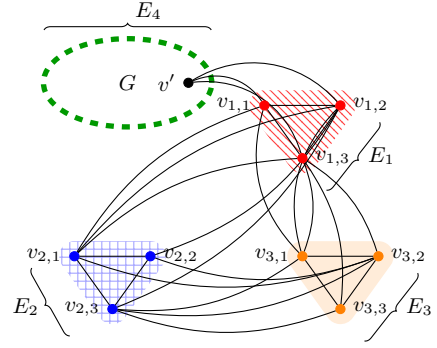	
	
\begin{compactitem}[--]
\item $G'[V_1],G'[V_2],G'[V_3]$ are all cliques of size $3$, i.e., triangles. We name the vertices in the following way: $V_1=\{v_{1,1},v_{1,2},v_{1,3}\}$, $V_2=\{v_{2,1},v_{2,2},v_{2,3}\}$, $V_3=\{v_{3,1},v_{3,2},v_{3,3}\}$.
\item $G[V_4]=(V,E)$, i.e., the organization has as its induced subgraph exactly the graph given in $I$. The vertices will keep the naming in~$I$. In particular, vertex~$v'$ will keep its name.
\item We add the edges $\{v_{i,j},v_{j,k}\}$ for every $i,j,k\in[3]$ for $i\neq j$. In other words for each triangle, there is one vertex that dominates all vertices in all other triangles. Finally, we add edges $\{v',v_{1,i}\}$ for $i\in[3]$. 
\end{compactitem}
	We claim that $I$ is a yes-instance of \textsc{Dominating Set Membership} if and only $I'$ is a yes-instance of \probPVCCE.
	
$(\Longrightarrow)$: Assume $I$ is a yes-instance of \textsc{Dominating Set Membership}. Let $\minvc$ be the minimum dominating set size of $G[V_1]$. We claim that $\alloc=(0,0,1,\minvc)$ is an allocation in the core. We start by noting that the minimum dominating set size of the graph must be of size at least $\minvc+1$, as it is not possible to dominate the vertices in $V_1\cup V_2\cup V_3$ using any single vertex.
As only $v'$ in $V_4$ can be used to dominate vertices $V_1\cup V_2\cup V_3$, namely $V_1$, it follows that the minimum dominating set size is at least $\minvc+1$.

Let $V'$ be a minimum dominating set of size $\minvc$ containing $v'$. The set $V'\cup\{v_{3,2}\}$ dominates all vertices in the graph, as vertices in $V_4$ are dominated by $V'$, vertices in $V_1$ are dominated by $v'$, and vertices in $V_2$ and $V_3$ are dominated by $v_{3,2}$. We now show that the allocation $(0,0,1, \minvc)$ is not blocked.
Suppose agent~$4$ is part of an inclusion-wise minimal blocking coalition.
Then agent~$1$ must be part of the blocking coalition as well, as the other agents' vertex sets do not have vertices that are adjacent to vertices in $V_4$ and agent~4 does not block on her own.
However, agents~1 and~4 have a total \costtxt\ of~$\minvc$ and adding any additional agent would increase the \costtxt\ of the blocking coalition to $\minvc+1 = \sum_{i \in [4]}\allocel i$.
For every subset $\pset \subseteq [3]$, it holds that $\sum_{i \in [3]}\allocel i \leq 1 \leq \util(\pset)$, contradicting $\pset$ blocking.

$(\Longleftarrow)$: Assume $I'$ is a yes-instance of \probPDSCE.
Let \alloc\ be a \stable\ \allocation.
No pair of agents in $[3]$ can have a total \costtxt\ of more than~$1$, as the minimum dominating set size for each pair is exactly~$1$.
From this it follows that $\allocel 1+\allocel 2+\allocel 3<2$.
Let~$\hat k$ be the minimum dominating set size of $G'$.
Thereby it follows that $\allocel 4> \hat k -2$, and therefore the cardinality of a minimum dominating set of $G[V_4]$ is lower-bounded by $\hat k-1$.
This implies that $G$ must admit a minimum-cardinality dominating set containing $v'$, as the only vertex that can be used to cover vertices in $V_1\cup V_2\cup V_3$ is $v'$ and it is not possible to cover the vertices in $V_1\cup V_2\cup V_3$ using only a single vertex from $V_1\cup V_2\cup~V_3$.
\end{proof}
}

Next, we show that the problem remains \thetaC-hard even when $\Orgsize = 9$ by reducing from \probPVCCE. %

\begin{restatable}[\appsymb]{theorem}{thmDSthetaorgsize}
\label{thm:DSthetaorgsize}
\probPDSCE\ is \thetaC-hard even when $\Orgsize=~9$.
\end{restatable}

\newcommand{\GindDS}[1]{\ensuremath{\GDS[\bigcup_{\agent{} \in #1}\VDS_{\agent{}}]}}
\newcommand{\GindVCspec}[1]{\ensuremath{\GVC[\bigcup_{\agent{} \in #1}E_{\agent{}}]}}
\newcommand{\Uag}{\players^{\DSsym}_{\VVC}}
\newcommand{\UagS}{\hat{\players}^{\DSsym}_{\VVC}}
\newcommand{\DSsym}{\textsc{DS}}
\newcommand{\VCsym}{\textsc{VC}}
\newcommand{\GVC}{G^{\VCsym}}
\newcommand{\EVC}{E^{\VCsym}}
\newcommand{\VVC}{V^{\VCsym}}
\newcommand{\GDS}{G^{\DSsym}}
\newcommand{\EDS}{E^{\DSsym}}
\newcommand{\VDS}{V^{\DSsym}}

\newcommand{\NVC}{\players^{\VCsym}}
\newcommand{\NDS}{\players^{\DSsym}}

\newcommand{\iVC}{a_i^{\VCsym}}
\newcommand{\iDS}{a_i^{\DSsym}}
\newcommand{\ivDS}{a_v^{\DSsym}}

\newcommand{\VhVC}{\hat{V}^{\VCsym}}
\newcommand{\VpVC}{{V'}^{\VCsym}}
\newcommand{\VppVC}{{V''}^{\VCsym}}

\newcommand{\VhDS}{\hat{V}^{\DSsym}}
\newcommand{\VpDS}{{V'}^{\DSsym}}
\newcommand{\VppDS}{{V''}^{\DSsym}}

\newcommand{\UV}{U^{\DSsym}_{\VVC}}

\newcommand{\EhVC}{\hat{E}^{\VCsym}}

\newcommand{\allocVC}{\ensuremath{\alloc^{\VCsym}}}
\newcommand{\allocDS}{\ensuremath{\alloc^{\DSsym}}}
\newcommand{\allocelVC}[1]{\allocel{#1}^{\VCsym}}
\newcommand{\allocelDS}[1]{\allocel{#1}^{\DSsym}}

\newcommand{\psetVC}{\pset^{\VCsym}}
\newcommand{\psetDS}{\pset^{\DSsym}}

\appendixproofwithstatement{thm:DSthetaorgsize}{\thmDSthetaorgsize*}{
\begin{proof}
 	We reduce from \probPVCCE, which by \cref{thm:VCtheta} is \thetaC-hard even if $\Orgsize = 9$.
	Let $I^{\VCsym} = \GVC = (\VVC, \EVC)$ with the agent set $\NVC$ be an instance of \probPVCCE.
	We construct a reduced instance $I^{\DSsym} = \GDS = (\VDS, \EDS)$ with the agent set $\NDS$ of \probPDSCE\ that is inspired by the standard reduction from vertex cover to dominating set.
	The vertex set $\VDS$ is constructed as follows:
	\begin{compactitem}[--]
		\item For every $v \in \VVC$, we construct a vertex-vertex $u_v$ in $\GDS$.
		\item For every $e \in \EVC$, we construct an edge-vertex $u_e \in \GDS$.  
	\end{compactitem}
	Let $\UV \coloneqq \{u_v \mid v \in \VVC\}$ be the set of vertex-vertices.
	The edge set $\EDS$ is constructed as follows:
	\begin{compactitem}[--]
		\item For every $v, v' \in \VDS, v \neq v'$, we add the edge $\{u_v, u_{v'}\}$.
		\item For every $e = \{v, v'\} \in E$, we add the edges $\{u_v, u_e\}$ and $\{u_{v'}, u_e\}$.
	\end{compactitem}
	The agent set $\NDS$ is constructed as follows:
	\begin{compactitem}[--]
		\item For every $\iVC \in \NVC$, we create an agent $\iDS$ that owns the vertices $\{u_e \mid e \in E_{\iVC}\}$. Clearly $\iDS$ owns the same number of vertices as the number of edges owned by $\iVC$ in $I^{\VCsym}$.
		\item For every $v \in \VDS$, we create a vertex agent $\ivDS$ that owns the vertex set $\{u_v\}$.
	\end{compactitem}
	Let $\Uag \coloneqq \{\ivDS \mid v \in \VVC\}$.
	Observe that the maximum organization size of $I^{\DSsym}$ is equal to the maximum organization size of $I^{\VCsym}$, and is thus a constant.
	
\begin{claim}\label{clm:dsvertexprop}
Let $\VhDS \subseteq \VDS$ 
and let $\VpDS$ be a dominating set of
 $\GDS[\VpDS]$.
		Then there is a dominating set $\VppDS$ of $G[\VhDS \cup \UV]$ such that $|\VppDS| \leq |\VpDS|$ and $\VppDS \subseteq \UV$.
	\end{claim}
	\begin{proof}\renewcommand{\qedsymbol}{\hfill (end of the proof of~\cref{clm:dsvertexprop})~$\diamond$}
		We construct the dominating set $\VppDS$ as follows: For every $e \in \EDS$ such that $u_e \in \VpDS$, we remove $u_e$ from $\VpDS$ and replace it with a vertex $u_v$ such that $v \in e$, unless such $u_v$ is already in~$\VpDS$, in which case we just remove $u_e$. 
		Since $u_v$ dominates every vertex that $u_e$ dominates, if $\VpDS$ is a dominating set of $\GDS[\VhDS]$, so is~$\VppDS$; because $\VpDS$ is non-empty, set $\VppDS$ must contain at least one vertex from $\UV$, and thus every vertex in~$\UV$ is dominated.
		Clearly $|\VppDS| \leq |\VpDS|$ and $\VppDS \subseteq \UV$.
		
	\end{proof}
	
	\begin{claim}\label{clm:mindcvceq}
		Let $\EhVC \subseteq \EVC$ and $\minvc \in [|\VVC|]$.
		Then $\GVC[\EhVC]$ admits a vertex cover of size at most $\minvc$ if and only if $\GDS[\{u_e \mid e \in \EhVC\} \cup \UV]$ admits a dominating set of size at most $\minvc$.
	\end{claim}
	\begin{proof}\renewcommand{\qedsymbol}{\hfill (end of the proof of~\cref{clm:mindcvceq})~$\diamond$}
		Assume $\GVC[\EhVC]$ admits a vertex cover $\VpVC$ of size at most $\minvc$.
		Then we construct a dominating set $\VpDS \coloneqq \{u_v \mid v \in \VpVC\}$ of $\GDS[\{u_e \mid e \in \EVC\} \cup \UV]$.
		Clearly $|\VpDS| = |\VpVC|$.
		Every vertex in $\{u_e \mid e \in \EVC\} \cup \UV$ is also dominated: Because~$\VpVC$ is non-empty, set~$\VpDS$ must contain at least one vertex from $\UV$, and thus every vertex in $\UV$ is dominated.
		Because~$\VpVC$ is a vertex cover, for every $e \in \EVC$ there is $v \in e \cap \VpVC$. 
		Therefore $u_v \in \VpDS$ and vertex $u_e$ is dominated, as required.
		
		Now assume  $\GDS[\{u_e \mid e \in \EhVC\} \cup \UV]$ admits a dominating set $\VpDS$ of size at most $\minvc$.
		By \cref{clm:dsvertexprop}, we can assume that $\VpDS \subseteq \UV$.
Let us construct a vertex cover $\VpVC \coloneqq \{v \in \VVC \mid u_v \in \VpDS\}$.
		Clearly $|\VpVC| = |\VpDS| \leq \minvc$.
		Moreover, for every $e \in \EVC$ there must be a vertex $u_v \in \VpDS$ that dominates $u_e$.
		By construction, the corresponding vertex $v \in \VVC$ covers edge $e$, and we have that $v \in \VpVC$.
		Thus $\VpVC$ is a vertex cover.
	\end{proof}
	
	We now show the correctness of our reduction.
	
	$(\Longrightarrow)$: Assume that \probPVCCE-instance $I^{\VCsym}$ admits a core stable allocation $\allocVC$.
	We construct a \stable\ allocation $\allocDS$ of $I^{\DSsym}$ as follows: For every $\iVC\in \NVC$, $\allocelDS{\iDS} \coloneqq \allocelVC{\iVC}$ and for every $v \in \VVC$, $\allocelDS{\ivDS} \coloneqq 0$.

	We first show that $\GDS$ admits a dominating set of size at most $\sum_{\agent{} \in \NDS}\allocelDS{\agent{}} = \sum_{\agent{} \in \NVC}\allocelVC{i}$, and thus $\sum_{\agent{} \in \NDS}\allocelDS{\agent{}} \geq \util(\players)$.
	Let~$\VpVC$ be a minimum vertex cover of $\GVC$.
	Because \allocVC\ is an allocation, we have that $\sum_{\agent{} \in \NVC}\allocelVC{\agent{}} = \util(\NVC) = |\VpVC|$.
	By \cref{clm:mindcvceq}, graph $\GDS$ admits a dominating set of size at most $|\VpVC|$. Thus $\util(\NDS) \leq |\VpDS| =  \sum_{\agent{} \in \NDS}\allocelDS{\agent{}}$, as required.
	
	Next we show that for every $\psetDS \subseteq \NDS$, it holds that $\sum_{\agent{} \in \psetDS}\allocelDS{\agent{}} \leq \util(\psetDS)$.
Assume, towards a contradiction, that there is $\psetDS \subseteq \NDS$ such that  $\sum_{\agent{} \in \psetDS}\allocelDS{\agent{}} < \util(\psetDS)$.
	Let $\VpDS$ be a minimum dominating set of $\GindDS{\psetDS}$. 
	By \cref{clm:dsvertexprop}, there is a dominating set $\VppDS$ of $\psetDS \cup \Uag$ such that $|\VppDS| \leq |\VpDS|$.
	Let $\psetVC \coloneqq \{\iVC  \mid \iDS \in \psetDS\}$.
	By \cref{clm:mindcvceq}, graph $\GindVCspec{\psetVC}$ admits a vertex cover $\VpVC$ such that $|\VpVC| \leq |\VppDS|$.
	We have that \begin{align*}
		\sum_{\agent{} \in \psetVC}\allocelVC{\agent{}} &= \sum_{\agent{} \in \psetDS}\allocelDS{\agent{}} > \util(\psetDS) = |\VpDS| \\&\geq |\VppDS| \geq  |\VpVC|  \geq \util(\psetVC),
	\end{align*} a contradiction to \allocVC\ being core stable.
	
	$(\Longleftarrow)$:
	Assume $I^{\DSsym}$ admits a core stable allocation $\allocDS$. 
	We construct an allocation $\allocVC$ of $I^{\VCsym}$ as follows:
Let $f \colon \Uag \to \NVC$ be an arbitrary function such that for every $v \in \UV$ it holds that $f(\ivDS)$ owns an edge that is incident to $v$.  
In other words, the function $f$ assigns every vertex agent in $\Uag$ to some agent in $\NVC$ that has an edge that is incident to this vertex.
For every $\iVC \in \NVC$, let \[\allocelVC{\iVC} \coloneqq \allocelDS{\iDS} + \sum_{\ivDS \in f^{-1}(\iVC)}\allocelDS{\ivDS},\] i.e., every agent obtains the cost of the corresponding agent from $\NDS$ and the cost of any vertex agents from $\Uag$ that are assigned to it.

First we observe that $\sum_{\agent{} \in \NVC}\allocelVC{\agent{}} = \sum_{\agent{} \in \NDS}\allocelDS{\agent{}} \geq \util(\NDS)$.
Since $\allocDS$ is an allocation, a minimum dominating set of $\GDS$ is of size $\util(\NDS)$.
By \cref{clm:mindcvceq}, graph $\GVC$ must admit a vertex cover of size at most $\util(\NDS)$, and thus $\sum_{\iVC \in \NVC}\allocelVC{\iVC} \geq \util(\NVC)$.
	
Next we show that for every $\psetVC \subseteq \NVC$, it holds that  $\sum_{\agent{}\in \psetVC}\allocelVC{\agent{}} \leq \util(\psetVC)$.
Assume, towards a contradiction, that there is $\psetVC \subseteq \NVC$ such that  $\sum_{\agent{} \in \psetVC}\allocelVC{\agent{}} > \util(\psetVC)$.
Let~$\VpVC$ be a minimum vertex cover of $\GindVCspec{\psetVC}$.
Let $\psetDS \coloneqq \{\iDS \mid \iVC \in \psetVC\}$.
Let~$\VpDS$ be the minimum dominating set of $\GindDS{\psetDS \cup \UV}$.
By \cref{clm:dsvertexprop} we can assume that $\VpDS \subseteq \UV$ and by \cref{clm:mindcvceq} it holds that $|\VpDS| \leq |\VpVC|$.
We can also assume that every vertex in $\VpDS$ is adjacent to an edge vertex in $\bigcup_{\iVC \in \psetVC}U_{\iDS}$, as otherwise we could remove this vertex and obtain a smaller dominating set. %

Let $\UagS$ be the set of agents $\ivDS, v \in \VpVC$ such that~$v$ is incident to an edge in $\cup_{\agent{} \in \psetVC}\EVC_{\agent{}}$.
By construction and earlier reasoning we have that $\VpDS \subseteq \{u_v \mid v \in \VVC, \ivDS \in \UagS\}$.
Observe that by construction of \allocDS\ and the fact that $\allocelDS{\agent{}} \geq 0$ for every $\agent{} \in \NDS$
\begin{align}
		\sum_{\agent{} \in \psetDS \cup \UagS} \allocelDS{\agent{}}&=  \sum_{\iDS \in \psetDS}\allocelDS{\iDS} + \sum_{\agent{} \in \UagS}\allocelVC{\agent{}} \label{eq:vcdsbound1}\\
		& \geq \sum_{\iDS \in \psetDS}\allocelDS{\iDS} + \sum_{\iDS \in \psetDS}\sum_{v \in f^{-1}(\iVC)}\allocelDS{\ivDS} \nonumber\\&= \sum_{\agent{} \in \psetVC} \allocelVC{\agent{}}.\nonumber
	\end{align}
	
	Since $\VpDS$ is a dominating set of $\GDS[\psetDS \cup \Uag]$ and $\VpDS  \subseteq \{u_v \mid v \in \VVC, \ivDS \in \UagS\}$, set $\VpDS$ is also a dominating set of $\GDS[\psetDS \cup \UagS]$.
	We obtain that
	\begin{align}
		\util(\psetDS \cup \UagS) \leq |\VpDS| \leq |\VpVC| = \util(\psetVC).\label{eq:vcdsbound2}
	\end{align}
	
	By combining \cref{eq:vcdsbound1,eq:vcdsbound2} and $\sum_{\agent{} \in \psetVC}\allocelVC{\agent{}} > \util(\psetVC)$, we obtain that
	\begin{align*}
		\sum_{\agent{} \in \psetDS \cup \UagS} \allocelDS{\agent{}} \geq \sum_{\agent{} \in \psetVC} \allocelVC{\agent{}} > \util(\psetVC) \geq \util(\psetDS \cup \UagS),
	\end{align*}
	a contradiction to core stability of \allocDS.
	
This concludes the backward direction and hence the proof.
\end{proof}
}

Finally, we prove the conjecture of \citet{Velzen04} by showing that the unpartitioned variant \probDSCE\ is indeed \NP-hard. %
The correctness of the reduction is based on the characterization result of \citet{Deng2009COG} which  \citet{Velzen04} explicitly shows for \minDS; see \ifcr\else appendix \fi for more details.
The characterization result also implies \NP-containment.

\begin{restatable}[\appsymb, \cite{Velzen04}]{theorem}{thmDSdporgsize}\label{thm:DSdp:orgsize}
 \probDSCE\ is \NP-complete and hence, \probPDSCE\ is \NP-complete when $\Orgsize = 1$.
\end{restatable}

\newcommand{\varv}[1]{\ensuremath{y_{#1}}}

\appendixproofwithstatement{thm:DSdp:orgsize}{\thmDSdporgsize*}{

  As already mentioned, we will we use the following lemma from \citet{Velzen04}. %
To this end, define \myemph{\fmds} of a graph $G$ as the objective value of an optimal solution to the following linear program:
For every $v \in V$ we have a variable~$y_v \geq~0$.
\begin{align*}\tag{LP1} \label{fmdsLP}
\text{Minimize } &\sum_{v \in V} y_v \text{ subject to}\\
\forall~v \in V&\sum_{v' \in N_G(v)}y_{v'} + y_v \geq 1
\end{align*}
\begin{lemma}[\cite{Velzen04}, Theorem 4.2]\label{lem:velzen}
An instance $G = (V, E)$ of \probDSCE\ is a yes-instance if and only if the \fmds\ of $G$ is equal to the cardinality of a minimum dominating set of $G$.
\end{lemma}

We are ready to show \cref{thm:DSdp:orgsize}.
\begin{proof}
We use the standard reduction from \textsc{3SAT} to \DS.
Let $\varphi$ be a SAT-formula over the set of $\varnr{}$ variables \Vars.
We construct the input graph $G = (V,E)$ as follows:
	\begin{compactitem}[--]
		\item For each variable $\svar{}\in\Vars{}$, we add three vertices corresponding to the two literals, $\svar{}$ and $\snegvar{}$, and a dummy. For each clause $C\in\varphi$ we add the vertex $v_{C}$. 
		\item We add an edge between $v_\ell$ and $v_C$, if the literal $\ell$ appears in the clause $C$. We add an edge between each pair in $\{v_{\svar{}},v_{\snegvar{}},v_{\svar{}^d}\}$. 
	\end{compactitem} 
	
	Since $\Orgsize = 1$ (equivalently, the constructed instance is an instance of \probDSCE), every vertex is owned by its own agent.

	We obtain the following structural result:
\begin{claim}\label{clm:fracDSsize}
The \fmds\ of $G$ is $\varnr{}$.
\end{claim}
\begin{proof}\renewcommand{\qedsymbol}{\hfill (end of the proof of~\cref{clm:fracDSsize})~$\diamond$}
We first show that the \fmds\ of $G$ is at least \varnr{}.
Consider the linear program corresponding to $G$ as described in \ref{fmdsLP}.
For the ease of presentation, for every $x \in \Vars{}$, we denote the variable corresponding to $v_{\svar{}}$ by~\varv{\svar{}}, the variable corresponding to $v_{\snegvar{}}$ by~\varv{\snegvar{}}, and the variable corresponding to $v_{\svar{}^d}$ by~\varv{\svar{}^d}.
Similarly, for every clause $C \in \varphi$, we denote the variable corresponding to $v_C$ by~\varv{C}.

Observe that for every $\svar{} \in \Vars{}$, the constraint of $v_{\svar{}^d}$ states that $\varv{\svar{}} + \varv{\snegvar{}} + \varv{\svar{}^d} \geq 1$.
Thus $\sum_{v \in V}y_v \geq \sum_{\svar{} \in \Vars{}}(\varv{\svar{}} + \varv{\snegvar{}} + \varv{\svar{}^d}) \geq \varnr{}$.
		
Next we show that the linear program admits solution of size $\varnr{}$.
For every $\svar{} \in \Vars{}$, let $\varv{\svar{}} \coloneqq \frac{1}{2}$ and $\varv{\snegvar{}} \coloneqq \frac{1}{2}$.
We show this satisfies the solution:
For every $\svar{} \in \Vars{}$, the constraint of $v_{\svar{}^d}$ states that $\varv{\svar{}} + \varv{\snegvar{}} + \varv{\svar{}^d} \geq 1$. We have that $\frac{1}{2} + \frac{1}{2} \geq 1$, so the constraint is satisfied.
The constraint of $v_{\svar{}}$ states that $\varv{\svar{}} + \varv{\snegvar{}} + \varv{\svar{}^d} + \sum_{C \in \varphi, \svar{} \in C}\varv{C_j} \geq 1$.
Again $\frac{1}{2} + \frac{1}{2} \geq 1$.
We can show that the constraint of $v_{\snegvar{}}$ is satisfied analogously.
		
		For every $C$, we may assume that $C$ contains at least two literals.
		The constraint of $v_{C}$ is $\sum_{\ell \in C}\varv{\ell} \geq 1$.
		Observe that $\varv{\ell} = \frac{1}{2}$ for every $\ell \in C$.
		We have that $|C|\frac{1}{2} \geq 1$, and thus the constraint is satisfied.
		
		Thus the \fmds\ of $G$ is~\varnr{}. 
	\end{proof}
	
	$(\Longrightarrow)$: Assume $\varphi$ admits a satisfying assignment $\sigma$.
	By reasoning of the proof of \cref{clm:G1DOMSETSATUNSAT} from the proof of \cref{lem:verifDS}, graph~$G$ admits a dominating set of size \varnr{}: Observe that the process of constructing $G$ is the same as the process of constructing $G_1$, $G^{(1)}_2$, and~$G^{(2)}_2$ from the proof of \cref{lem:verifDS}, except for the lack of special vertex.
	As the dominating set constructed in the proof of \cref{clm:G1DOMSETSATUNSAT} does not contain a special vertex, we can obtain a dominating set of~$G$ by using this construction.
	
By \cref{clm:fracDSsize}, graph~$G$ admits a dominating set of size $\varnr{}$ and thus the cardinality of a minimum dominating set and the \fmds\ of $G$ are equal. By \cref{lem:velzen}, the instance admits a core stable allocation.

	$(\Longleftarrow)$:
	Assume $G$ admits a core stable allocation.
	Then by \cref{lem:velzen} the \fmds\ and the cardinality of a minimum dominating set of $G$ are equal, and by \cref{clm:fracDSsize}, graph~$G$ admits a dominating set $V'$ of size~$\varnr{}$.
	In the proof of \cref{clm:G1DOMSETSATUNSAT} from the proof of \cref{lem:verifDS} we construct a satisfying assignment $\sigma$ when $G_1$ admits a dominating set of size~$\varnr{1}$.
Since this dominating set does not contain the special vertex, we can use this construction to construct a satisfying assignment of $\varphi$ from the dominating set $V'$.
	This concludes the proof.
\end{proof}
}

\section{\ParMgame\ and \ParSTgame
}\label{sec:PMGPSTG}
\appendixsection{sec:PMGPSTG}

In this section we discuss the computational complexity of partitioned matching and spanning tree games.

For \minST, \citet{faigle1997complexity} show that the unpartitioned variant \STgameshort-\probCV\ is \coNP-complete. %
We observe that their approach for showing \coNP-containment can be adapted for our partitioned variant:
\begin{restatable}[\appsymb]{independentobservation}{proMSTconpc}\label{pro:MSTconpc}
\probPSTCV\ is contained in \coNP.
\end{restatable}

\appendixproofwithstatement{pro:MSTconpc}{\proMSTconpc*}{
\begin{proof}
Given an \allocation\ \alloc\ and blocking coalition $\pset$, we can compute the minimum weight spanning tree on subgraph induced by the vertices of the agents in \pset\ and verify whether it is less than $\util(\pset)$.
Thus $\pset$ is a polynomial-time verifiable witness for a no-instance, meaning that \probPSTCV\ is contained in \coNP.
\end{proof}
}

As for the existence question, \cref{cor:parSTalwaysY} implies that every \probPSTCE-instance is a yes instance and a \stable\ \allocation\ can be found in polynomial time. %

Next, we turn to \maxMatching.
\citet{Biro19} show that \probPMCE\ is \coNP-hard and it is polynomial-time solvable when $\Orgsize=2$.
We complement the hardness result by proving \coNP-containment by utilizing the ellipsoid method.

\begin{restatable}[\appsymb]{proposition}{lemPMGcoNPcontain}\label{lem:PMGcoNPcontain}
  \probPMCE\ is contained in \coNP.
\end{restatable}

\appendixproofwithstatement{lem:PMGcoNPcontain}{\lemPMGcoNPcontain*}{
\begin{proof}
  We show \coNP-containment using the ellipsoid method; see \cref{sec:ellipsoid} for more details.
We show a polynomial-time verifiable certificate for a no-instance.
Our proof proceeds in two parts: (1) Given a no-instance $G$ of \probPMCE, we show that a polynomial-space certificate exists, and (2) we show that given a certificate constructed in Part (1), we can verify that $G$ is a no-instance in polynomial time.  

\mypara{Part 1.}
Let $G$ be a no-instance of \probPMCE.
We show that a certificate for no-instance must exist, although we do not yet show how to verify it.
Observe that an \allocation\ \alloc\ is \stable\ if and only if it is satisfies the linear constraints described in \cref{subsec:coopgamesstable}.
Thus we can formulate \probPMCE\ as a linear program.

Let $Q$ be the corresponding separation oracle. 
Recall that $Q$ does not necessarily have to run in polynomial time, but since we can brute force all the possible blocking coalitions, it must exist.
Since $G$ does not have a \stable\ allocation, given an allocation $\alloc$,
oracle $Q$ either returns a blocking coalition $\pset \subseteq \players$,
or answers that $\alloc$ is not a \preimputation.
Let $\hat{Q}$ be the set that contains a pair $(\alloc, \pset)$ for every call to $Q$ over the course of solving \probPMCE\ on $G$, where (1) $\alloc$ is the allocation that is queried for $Q$ and (2) $\pset$ is a coalition that blocks $\alloc$, or $\pset = \players$ if $\alloc$ is not a \preimputation.
Recall that since $G$ does not admit a \stable\ \allocation, oracle $Q$ never returns that \alloc\ is in the core.
Since the ellipsoid method runs in time $T \cdot \operatorname{poly}|V|$ where $T$ is the running time of $Q$, oracle $Q$ is called polynomially many times over the execution of the ellipsoid method, and thus the size of $\hat{Q}$ is polynomial on the input size.

\textbf{Part 2}.
Assume we are given $\hat{Q}$ as a witness of a no-instance.
We can verify that $\hat{Q}$ only contains violated constraints in polynomial time:
Because we can compute a maximum matching in polynomial time~\cite{Edmonds65}, we can verify in polynomial time that for every $(\alloc, \pset) \in \hat{Q}$ conditions (1) and (2) are not satisfied.

Next, we solve \probPMCE\ on $G$ using $\hat{Q}$ as an oracle.
By construction, set $\hat{Q}$ contains a violated constraint for every $\alloc$ that is queried from $Q$ over the course of the execution of the ellipsoid method on $G$, and thus we can use $\hat{Q}$ as an oracle.
Since we can query $\hat{Q}$ in polynomial time, the ellipsoid method runs in polynomial time.
As the ellipsoid method and~$\hat{Q}$ are both correct, the ellipsoid method returns no if and only if $G$ does not admit a \stable\ \allocation, as required.
This verifies witness $\hat{Q}$ in polynomial-time.
\end{proof}
}

Finally, we consider parameterized complexity.
It is straightforward that both core verification and core existence are \FPT\ wrt.\ the number of agents $\nbOrg$, because the number of possible coalitions is bounded by $2^{\nbOrg}$, and their value or cost can be computed in polynomial time.

\begin{restatable}[\appsymb]{proposition}{thmMSTMVerifk}\label{thm:MSTMVerifk}
	\probPMCV\ and \probPSTCV\ are \FPT\ \wrt\ \nbOrg.
\end{restatable}

\appendixproofwithstatement{thm:MSTMVerifk}{\thmMSTMVerifk*}{
\begin{proof}
Recall that \allocation\ \alloc\ is \stable\ if and only if it satisfies  $\displaystyle\sum_{i \in \players} \allocel i = \posutil(\players)$ and  for every $\pset \subseteq \players$, it holds that $\displaystyle\sum_{i \in \pset}\allocel i \geq \posutil(\pset)$.
As both a minimum spanning tree and a maximum matching can be computed in polynomial time, we can verify these conditions in \FPT-time wrt.\ $\nbOrg$. This concludes the proof.
\end{proof}
}

\begin{restatable}[\appsymb]{proposition}{thmPMGCFkFPT}\label{thm:PMGCFkFPT}
	\probPMCE\ is \FPT\ \wrt\ $\nbOrg$.
\end{restatable}

\appendixproofwithstatement{thm:PMGCFkFPT}{\thmPMGCFkFPT*}{
\begin{proof}
Recall that \allocation\ \alloc\ is \stable\ if and only if it satisfies  $\displaystyle\sum_{i \in \players} \allocel i = \posutil(\players)$ and  for every $\pset \subseteq \players$, it holds that $\displaystyle\sum_{i \in \pset}\allocel i \geq \posutil(\pset)$.
If we consider every element of \alloc\ as a variable in a linear program,
this is clearly a linear program with $\nbOrg$ variables and $2^{\nbOrg}$ constraints.
Recall that as a maximum cardinality matching can be found in polynomial time~\cite{Edmonds65}, this linear program can be constructed in \FPT-time wrt.\ $\nbOrg$.
As its size is bounded by $\nbOrg$, it can also be solved in \FPT-time wrt.\ $\nbOrg$.
This concludes the proof.
\end{proof}
}

\section{Conclusion}\label{sec:conclude}

We establish a comprehensive computational complexity landscape for partitioned combinatorial optimization games (\PCOG{s}), a natural generalization of classical combinatorial optimization games (\COG{s}),
in which each agent owns a \emph{portion} of the input structure rather than a single element.
Our analysis of four fundamental optimization problems—\minVC, \minDS, \minST, and \maxMatching—reveals intriguing complexity patterns across core verification (\probCV) and core existence (\probCE) problems.

A natural next step is to pin down the algorithmic complexity of computing a \stable\ allocation for \PCOG{s}.
Note that the decision version, \probCE, is already $\thetaC$-complete for  \ParVCgame{s} and \ParDSgame{s} and \coNP-complete for \ParMgame{s}.
This implies that the two former problems are hard for the search complexity class~$\FPNPlog$,\footnote{$\FPNPlog$\cite{Selman94,BuhrmanT96,BuhrmanKT98} is a complexity class that contains all search problems where a desirable solution (if it exists) can be found in polynomial time with logarithmically many queries to an $\NP$-oracle.}
but it is unclear whether this is tight since it is not clear how to find an optimal solution via logarithmically many calls to \NP-oracle.
A complementary line of work is to identify additional parameters that yield fixed-parameter tractability. %
Finally, our framework also invites the study of other optimization objectives (e.g., flows, cuts, facility-location) and of stability notions beyond the core. 

We conclude with a few preliminary results on individually rational (IR) allocations—pre-imputations where no single agent can obtain strictly smaller utility by acting alone.
First, it is straightforward that both verification and existence of IR 
for \minST{} and \maxMatching{} are polynomial-time solvable. 
Second, we observe that for the four studied optimization goals, an IR allocation always exists since the corresponding games are super-additive and super-additive games always admit an  IR \allocation; we provide a proof in the \ifcr\else appendix \fi for the sake of completeness.
However, finding an IR allocation is \NP-hard whenever the underlying optimization problem is \NP-hard, as finding a pre-imputation becomes \NP-hard.
Finally, our \DP-completeness results (\cref{thm:VCverif,lem:verifDS}) for \PCOG-\probCV{}
extend to IR verification through a two-fold observation: (1) the hardness holds with only one agent in which case IR is equivalent to core stability and (2) the \DP-containment proof adapts directly to IR.

\begin{ack}
This work and all the authors have been funded by the Vienna Science and Technology Fund (WWTF)~[10.47379/VRG18012].
\end{ack}

\bibliography{bib}

\begin{thebibliography}{44}
\providecommand{\natexlab}[1]{#1}
\providecommand{\url}[1]{\texttt{#1}}
\expandafter\ifx\csname urlstyle\endcsname\relax
  \providecommand{\doi}[1]{doi: #1}\else
  \providecommand{\doi}{doi: \begingroup \urlstyle{rm}\Url}\fi

\bibitem[Arora and Barak(2009)]{arora2009computational}
S.~Arora and B.~Barak.
\newblock \emph{Computational complexity: a modern approach}.
\newblock Cambridge University Press, 2009.

\bibitem[Berman et~al.(2003)Berman, Karpinski, and Scott]{BKS-2bal3sat-2003}
P.~Berman, M.~Karpinski, and A.~D. Scott.
\newblock Approximation hardness of short symmetric instances of {MAX-3SAT}.
\newblock Technical Report TR03-049, Electronic Colloquium on Computational
  Complexity, 2003.

\bibitem[Bird(1976)]{Bird76}
C.~G. Bird.
\newblock On cost allocation for a spanning tree: A game theoretic approach.
\newblock \emph{Networks}, 6\penalty0 (4):\penalty0 335--350, 1976.

\bibitem[Bir{\'o} et~al.(2018)Bir{\'o}, Kern, Paulusma, and
  Wojuteczky]{biro2018stable}
P.~Bir{\'o}, W.~Kern, D.~Paulusma, and P.~Wojuteczky.
\newblock The stable fixtures problem with payments.
\newblock \emph{Games and Economic Behavior}, 108:\penalty0 245--268, 2018.

\bibitem[Bir{\'{o}} et~al.(2019)Bir{\'{o}}, Kern, P{\'{a}}lv{\"{o}}lgyi, and
  Paulusma]{Biro19}
P.~Bir{\'{o}}, W.~Kern, D.~P{\'{a}}lv{\"{o}}lgyi, and D.~Paulusma.
\newblock Generalized matching games for international kidney exchange.
\newblock In \emph{Proceedings of the 18th International Conference on
  Autonomous Agents and MultiAgent Systems (AAMAS 2019)}, pages 413--421, 2019.

\bibitem[Borm et~al.(2001)Borm, Hamers, and Hendrickx]{BHH2001ORgames}
P.~Borm, H.~Hamers, and R.~Hendrickx.
\newblock Operations research games: {A} survey.
\newblock \emph{TOP: {A}n Official Journal of the Spanish Society of Statistics
  and Operations Research,}, 9:\penalty0 139--199, 2001.

\bibitem[Buhrman and Thierauf(1996)]{BuhrmanT96}
H.~Buhrman and T.~Thierauf.
\newblock The complexity of generating and checking proofs of membership.
\newblock In \emph{Proceedings of 13th Annual Symposium on Theoretical Aspects
  of Computer Science ({STACS}'96)}, pages 75--86, 1996.

\bibitem[Buhrman et~al.(1998)Buhrman, Kadin, and Thierauf]{BuhrmanKT98}
H.~Buhrman, J.~Kadin, and T.~Thierauf.
\newblock Functions computable with nonadaptive queries to {NP}.
\newblock \emph{Theory of Computing Systems}, 31\penalty0 (1):\penalty0 77--92,
  1998.

\bibitem[Chen et~al.(2025)Chen, Hatschka, and Simola]{fullpaper}
J.~Chen, C.~Hatschka, and S.~Simola.
\newblock Partitioned combinatorial optimization games.
\newblock Technical report, arXiv, 2025.
\newblock Full version of this paper.

\bibitem[Chv{\'a}tal(1983)]{chvatal83}
V.~Chv{\'a}tal.
\newblock \emph{Linear Programming}.
\newblock Series of books in the mathematical sciences. W. H. Freeman, 1983.

\bibitem[Claus and Kleitman(1973)]{claus1973cost}
A.~Claus and D.~Kleitman.
\newblock Cost allocation for a spanning tree.
\newblock \emph{Networks}, 3\penalty0 (4):\penalty0 289--304, 1973.

\bibitem[Cohen et~al.(2011)Cohen, Cordeiro, Trystram, and
  Wagner]{cohen2011multi}
J.~Cohen, D.~Cordeiro, D.~Trystram, and F.~Wagner.
\newblock Multi-organization scheduling approximation algorithms.
\newblock \emph{Concurrency and computation: Practice and experience},
  23\penalty0 (17):\penalty0 2220--2234, 2011.

\bibitem[Cohen et~al.(2014)Cohen, Cordeiro, and Raphael]{cohen2014energy}
J.~Cohen, D.~Cordeiro, and P.~L.~F. Raphael.
\newblock Energy-aware multi-organization scheduling problem.
\newblock In \emph{Proceedings of the 20th International Conference on Parallel
  and Distributed Computing (Euro-Par 2014)}, pages 186--197. Springer, 2014.

\bibitem[Cygan et~al.(2015)Cygan, Fomin, Kowalik, Lokshtanov, Marx, Pilipczuk,
  Pilipczuk, and Saurabh]{CyFoKoLoMaPiPiSa2015}
M.~Cygan, F.~V. Fomin, L.~Kowalik, D.~Lokshtanov, D.~Marx, M.~Pilipczuk,
  M.~Pilipczuk, and S.~Saurabh.
\newblock \emph{Parameterized Algorithms}.
\newblock Springer, 2015.

\bibitem[Deng(2009)]{Deng2009COG}
X.~Deng.
\newblock Combinatorial optimization games.
\newblock In \emph{Encyclopedia of Optimization, Second Edition}, pages
  387--391. Springer, 2009.

\bibitem[Deng et~al.(1999)Deng, Ibaraki, and Nagamochi]{deng1999algorithmic}
X.~Deng, T.~Ibaraki, and H.~Nagamochi.
\newblock Algorithmic aspects of the core of combinatorial optimization games.
\newblock \emph{Mathematics of Operations Research}, 24\penalty0 (3):\penalty0
  751--766, 1999.

\bibitem[Durand and Pascual(2021)]{durand2021efficiency}
M.~Durand and F.~Pascual.
\newblock Efficiency and equity in the multi organization scheduling problem.
\newblock \emph{Theoretical Computer Science}, 864:\penalty0 103--117, 2021.

\bibitem[Edmonds(1965)]{Edmonds65}
J.~Edmonds.
\newblock Paths, trees, and flowers.
\newblock \emph{Canadian Journal of Mathematics}, 17:\penalty0 449–467, 1965.

\bibitem[Est{\'e}vez-Fern{\'a}ndez and Hamers(2020)]{estevez2020chinese}
A.~Est{\'e}vez-Fern{\'a}ndez and H.~Hamers.
\newblock Chinese postman games with multi-located players.
\newblock \emph{European Journal of Operational Research}, 285\penalty0
  (2):\penalty0 458--469, 2020.

\bibitem[Faigle and Kern(1993)]{FK93CCG}
U.~Faigle and W.~Kern.
\newblock On some approximately balanced combinatorial cooperative games.
\newblock \emph{{ZOR} -- Methods and Models of Operation Research}, 38\penalty0
  (2):\penalty0 141--152, 1993.

\bibitem[Faigle et~al.(1997)Faigle, Kern, Fekete, and
  Hochst{\"a}ttler]{faigle1997complexity}
U.~Faigle, W.~Kern, S.~P. Fekete, and W.~Hochst{\"a}ttler.
\newblock On the complexity of testing membership in the core of min-cost
  spanning tree games.
\newblock \emph{International Journal of Game Theory}, 26:\penalty0 361--366,
  1997.

\bibitem[Fang and Kim(2005)]{Fang05}
Q.~Fang and H.~K. Kim.
\newblock A note on balancedness of dominating set games.
\newblock \emph{Journal of Combinatorial Optimization}, 10\penalty0
  (4):\penalty0 303--310, 2005.

\bibitem[Fang et~al.(2008)Fang, Kong, and Zhao]{Fang08}
Q.~Fang, L.~Kong, and J.~Zhao.
\newblock Core stability of vertex cover games.
\newblock \emph{Internet Mathematics}, 5\penalty0 (4):\penalty0 383--394, 2008.

\bibitem[Garey and Johnson(1979)]{GJ79}
M.~R. Garey and D.~S. Johnson.
\newblock \emph{Computers and Intractability: {A} Guide to the Theory of
  NP-completeness}.
\newblock Mathematical Sciences Series. W. H. Freeman and Company, 1979.

\bibitem[Gourv{\`{e}}s et~al.(2012)Gourv{\`{e}}s, Monnot, and
  Pascual]{Gourves12}
L.~Gourv{\`{e}}s, J.~Monnot, and F.~Pascual.
\newblock Cooperation in multiorganization matching.
\newblock \emph{Algorithmic Operations Research}, 7\penalty0 (2), 2012.

\bibitem[Granot and Huberman(1981)]{Granot81}
D.~Granot and G.~Huberman.
\newblock Minimum cost spanning tree games.
\newblock \emph{Mathematical Programming}, 21\penalty0 (1):\penalty0 1--18,
  1981.

\bibitem[Gr{\"{o}}tschel et~al.(1988)Gr{\"{o}}tschel, Lov{\'{a}}sz, and
  Schrijver]{GLS1988}
M.~Gr{\"{o}}tschel, L.~Lov{\'{a}}sz, and A.~Schrijver.
\newblock \emph{Geometric Algorithms and Combinatorial Optimization}, volume~2
  of \emph{Algorithms and Combinatorics}.
\newblock Springer, 1988.

\bibitem[Hafezalkotob and Naseri(2016)]{hafezalkotob2016}
A.~Hafezalkotob and F.~Naseri.
\newblock Cooperative network flow problem with pricing decisions and
  allocation of benefits: A game theory approach.
\newblock \emph{Journal of Industrial and Systems Engineering}, 9, Special
  issue on supply chain:\penalty0 73--87, 2016.

\bibitem[Hamers et~al.(1999)Hamers, Borm, van~de Leensel, and
  Tijs]{hamers1999cost}
H.~Hamers, P.~Borm, R.~van~de Leensel, and S.~Tijs.
\newblock Cost allocation in the chinese postman problem.
\newblock \emph{European Journal of Operational Research}, 118\penalty0
  (1):\penalty0 153--163, 1999.

\bibitem[Hemaspaandra et~al.(2005)Hemaspaandra, Spakowski, and
  Vogel]{HSV05Kemeny}
E.~Hemaspaandra, H.~Spakowski, and J.~Vogel.
\newblock The complexity of {K}emeny elections.
\newblock \emph{Theoretical Computer Science}, 349\penalty0 (3):\penalty0
  382--391, 2005.

\bibitem[Le et~al.(2016)Le, Nguyen, and Bektas]{Le16}
P.~H. Le, T.~Nguyen, and T.~Bektas.
\newblock Generalized minimum spanning tree games.
\newblock \emph{EURO Journal on Computational Optimization}, 4\penalty0
  (2):\penalty0 167--188, 2016.

\bibitem[Miquel et~al.(2006)Miquel, van Velzen, Hamers, and
  Norde]{miquel2006fixed}
S.~Miquel, B.~van Velzen, H.~Hamers, and H.~Norde.
\newblock Fixed tree games with multilocated players.
\newblock \emph{Networks: An International Journal}, 47\penalty0 (2):\penalty0
  93--101, 2006.

\bibitem[Nguyen et~al.(2014)Nguyen, Nguyen, Roos, and Rothe]{NNRR2014}
N.~Nguyen, T.~T. Nguyen, M.~Roos, and J.~Rothe.
\newblock Computational complexity and approximability of social welfare
  optimization in multiagent resource allocation.
\newblock \emph{Autonomous Agents and Multi-Agent Systems}, 28\penalty0
  (2):\penalty0 256--289, 2014.

\bibitem[Papadimitriou(1994)]{papadimitrioubook}
C.~H. Papadimitriou.
\newblock \emph{Computational complexity}.
\newblock Addison-Wesley, 1994.

\bibitem[Papadimitriou and Yannakakis(1984)]{PY84}
C.~H. Papadimitriou and M.~Yannakakis.
\newblock The complexity of facets (and some facets of complexity).
\newblock \emph{Journal of Computer and System Sciences}, 28\penalty0
  (2):\penalty0 244--259, 1984.

\bibitem[Pascual et~al.(2009)Pascual, Rzadca, and
  Trystram]{pascual2009introduction}
F.~Pascual, K.~Rzadca, and D.~Trystram.
\newblock Cooperation in multi-organization scheduling.
\newblock \emph{Concurrency and Computation: Practice and Experience},
  21\penalty0 (7):\penalty0 905--921, 2009.

\bibitem[Potters et~al.(1992)Potters, Curiel, and Tijs]{potters1992traveling}
J.~A. Potters, I.~J. Curiel, and S.~H. Tijs.
\newblock Traveling salesman games.
\newblock \emph{Mathematical Programming}, 53:\penalty0 199--211, 1992.

\bibitem[Prim(1957)]{Prim}
R.~C. Prim.
\newblock Shortest connection networks and some generalizations.
\newblock \emph{The Bell System Technical Journal}, 36\penalty0 (6):\penalty0
  1389--1401, 1957.

\bibitem[Rzadca(2007)]{Rzadca2007MultiScheduling}
K.~Rzadca.
\newblock Scheduling in multi-organization grids: {M}easuring the inefficiency
  of decentralization.
\newblock In \emph{Proceedings of the 7th International Conference on Parallel
  Processing and Applied Mathematics (PPAM'07)}, volume 4967, pages 1048--1058,
  2007.

\bibitem[Schrijver(1999)]{Schrijver99book}
A.~Schrijver.
\newblock \emph{Theory of linear and integer programming}.
\newblock Wiley-Interscience series in discrete mathematics and optimization.
  Wiley, 1999.

\bibitem[Selman(1994)]{Selman94}
A.~L. Selman.
\newblock A taxonomy of complexity classes of functions.
\newblock \emph{Journal of Computer and System Sciences}, 48\penalty0
  (2):\penalty0 357--381, 1994.

\bibitem[Shapley(1967)]{Shapley1967}
L.~S. Shapley.
\newblock On balanced sets and cores.
\newblock \emph{Naval Research Logistics Quarterly}, 14\penalty0 (4):\penalty0
  453--460, 1967.

\bibitem[van Velzen(2004)]{Velzen04}
B.~van Velzen.
\newblock Dominating set games.
\newblock \emph{Operation Research Letter}, 32\penalty0 (6):\penalty0 565--573,
  2004.

\bibitem[Xiao et~al.(2021)Xiao, Wang, and Fang]{Xiao21}
H.~Xiao, Y.~Wang, and Q.~Fang.
\newblock On the convexity of independent set games.
\newblock \emph{Discrete Applied Mathematics}, 291:\penalty0 271--276, 2021.

\end{thebibliography}

\ifcr
\else
  \clearpage

  \begin{appendices}
  \appendixtext
  \end{appendices}
\fi

\end{document}